\documentclass{jfm}
\usepackage{amsmath}
\usepackage{amssymb}
\usepackage{yhmath}
\usepackage[english]{babel}
\usepackage{epstopdf, epsfig}
\usepackage{comment}
\usepackage{color}
\usepackage{setspace}
\usepackage{siunitx}
\usepackage{booktabs}
\usepackage{units}
\usepackage[normalem]{ulem}
\usepackage{comment}
\usepackage[makeroom]{cancel}
\usepackage{xspace}
\usepackage{float}
\usepackage{adjustbox}
\usepackage{makecell}
\usepackage{mathtools}
\usepackage{bbold}
\usepackage{siunitx}
\usepackage{xcolor}
\usepackage{graphicx}
\usepackage{amsmath}

\renewcommand{\theequation}{\thesection.\arabic{equation}}
\usepackage{mathrsfs}
\newcommand{\Peclet}{P\'eclet\xspace}
\newcommand{\Damkohler}{Damk{\"o}hler\xspace}

\shorttitle{Surfactant-contaminated superhydrophobic channels}

\shortauthor{S. D. Tomlinson and others}

\title{Laminar drag reduction in surfactant- contaminated superhydrophobic channels}

\author{Samuel D. Tomlinson\aff{1}
  \corresp{\email{samuel.tomlinson@manchester.ac.uk}},
  Fr\'{e}d\'{e}ric Gibou\aff{2},
  Paolo Luzzatto-Fegiz\aff{2},
  Fernando Temprano-Coleto\aff{3},
  Oliver E. Jensen\aff{1}
  \and Julien R. Landel\aff{1}}

\affiliation{
\aff{1}Department of Mathematics, University of Manchester, Oxford Road, Manchester M13 9PL, UK
\aff{2}Department of Mechanical Engineering, University of California, Santa Barbara, CA 93106, USA
\aff{3}Andlinger Center for Energy and the Environment, Princeton University, Princeton, NJ 08544, USA
}

\begin{document}

\maketitle

\begin{abstract}

While superhydrophobic surfaces (SHSs) show promise for drag reduction applications, their performance can be compromised by traces of surfactant, which generate Marangoni stresses that increase drag.
This question is addressed for soluble surfactant in a three-dimensional laminar channel flow, with periodic SHSs made of long finite-length longitudinal grooves located on both walls. 
We assume that bulk diffusion is sufficiently strong for cross-channel concentration gradients to be small.
Exploiting long-wave theory and accounting for the difference between the rapid transverse and slower longitudinal Marangoni flows, we derive a one-dimensional model for surfactant transport from the full three-dimensional transport equations. 
Our one-dimensional model allows us to predict the drag reduction and surfactant distribution across the parameter space. 
The system exhibits multiple regimes, involving competition between Marangoni effects, bulk and interfacial diffusion, bulk and interfacial advection, shear dispersion and surfactant exchange between the bulk and the interface. 
We map out asymptotic regions in the high-dimensional parameter space, and derive explicit closed-form approximations of the drag reduction, without any fitting or empirical parameters. 
The physics underpinning the drag reduction effect and the negative impact of surfactant is discussed through analysis of the velocity field and surfactant concentrations, which show both uniform and non-uniform stress distributions. 
Our theoretical predictions of the drag reduction compare well with results from the literature solving numerically the full three-dimensional transport problem.
Our atlas of maps provides a comprehensive analytical guide for designing surfactant-contaminated channels with SHSs, to maximise the drag reduction in applications.

\end{abstract}


\section{Introduction}\label{sec:introduction}

When a fluid flows over a superhydrophobic surface (SHS), surface chemistry and microscopic roughness combine to entrap an array of microscopic gas pockets at the SHS. 
In the non-wetted or Cassie--Baxter state \cite[see, e.g., the reviews of ][]{rothstein2010slip,lee2016superhydrophobic,park2021superhydrophobic}, the reduced liquid--solid contact area at the SHS reduces viscous drag at the surface when compared to a solid wall, which leads to an overall decrease in drag.
Owing to these drag-reducing capabilities, SHSs have been considered in a number of applications, such as low-Reynolds-number laminar flows \cite[e.g.][]{ou2005direct,sbragaglia2007note,rothstein2010slip,schonecker2014influence,lee2016superhydrophobic,landel2020theory}, high-Reynolds-number turbulent flows \cite[e.g.][]{park2014superhydrophobic,turk2014turbulent,golovin2016bioinspired,seo2018effect,rastegari2019drag,park2021superhydrophobic} and the thermal management of microelectronics \cite[e.g.][]{baier2010thermocapillary,cheng2015numerical,lam2015analysis,kirk2020thermocapillary}. 

Although the physical mechanism behind laminar drag reduction is well understood, SHSs seldom achieve the high drag-reduction performance predicted by theory \citep{lee2016superhydrophobic,park2021superhydrophobic}. 
A number of practical difficulties can explain the lower performance measured experimentally, such as displacement of the liquid--gas interfaces \citep{biben2008wetting, ng2009stokes}, contact-angle effects \citep{sbragaglia2007note, teo2010flow}, viscous drag from the gas phase \citep{schonecker2014influence, game2017physical}, as well as surfactant-induced Marangoni stresses \citep{peaudecerf2017traces,landel2020theory}.
In this study, we investigate in detail the impact of soluble surfactant on the drag reduction of SHSs in laminar channel flows.
Surfactants can be transported by a liquid, adsorb at liquid--gas interfaces and lower the surface tension of these interfaces \citep{manikantan2020surfactant}.
Transported by the flow along the liquid--gas interface, surfactants can accumulate at stagnation points (liquid--gas--solid contact lines), inducing an adverse Marangoni stress at the interface which increases the drag, thereby negating the drag-reducing effect of the nominally shear-free interface. 

The impact of surfactants on SHS drag reduction was suggested by experimental studies in the last decade, which reported only a modest drag reduction when compared to solid walls.
For example, \citet{kim2012pressure} used microscopy to analyse the location of the liquid--gas interface in a channel flow over transverse ridges.
Combined with flow-rate measurements, they found that the frictional properties of the channel were insensitive to the degree of microtexture wetting, and hence closely resembled solid walls.
Likewise, using fluorescence microscopy and passive tracers, \citet{bolognesi2014evidence} studied the  flow field and shape of the liquid--gas interfaces for a channel flow over longitudinal ridges.
They found non-zero shear stresses at the interfaces, whereas most theoretical and numerical studies normally employ shear-free conditions to model SHSs \citep{rothstein2010slip}.

Following these observations, several studies have explored the effects of surfactants on SHS drag reduction.
\citet{schaffel2016local} combined fluorescence correlation spectroscopy with numerical simulations to examine the effective slip length in a channel flow with cylindrical pillars.
They found that the effective slip length reduced when compared to simulations of a surfactant-free channel. 
\citet{peaudecerf2017traces} performed experiments and numerical simulations in a channel with longitudinal ridges, demonstrating that low levels of surfactant could yield large changes in interfacial conditions. 
They showed that liquid--gas interfaces could be rendered no-slip at surfactant concentrations well below typical environmental values.
They used channel-flow experiments where the imposed pressure gradient was removed abruptly to show that the observed reverse flow was only possible due to adverse out-of-equilibrium surfactant gradients.
\citet{song2018effect} performed experiments for finite longitudinal ridges and effectively infinite concentric, annular ridges.
For finite longitudinal ridges, surface tension gradients arose due to the presence of downstream transverse contact lines, increasing the drag when compared to infinite concentric ridges, where there were no stagnation points.

The experiments of \citet{peaudecerf2017traces} and \citet{song2018effect} were conducted in nominally ``clean'' channels without added surfactant.
\citet{schaffel2016local} also performed experiments with added surfactant, finding barely measurable increases in drag relative to their nominally ``clean'' experiments. 
As noted by \citet{peaudecerf2017traces}, the counterintuitive result reported by \citet{schaffel2016local} was most likely due to the fact that traces of surfactants were already present in their experiment. 
Surfactant traces are inherently present in most engineered systems, due to general manufacturing conditions, materials, and the high surface-to-volume ratio of the SHS textures. 
For instance, microfluidic experimental devices are often made of polydimethylsiloxane (PDMS), which is known to lead to surfactant effects \citep{hourlier2018extraction}.
In nature, surfactant traces have been measured in sea water \citep{pereira2018reduced}, rivers, estuaries and fog \citep{lewis1991chronic,facchini2000surface}.

Due to the potentially strong adverse effect of surfactants on the drag reduction performance of SHSs, it is crucial to model their impact.
Theoretical models can explain experimental observations, and also provide predictions for the design of SHSs where surfactant effects are mitigated. 
However, as discussed by \citet{landel2020theory}, the theoretical modelling of flows inclusive of surfactant over SHSs is complex.
To make progress, \citet{landel2020theory} assumed that the surfactant concentration is small. 
This assumption is consistent with normal environmental conditions where surfactants are found in trace amounts \citep{peaudecerf2017traces}.
\citet{landel2020theory} constructed a scaling theory to model the slip and drag in steady two-dimensional (2D) pressure-driven channel flows bounded by SHSs, made of long transverse gratings, in the low-Reynolds-number flow regime.
They also performed finite-element 2D numerical simulations to compute the constants in the scaling theory, and validated their scaling predictions. 
Their results showed that the slip length and drag reduction are affected by surfactants across a broad range of the parameter space. 
They focused on parameter combinations associated with microfluidic applications; it remains of interest to perform a comprehensive asymptotic analysis of the parameter space.
Moreover, their theory did not consider non-uniform interfacial surfactant distributions associated with the ``stagnant cap'' regime, where the upstream region of the interface has a weak surfactant gradient and is almost shear-free, whilst the downstream region of the interface has a strong surfactant gradient and is effectively no-slip \cite[see, e.g.,][which describe similar behaviour for air bubbles rising in surfactant-contaminated water] {bond1928lxxxii, frumkin1947effect, levich1962physicochemical, he1991size}.
In this regime, the advection of surfactant at the interface dominates relative to surface diffusion and bulk--surface exchange.

The scaling theory of \citet{landel2020theory} was extended to three-dimensional (3D) low-Reynolds number channel flows with long but finite longitudinal gratings by \citet{temprano2023single}. 
They used a long-wave limit and assumed low surfactant concentrations to show that the slip velocity and the slip length scale in a similar fashion as in 2D flows. 
In the limit of large \Damkohler numbers, as found for common surfactants, and sufficiently small bulk \Peclet numbers, they predicted that significant slip can be achieved provided that the grating length is longer than both a modified depletion length and a mobilization length. 
The modified depletion length depends on surfactant properties and the height of the channel, and is generally small, of the order of 1 mm, for small-scale applications. 
The mobilization length depends on the normalised surfactant concentration, Marangoni number, \Damkohler number and Biot number. 
Since the mobilization length can be much longer for typical small-scale applications, they concluded that the mobilization length alone controls the slip in most small-scale applications. 
If the grating length is smaller than the mobilization length, they showed that the slip velocity increases with the square of the grating length, which is in agreement with their numerical simulations and experiments, as well as with the experimental results from \citet{peaudecerf2017traces} and \citet{song2018effect}.

The contamination of SHS channels with surfactant may be simplified if one considers only the effect of insoluble surfactant at the liquid--gas interface, neglecting the exchange with bulk surfactant which remains at some background concentration.
\citet{baier2021influence} found analytical expressions for the velocity field and interfacial surfactant distribution in a 2D channel flow over transverse grooves in an advection-dominated regime (i.e., large interfacial \Peclet numbers), under the assumption that the surface tension depends linearly on the surface concentration. 
Under the assumption of insoluble surfactant, drag is not reduced by increasing streamwise groove length, in contrast with findings from experiments and soluble-surfactant models, where slip increases with the square of the groove length \citep{peaudecerf2017traces,landel2020theory,temprano2023single}. 
For periodic transverse grooves, \citet{baier2021influence} determined the effective slip length, which is strongly dependent on the Marangoni number for large gas fractions, decreasing rapidly as the Marangoni number increases from zero. 
\citet{mayer2022superhydrophobic} also considered surface immobilisation due to insoluble surfactant in a shear flow over periodic transverse SHSs; however, unlike \citet{baier2021influence}, the authors varied the surfactant load, interfacial \Peclet and Marangoni number, using a non-linear equation of state. 
They combined asymptotic theory with numerical solutions to identify two distinct mechanisms behind surface immobilisation. 
The first is the previously discussed stagnant cap mechanism.
The second immobilisation mechanism emerges when there is a region of near-maximal surfactant concentration close to the downstream stagnation point.
In contrast to the stagnant cap mechanism, surfactant gradients need not be large for appreciable Marangoni stresses to develop.
The near-maximal surfactant concentration mechanism is captured using a non-linear equation of state and thus can appear outside of the advection-dominated region, i.e. for small \Peclet and Marangoni numbers, provided there is sufficient surfactant present at the interface.

Liquid-infused surfaces (LISs) offer an alternative to SHSs where the SHS pockets are filled with a lubricating immiscible fluid instead of gas \citep{wong2011bioinspired,wexler2015shear}.
They can self-repair and are more robust than SHSs if properly designed. However, the drag reduction of LISs  decreases as the viscosity ratio between internal and external fluids increases \citep{schonecker2014influence}.
Furthermore, recent numerical simulations by \citet{sundin2022slip} have indicated that LISs may be more susceptible to surfactant effects than SHSs.
\citet{sundin2022slip} extended the theory introduced in \citet{landel2020theory} to account for surfactants in a 2D shear flow, predicting the critical surfactant concentration for the slip to be reduced appreciably: $\hat{C}_c =4\times10^{-4}\,$mol/m$^3$ for water--air SHSs and $\hat{C}_c = 5\times10^{-5}\,$mol/m$^3$ for water--dodecane LISs.
For low applied shear stresses, \citet{sundin2022slip} found that the distribution of surfactant at the interface is approximately uniform and a scaling theory was used to derive an expression for the slip length.
For high applied shear stresses, surfactant accumulates at the downstream stagnation point and forms a stagnant cap. 
\citet{sundin2022slip} employed numerical simulations to find that the stagnant cap regime only exists below a particular bulk concentration (above which the scaling theory once again becomes valid).
Considering a typical surfactant, e.g., sodium dodecyl sulfate (SDS), the critical bulk concentration is shown to be proportional to the inverse Marangoni number. 

In this paper, we show that a one-dimensional (1D) asymptotic theory, derived as coupled nonlinear ordinary differential equations (ODEs) from the full 3D transport problem, can capture the impact of surfactant in 3D channels bounded by periodic SHSs made of long finite-length longitudinal grooves.
The main assumption behind our theory is that bulk diffusion is strong enough to suppress cross-channel concentration gradients. 
We thereby sidestep the need for introducing empirical or fitting coefficients associated with scaling analyses, as done in previous studies \citep{landel2020theory,sundin2021roughness,temprano2023single}.
This allows us to address non-uniform shear stresses at the liquid--gas interface. 
By mapping the drag reduction across a large part of the high-dimensional parameter space, we identify a multitude of asymptotic regions and their boundaries, unexplored by previous studies.
Explicit closed-form asymptotic solutions predict the drag reduction in all the regions of the parameter space studied, offering analytical predictions for practical use where numerical simulation of the fluid and surfactant equations are computationally expensive.
By addressing the role of shear dispersion, the theory developed here constitutes also a stepping stone towards a wider class of laminar flows with weaker cross-channel diffusion, where the bulk concentration field varies in three-dimensions and must be resolved numerically.

The paper is arranged as follows. 
In \S\ref{sec:formulation}, the full 3D transport problem is formulated in terms of nine dimensionless parameters. 
In \S\ref{sec:model}, an asymptotic model is derived for the flow and surfactant transport; cross-channel integration reduces the number of independent dimensionless parameters to six.
In \S\ref{sec:results}, key results are presented for the drag reduction using 2D maps that illustrate the structure of the parameter space; the underlying physics is described using the surfactant distribution and the 3D velocity field of the channel flow.
In \S\ref{sec:discussion}, the implications and extensions of this study are discussed; a table of five dimensionless groups that control drag and the negative impact of surfactant is presented, expressing these dimensionless groups in terms of the dimensional parameters of the problem. 

\section{Formulation} \label{sec:formulation}

\subsection{Governing equations}

\begin{figure}
    \centering
    (\textit{a}) \hfill \hspace{.8cm} (\textit{b}) \hfill \hfill \hfill \\
    \includegraphics[width=.99\textwidth]{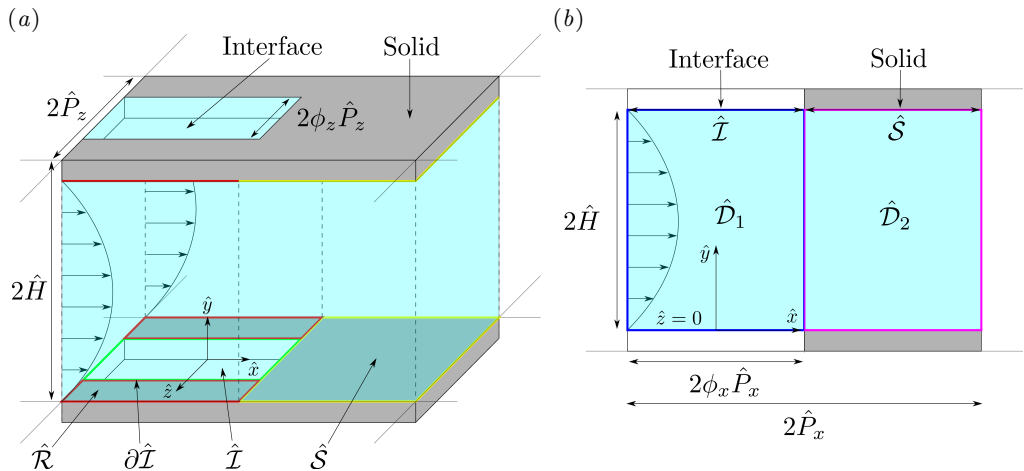}
    \caption{(\textit{a}) Schematic depicting a plane periodic streamwise channel flow (illustrated by the array of arrows) of a liquid transporting a soluble surfactant. The origin of the Cartesian coordinate system, $\hat{\boldsymbol{x}} = \boldsymbol{0}$, is located in the middle of the bottom interface. The channel height in the wall-normal direction is $2 \hat{H}$, the transverse gas fraction, $\phi_z$, and the transverse period, $2\hat{P}_z$. 
    On the top and bottom SHSs are no-slip ridges, $\hat{\mathcal{R}}$ (outlined in red), solid surfaces, $\hat{\mathcal{S}}$ (yellow), and liquid--gas interfaces, $\hat{\mathcal{I}}$ (green) onto which surfactants can adsorb and desorb, modifying the interfacial stress through the Marangoni effect.
    (\textit{b}) Cross-sectional view of the periodic domain at $\hat{z}=0$, showing the streamwise gas fraction, $\phi_x$, streamwise period, $2\hat{P}_x$, and highlighting domains $\hat{\mathcal{D}}_1$ and $\hat{\mathcal{D}}_2$ (magenta).}
    \label{fig:dimensional_shematics}
\end{figure}

Consider a steady 3D laminar channel flow contaminated with a soluble surfactant and bounded between two SHSs that are separated by a distance $2\hat{H}$, as illustrated in figure \ref{fig:dimensional_shematics}.
Hats indicate dimensional quantities.
The $\hat{x}$-, $\hat{y}$- and $\hat{z}$-coordinates are oriented in the streamwise, wall-normal and transverse directions, with $\hat{\boldsymbol{x}} = (\hat{x},\,\hat{y},\,\hat{z})$.
The liquid is assumed to be incompressible and Newtonian with dynamic viscosity $\hat{\mu}$, velocity $\hat{\boldsymbol{u}}=(\hat{u}(\hat{\boldsymbol{x}}),\,\hat{v}(\hat{\boldsymbol{x}}),\,\hat{w}(\hat{\boldsymbol{x}}))$, pressure $\hat{p}(\hat{\boldsymbol{x}})$, bulk surfactant distribution $\hat{c}(\hat{\boldsymbol{x}})$ and interfacial surfactant distribution $\hat{\Gamma}(\hat{x},\hat{z})$.
Owing to the periodicity of the geometry, we restrict attention to a single periodic cell with streamwise (transverse) period length $2 \hat{P}_x$ ($2 \hat{P}_z$), liquid--gas interface length (width) $2 \phi_x \hat{P}_x $ ($2 \phi_z \hat{P}_z$) and gas fraction $\phi_x$ ($\phi_z$). 
Liquid--gas interfaces, or plastrons, are assumed to be flat.
The domain is partitioned into two subdomains per period, that are bounded by the plastron and solid ridge, namely
\begin{subequations} \label{eq:dimensional_domains}
    \begin{align}
    \hat{\mathcal{D}}_1 &= \{\hat{x}\in [-\phi_x\hat{P}_x,\, \phi_x\hat{P}_x]\} \times \{\hat{y}\in [0, \,2 \hat{H}]\} \times \{\hat{z} \in [- \hat{P}_z, \,\hat{P}_z]\}, \\
    \hat{\mathcal{D}}_2 &= \{\hat{x}\in [\phi_x \hat{P}_x, \,(2 - \phi_x )\hat{P}_x]\} \times \{\hat{y}\in [0,\, 2 \hat{H}]\} \times \{\hat{z} \in [- \hat{P}_z, \,\hat{P}_z]\},
    \end{align}
\end{subequations}
as outlined in figure \ref{fig:dimensional_shematics}(\textit{b}). 
At the SHSs, $\hat{y} =0$ and $\hat{y} = 2\hat{H}$, we define the liquid--gas interfaces, the ridge surfaces and the solid surfaces, respectively, as
\begin{subequations} \label{eq:dimensional_interface}
    \begin{align}
    \hat{\mathcal{I}} &= \{\hat{x}\in [-\phi_x\hat{P}_x, \,\phi_x\hat{P}_x]\} \times \{\hat{y}\in\{0,\, 2\hat{H}\}\} \times \{\hat{z} \in [- \phi_z \hat{P}_z, \,\phi_z \hat{P}_z]\}, \\
    \hat{\mathcal{R}} &= \{\hat{x}\in [-\phi_x\hat{P}_x,\,\phi_x\hat{P}_x]\} \times \{\hat{y}\in\{0,\,2\hat{H}\}\} \times \{\hat{z} \in [- \hat{P}_z,\,-\phi_z \hat{P}_z]\cup [\phi_z \hat{P}_z,\,\hat{P}_z]\}, \\
    \hat{\mathcal{S}} &= \{\hat{x}\in [\phi_x \hat{P}_x,\, (2 - \phi_x)\hat{P}_x]\} \times \{\hat{y}\in\{0, \,2\hat{H}\}\} \times \{\hat{z} \in [- \hat{P}_z, \,\hat{P}_z]\},
    \end{align}
\end{subequations}
as outlined in figure \ref{fig:dimensional_shematics}(\textit{a}). 
Note that placing the origin of the coordinate system at the centre of the domain (common practice in the literature for flows over SHSs) would increase the number of matching conditions between subdomains from two to three.

To model the fluid we use the steady Stokes equations, neglecting inertia and any body forces.
The bulk surfactant is coupled to the flow field by a steady advection--diffusion equation. 
In $\hat{\mathcal{D}}_1$ and $\hat{\mathcal{D}}_2$, we therefore have
\refstepcounter{equation} \label{eq:dimensional_equations}
\begin{equation}
	\hat{\boldsymbol{\nabla}} \cdot \hat{\boldsymbol{u}} = 0, \quad \hat{\mu} \hat{\nabla}^2 \hat{\boldsymbol{u}}- \hat{\boldsymbol{\nabla}} \hat{p} = \boldsymbol{0}, \quad
	\hat{D} \hat{\nabla}^2 \hat{c} - \hat{\boldsymbol{u}}\cdot \hat{\boldsymbol{\nabla}} \hat{c} = 0, \tag{\theequation\textit{a--c}}
\end{equation}
where $\hat{D}$ is the surfactant bulk diffusivity. 
The interfacial surfactant is coupled to the flow by a steady advection--diffusion equation and an equation of state.
The bulk concentration is coupled to the interfacial surfactant by continuity of flux, where exchange at the interface is modelled using a source--sink term consistent with the Henry isotherm \citep{chang1995adsorption}.
The equation of state and adsorption--desorption kinetics are linearised, which is valid for small deviations in the concentration of surfactant away from some reference value \citep{manikantan2020surfactant}.
On $\hat{\mathcal{I}}$, we balance the tangential components of the stress, $\hat{\mathsfbi{T}\,}\cdot \boldsymbol{n}$ where $\hat{\mathsfbi{T}\,} = -\hat{p} \mathsfbi{I} + \hat{\mu}(\hat{\boldsymbol{\nabla}}\hat{\boldsymbol{u}} + (\hat{\boldsymbol{\nabla}}\hat{\boldsymbol{u}})^{T})$ and $\boldsymbol{n}$ is the unit normal to $\hat{\mathcal{I}}$ (pointing into the channel), with tangential gradients of surface tension $\hat{\sigma}$. 
We assume that (\textit{i}) the effects arising from the gas trapped in the SHSs are negligible, (\textit{ii}) $\hat{\sigma} = \hat{\sigma}_0 - \hat{A}(\hat{\Gamma} - \hat{\Gamma}_0)$ where $\hat{\sigma}_0$ is the reference surface tension, $\hat{A}$ is the surface activity and $\hat{\Gamma}_0$ is the reference surface concentration of surfactant, and (\textit{iii}) the surface tension remains large enough to suppress deflections of the interface from its assumed flat state.
Thus, along $\hat{\mathcal{I}}$, we impose the tangential stress balance in the streamwise and transverse directions, no-penetration of velocity, continuity of surfactant flux and the transport equation for interfacial surfactant
\refstepcounter{equation} \label{eq:dimensional_interface_bcs}
\begin{multline} 
	\hat{\mu} \boldsymbol{n}\cdot \hat{\boldsymbol{\nabla}} \hat{u} - \hat{A} \hat{\Gamma}_{\hat{x}} = 0, \quad \hat{\mu} \boldsymbol{n}\cdot \hat{\boldsymbol{\nabla}} \hat{w} - \hat{A} \hat{\Gamma}_{\hat{z}} = 0, \quad \hat{v}=0, \quad \hat{D} \boldsymbol{n}\cdot \hat{\boldsymbol{\nabla}} \hat{c} - \hat{K}_a \hat{c} + \hat{K}_d \hat{\Gamma} =0, \\ \quad \hat{D}_I ( \hat{\Gamma}_{\hat{x}\hat{x}} + \hat{\Gamma}_{\hat{z}\hat{z}} ) + \hat{K}_a \hat{c} - \hat{K}_d \hat{\Gamma} -(\hat{u} \hat{\Gamma})_{\hat{x}} - (\hat{w} \hat{\Gamma})_{\hat{z}} = 0,
    \tag{\theequation\textit{a--e}}
\end{multline}
with $\hat{D}_I$ the surfactant interfacial diffusivity, $\hat{K}_a$  the adsorption rate and $\hat{K}_d$  the desorption rate. 
On $\partial \hat{\mathcal{I}}$ (the contact line bounding the interfaces), no-flux of surfactant requires that
\refstepcounter{equation} \label{eq:dimensional_interface_bnd_bcs}
\begin{equation} 
    \hat{u}\hat{\Gamma} - \hat{D}_I \hat{\Gamma}_{\hat{x}} = 0 \quad \text{at} \quad \hat{x} = \pm \phi_x \hat{P}_x, \quad \hat{w}\hat{\Gamma} - \hat{D}_I \hat{\Gamma}_{\hat{z}} = 0 \quad \text{at} \quad \hat{z} = \pm \phi_z \hat{P}_z.
    \tag{\theequation\textit{a,\,b}}
\end{equation}
Along $\hat{\mathcal{R}}$ and $\hat{\mathcal{S}}$, we impose no-slip, no-penetration of velocity and no-flux of bulk surfactant
\refstepcounter{equation} \label{eq:dimensional_solid_bcs} 
\begin{equation} 
	\hat{u} = 0, \quad \hat{w} =0, \quad \hat{v} = 0, \quad \hat{c}_{\hat{y}} =0.
	\tag{\theequation\textit{a--d}}
\end{equation}
Defining $\hat{\boldsymbol{q}} = (\hat{\boldsymbol{u}}, \,\hat{p}_{\hat{x}},\,\hat{c})$, periodicity across the unit cell $\hat{\mathcal{D}}_1\cup\hat{\mathcal{D}}_2$ means that
\refstepcounter{equation} \label{eq:dimensional_periodicity} 
\begin{equation}
    \hat{\boldsymbol{q}}(-\phi_x\hat{P}_x,\,\hat{y},\,\hat{z}) = \hat{\boldsymbol{q}}((2- \phi_x)\hat{P}_x,\,\hat{y},\,\hat{z} ), \quad \hat{\boldsymbol{q}}(\hat{x},\,\hat{y},\,-\hat{P}_z) = \hat{\boldsymbol{q}}(\hat{x},\,\hat{y},\,\hat{P}_z).
	\tag{\theequation\textit{a,\,b}}
\end{equation}
The bulk flow and concentration are continuous at the boundary between $\hat{\mathcal{D}}_1$ and $\hat{\mathcal{D}}_2$
\begin{equation} \label{eq:dimensional_continuity}
    \hat{\boldsymbol{q}}((\phi_x\hat{P}_x)^-,\,\hat{y},\,\hat{z}) = \hat{\boldsymbol{q}}((\phi_x\hat{P}_x)^+,\,\hat{y},\,\hat{z} ),
\end{equation}
where the superscripts $-$ and $+$ mean that the boundary condition is evaluated in  $\hat{\mathcal{D}}_1$ and $\hat{\mathcal{D}}_2$, respectively.

Owing to the symmetry about $\hat{y}=\hat{H}$, the top and bottom interfaces are assumed to have the same distribution of surfactant.  
We can integrate \eqref{eq:dimensional_equations}--\eqref{eq:dimensional_continuity} across the channel to show how the streamwise bulk surfactant flux changes as surfactants  adsorb and desorb
\begin{equation} \label{eq:dim_cons_1}
    \frac{\text{d}}{\text{d}\hat{x}}\int_{\hat{z}=-\hat{P}_z}^{\hat{P}_z} \int_{\hat{y}=0}^{2 \hat{H}} (\hat{u}\hat{c} - \hat{D} \hat{c}_{\hat{x}}) \, \text{d}\hat{y}\, \text{d}\hat{z}  = 2 \int_{\hat{z}=- \phi_z \hat{P}_z, \, \hat{y}=0}^{\phi_z \hat{P}_z} (\hat{K}_d \hat{\Gamma} - \hat{K}_a \hat{c}) \, \text{d}\hat{z};
\end{equation}
and likewise how the streamwise interfacial surfactant flux changes along the plastron
\begin{equation} \label{eq:dim_cons_2}
    \frac{\text{d}}{\text{d}\hat{x}}\int_{\hat{z}=-\phi_z \hat{P}_z, \, \hat{y}=0}^{\phi_z \hat{P}_z} (\hat{u}\hat{\Gamma} - \hat{D}_I \hat{\Gamma}_{\hat{x}}) \, \text{d}\hat{z} = -    \int_{\hat{z}=- \phi_z \hat{P}_z, \, \hat{y}=0}^{\phi_z \hat{P}_z} (\hat{K}_d \hat{\Gamma} - \hat{K}_a \hat{c}) \, \text{d}\hat{z}.
\end{equation}
The boundary-value problem \eqref{eq:dimensional_equations}--\eqref{eq:dimensional_continuity} can also be integrated to show that the total flux of liquid $\hat{Q}$ is uniform along the streamwise length of the channel
\begin{equation} \label{eq:dimensional_velocity_flux}
    \hat{Q} = \int_{\hat{z}=-\hat{P}_z}^{\hat{P}_z} \int_{\hat{y}=0}^{2 \hat{H}} \hat{u} \, \text{d}\hat{y} \, \text{d}\hat{z},
\end{equation}
the total flux of surfactant $\hat{K}$ is uniform along the streamwise length of the channel
\begin{equation} \label{eq:dimensional_surfactant_flux}
    \hat{K} = \int_{\hat{z}=-\hat{P}_z}^{\hat{P}_z} \int_{\hat{y}=0}^{2 \hat{H}} (\hat{u}\hat{c} - \hat{D} \hat{c}_{\hat{x}}) \, \text{d}\hat{y}\, \text{d}\hat{z} + 2 \int_{\hat{z}=-\phi_z \hat{P}_z, \, \hat{y}=0}^{\phi_z \hat{P}_z} (\hat{u}\hat{\Gamma} - \hat{D}_I \hat{\Gamma}_{\hat{x}}) \, \text{d}\hat{z},
\end{equation}
where the factor of $2$ in front of the second integral accounts for both top and bottom symmetrical SHSs, and the adsorption--desorption flux of surfactant between the interface and the bulk integrates to zero over the whole  interface owing to conservation of mass,
\begin{equation} \label{eq:dimensional_net_flux}
    \int_{\hat{x}=-\phi_x \hat{P}_x}^{\phi_x \hat{P}_x}     \int_{\hat{z}=- \phi_z \hat{P}_z, \, \hat{y}=0}^{\phi_z \hat{P}_z} (\hat{K}_d \hat{\Gamma} - \hat{K}_a \hat{c}) \, \text{d}\hat{z} \, \text{d}\hat{x} = 0.
\end{equation}
Integrating \eqref{eq:dim_cons_1} plus twice \eqref{eq:dim_cons_2} with respect to $\hat{x}$ recovers \eqref{eq:dimensional_surfactant_flux}.

A key quantity of interest in the present study is drag reduction. 
The flow is driven in the streamwise $\hat{x}$-direction by a cross-channel-averaged pressure drop per period given by $\Delta \hat{p} \equiv \langle \hat{p}\rangle(-\phi_x \hat{P}_x) - \langle \hat{p}\rangle((2-\phi_x)\hat{P}_x) > 0$, where $\langle \cdot \rangle \equiv \int_{\hat{z}=-\hat{P}_z}^{\hat{P}_z} \int_{\hat{y}=0}^{2 \hat{H}} \cdot \, \text{d}\hat{y} \, \text{d}\hat{z} / (4 \hat{P}_z \hat{H})$ is the cross-channel average. 
In the limit when $\hat{\mathcal{I}}$ is immobilized (yielding an effective no-slip boundary condition for the velocity), we have $\Delta \hat{p} = \Delta \hat{p}_R$, say, and in the uncontaminated (or surfactant-free) limit $\Delta \hat{p} = \Delta \hat{p}_U$ (when $\hat{\mathcal{I}}$ is a shear-free surface).
We define the normalised drag reduction as
\begin{equation}  \label{eq:dimensional_drag}
{DR} = \frac{\Delta \hat{p}_R - \Delta \hat{p}}{\Delta \hat{p}_R - \Delta \hat{p}_U},
\end{equation}
which varies from ${DR}=0$ to $1$ in the cases of minimum and maximum drag reduction, depending on the surfactant-induced Marangoni stresses. 
Equation \eqref{eq:dimensional_drag} differs from other definitions of the drag reduction which compare the flow over a SHS to a solid wall, such as $1 - \Delta \hat{p}/\Delta \hat{p}_R$ \cite[e.g.,][]{lee2016superhydrophobic}.

\subsection{Non-dimensionalisation} \label{subsection:nondimensionalisation}

Non-dimensionalising the governing equations \eqref{eq:dimensional_domains}--\eqref{eq:dimensional_drag} using $\epsilon \hat{U} = \hat{Q}/(\hat{H}^2)$ for the velocity scale, with $\epsilon = \hat{H}/ \hat{P}_x$ for the slenderness parameter, $\hat{P} = \hat{\mu} \hat{U}/\hat{H}$ for the pressure scale, $\hat{C} = \hat{K}/\hat{Q}$ for the bulk concentration scale and $\hat{G} = \hat{K}_a \hat{C}/\hat{K}_d$ for the interface concentration scale, we write
\refstepcounter{equation} \label{eq:nondimensionalisation}
\begin{equation}
    x = \frac{\hat{x}}{\hat{P}_x}, \quad \boldsymbol{x}_\perp = \frac{\hat{\boldsymbol{x}}_\perp}{\epsilon \hat{P}_x}, \quad u = \frac{\hat{u}}{\epsilon \hat{U}}, \quad \boldsymbol{u}_\perp = \frac{\hat{\boldsymbol{u}}_\perp}{\hat{U}}, \quad p = \frac{\hat{p}}{\hat{P}}, \quad c = \frac{\hat{c}}{\hat{C}}, \quad \Gamma = \frac{\hat{\Gamma}}{\hat{G}},
    \tag{\theequation\textit{a--g}}
\end{equation}
where  $\hat{\boldsymbol{x}}_\perp \equiv (\hat{y}, \,\hat{z})$ and $\hat{\boldsymbol{u}}_\perp \equiv (\hat{v},\, \hat{w})$. 
Assuming that $\epsilon \ll 1$, this normalisation yields a long-wave theory for steady flow in the streamwise direction and introduces a velocity scaling which captures the rapid cross-channel transport that acts to eliminate cross-channel gradients of surfactant.
The transverse flow decays exponentially quickly \citep{mcnair2022surfactant}, but should formally be retained to develop consistent expansions.
Here we only consider channels with an order-one aspect ratio (such that $\hat{H} \sim \hat{P}_z$); other asymptotic scalings are left for future work.
The longitudinal subdomains \eqref{eq:dimensional_domains} become
\begin{subequations}
    \begin{align}
    \mathcal{D}_1 &= \{x\in [-\phi_x, \,\phi_x]\} \times \{y\in [0, \,2]\} \times \{z \in [- P_z, \,P_z]\}, \\ 
    \mathcal{D}_2 &= \{x\in [\phi_x, \,2 - \phi_x]\} \times \{y\in [0,\, 2]\} \times \{z \in [- P_z,\, P_z]\},
    \end{align}
\end{subequations}
where $P_z = \hat{P}_z/\hat{H}$ is the non-dimensional pitch, and interfaces \eqref{eq:dimensional_interface} are
\begin{subequations}
    \begin{align}
    \mathcal{I} &= \{x\in [-\phi_x, \,\phi_x]\} \times \{ y \in\{0, \,2\}\} \times \{z \in [- \phi_z P_z, \,\phi_z P_z]\}, \\
    \mathcal{R} &= \{x\in [-\phi_x, \,\phi_x]\} \times \{ y \in\{0, \,2\}\} \times \{z \in [- P_z, \,-\phi_z P_z]\cup [\phi_z P_z, \,P_z]\}, \\
    \mathcal{S} &= \{x\in [\phi_x,\, 2 - \phi_x]\} \times \{ y \in\{0, \,2\}\} \times \{z \in [- P_z, \,P_z]\}.
    \end{align}
\end{subequations}

We substitute the non-dimensionalisation \eqref{eq:nondimensionalisation} into the governing equations \eqref{eq:dimensional_equations}--\eqref{eq:dimensional_drag} to acquire rescaled governing equations in terms of the non-dimensional variables given in \eqref{eq:nondimensionalisation} and $\epsilon^2$. 
In $\mathcal{D}_1$ and $\mathcal{D}_2$,
\refstepcounter{equation} \label{eq:nondimensional_equations} 
\begin{multline}
	\epsilon^2 u_x + \boldsymbol{\nabla}_\perp\cdot \boldsymbol{u}_\perp = 0, \quad \epsilon^2 \boldsymbol{u}_{xx} + \nabla^2_\perp \boldsymbol{u} -  \boldsymbol{\nabla}p = \boldsymbol{0}, \\ 
	\quad
	\Pen^{-1} (\epsilon^2 c_{xx}+ \nabla^2_\perp c) - \epsilon^2 u c_x - \boldsymbol{u}_\perp \cdot \boldsymbol{\nabla}_{\perp} c = 0,
    \tag{\theequation\textit{a--c}}
\end{multline}
with $\Pen = \hat{U} \hat{H}/\hat{D}$ the bulk P\'{e}clet number; $\boldsymbol{\nabla}_\perp \equiv (\partial_y,\, \partial_z)$ and $\nabla^2_\perp \equiv \partial_{yy} + \partial_{zz}$ are cross-channel differential operators. 
Along $\mathcal{I}$,
\refstepcounter{equation}  \label{eq:nondimensional_interface_bcs}
\begin{multline}
	\boldsymbol{n}\cdot \boldsymbol{\nabla} u - \Ma \Gamma_{x} = 0, \quad \boldsymbol{n}\cdot \boldsymbol{\nabla} w - \Ma \Gamma_{z} = 0, \quad v =0, \quad \boldsymbol{n}\cdot \boldsymbol{\nabla} c - Da (c -\Gamma)  =0, \\ \quad \Pen^{-1}_I (\epsilon^2 \Gamma_{xx} + \Gamma_{zz} ) + \Bi( c - \Gamma)  -	\epsilon^2 (u \Gamma)_{x} - (w \Gamma)_{z} = 0,
    \tag{\theequation\textit{a--e}}
\end{multline}
with $\Ma = \hat{A}\hat{G}/\hat{\mu}\hat{U}$ the Marangoni number, $\Da= \hat{K}_a \hat{H}/\hat{D}$ the \Damkohler number, $\Pen_I= \hat{H} \hat{U}/\hat{D}_I$ the interfacial P\'{e}clet number and $\Bi = \hat{K}_d \hat{H}/\hat{U}$ the Biot number. On $\partial \mathcal{I}$,
\refstepcounter{equation}  \label{eq:nondimensional_noflux}
\begin{equation}
    u \Gamma - \Pen^{-1}_I \Gamma_{x} = 0 \quad \text{at} \quad x = \pm \phi_x, \quad
    w \Gamma - \Pen^{-1}_I \Gamma_{z} = 0 \quad \text{at} \quad z = \pm \phi_z P_z.
    \tag{\theequation\textit{a,\,b}}
\end{equation}
On $\mathcal{R}$ and $\mathcal{S}$,
\refstepcounter{equation} \label{eq:nondimensional_solid_bcs}
\begin{equation}
	u = 0, \quad w =0, \quad v = 0, \quad \ c_{y} =0.
    \tag{\theequation\textit{a--d}}
\end{equation}
Periodicity and continuity between subdomains require that, for $\boldsymbol{q} = (\boldsymbol{u},\,p_x,\,c)$,
\refstepcounter{equation} \label{eq:nondimensional_periodicity}
\begin{equation} 
    \boldsymbol{q}(-\phi_x,\,y,\,z) = \boldsymbol{q}(2- \phi_x,\,y,\,z), \quad \boldsymbol{q}(x,\,y,\,-P_z) = \boldsymbol{q}(x,\,y,\,P_z),
    \tag{\theequation\textit{a,\,b}}
\end{equation}
and
\begin{equation} \label{eq:nondimensional_continuity}
   \boldsymbol{q}(\phi_x^-,\,y,\,z) = \boldsymbol{q}(\phi_x^+,\,y,\,z).
\end{equation}
The total liquid flux, total surfactant flux and net flux of surfactant from interface to bulk are given respectively by
\begin{equation} \label{eq:dimensionless_velocity_flux}
    \int_{z=-P_z}^{P_z} \int_{y=0}^{2} u \, \text{d} y \, \text{d} z = 1,
\end{equation}
\begin{equation} \label{eq:dimensionless_surfactant_flux}
    \int_{z=- P_z}^{ P_z} \int_{y=0}^{2} \left( u c -  \frac{c_{ x}}{\Pen} \right) \, \text{d}y \, \text{d}z + \frac{2 \Da}{\Bi \Pen}\int_{z=- \phi_z P_z ,\, y=0}^{\phi_z P_z} \left(u \Gamma - \frac{\Gamma_x}{\Pen_I}\right) \, \text{d} z = 1,
\end{equation}
\begin{equation} \label{eq:dimensionless_net_flux}
    \int_{x=- \phi_x}^{\phi_x} \int_{z=- \phi_z P_z,\, y=0}^{\phi_z P_z} (\Gamma - c) \, \text{d} z \, \text{d} x = 0,
\end{equation}
where bulk and surface fluxes are related to adsorption--desorption at the interface via
\begin{subequations} \label{eq:transport_eqs}
\begin{align}
    \frac{\text{d}}{\text{d} x}\int_{z=- P_z}^{P_z} \int_{ y=0}^{2}\left( u c -  \frac{c_{ x}}{\Pen} \right) \, \text{d} y \, \text{d} z &= \frac{2 \Da}{\epsilon^2 \Pen} \int_{z=- \phi_z P_z, \, y=0}^{\phi_z P_z} (\Gamma - c) \, \text{d} z , \\ 
    \frac{\text{d}}{\text{d} x} \int_{z=-\phi_z P_z, \, y=0}^{\phi_z P_z} \left(u \Gamma - \frac{\Gamma_x}{\Pen_I}\right) \, \text{d}z &= - \frac{\Bi}{\epsilon^2} \int_{z=- \phi_z P_z,\, y=0}^{\phi_z P_z} (\Gamma - c) \, \text{d} z .
\end{align}
\end{subequations}
The group $\Da/(\Bi\Pen) = \hat{K}_a/(\hat{K}_d \hat{H})$ in \eqref{eq:dimensionless_surfactant_flux} appears frequently in this problem.
The depletion length $\hat{L}_d = \hat{K}_a/\hat{K}_d$ \citep{manikantan2020surfactant} corresponds to the depth into the liquid necessary to balance the adsorption and desorption fluxes between the interface and the bulk, at equilibrium. 
Hence, the group $\Da/(\Bi\Pen)$ compares $\hat{L}_d$ to the channel height $\hat{H}$.
The drag reduction \eqref{eq:dimensional_drag} becomes
\begin{equation}  \label{eq:nondimensional_drag}
{DR} = \frac{\Delta p_R - \Delta p}{\Delta p_R - \Delta p_U},
\end{equation}
where $\Delta {p} \equiv \langle p\rangle(-\phi_x) - \langle p\rangle(2-\phi_x)$ and $\langle \cdot \rangle \equiv \int_{z=- P_z}^{ P_z} \int_{y=0}^{2} \cdot \, \text{d} y \text{d} z/ (4 P_z)$. 

The non-dimensional governing equations \eqref{eq:nondimensional_equations}, boundary conditions \eqref{eq:nondimensional_interface_bcs}--\eqref{eq:nondimensional_continuity} and flux constraints \eqref{eq:dimensionless_velocity_flux}--\eqref{eq:dimensionless_net_flux}, define a 3D boundary-value problem for $\boldsymbol{u}$, $p$, $c$ and $\Gamma$.
The solution depends on 9 dimensionless groups ($\epsilon$, $\Pen$, $\Ma$, $\Pen_I$, $\Bi$, $\Da$, $\phi_x$, $\phi_z$ and $P_z$) that characterise the geometry, flow, liquid and surfactant.
The number of dimensionless groups differs from \citet{landel2020theory} (where there are 8), as the authors considered 2D geometries, and \citet{temprano2023single} (where there are 10), as we have absorbed the non-dimensional background concentration into $\Ma$.
Our aim is to construct a reduced model in the limit $\epsilon \ll 1$ to predict the drag reduction \eqref{eq:nondimensional_drag}. 


\section{Model}\label{sec:model}

With 9 non-dimensional parameters in the problem, we choose distinguished limits to reveal different dominant physical balances. 
We take $\epsilon \rightarrow 0$ with some parameters held fixed and others varying proportionally to $\epsilon^2$.
Specifically, in \S\ref{subsec:strong_exchange}, we take $\Da= \Bi = O(1)$ as $\epsilon \rightarrow 0$, allowing the leading-order bulk ($c_0$) and interfacial ($\Gamma_0$) surfactant concentrations to remain in equilibrium, such that $c_0=\Gamma_0$; we call this the ``strong exchange limit''.
In \S\ref{subsec:weak_exchange}, we treat smaller $\Da$ and $\Bi$, so that $\Gamma_0$ decouples from $c_0$; we call this the ``moderate exchange limit''.
First, however (in \S\ref{subsec:Surfactant transport equations}), we introduce the resulting 1D surfactant transport equations coupling the bulk and the interface, which depend on 5 non-dimensional parameters that characterise the strength of bulk and interfacial advection, Marangoni effects, diffusion and bulk--interface exchange and one geometrical parameter.

\subsection{Surfactant transport equations} \label{subsec:Surfactant transport equations}

To derive the 1D surfactant transport equations from \eqref{eq:transport_eqs}, we assume that $\Pen = \Pen_I = \Ma = O(1)$ in the limit $\epsilon \ll 1$, which implies that cross-channel concentration gradients are small, so that $c\approx c_0(x)$ and $\Gamma \approx \Gamma_0(x)$.
A detailed derivation is provided in \S\ref{subsec:weak_exchange}; however, in this subsection, we introduce  the 1D surfactant transport equations that we solve to generate the results in \S\ref{sec:results}. We  briefly describe their physical meaning to provide the reader with an overview of the different physical processes at play in the problem studied.
The 1D surfactant transport equations relate  bulk and surface fluxes of surfactant through adsorption--desorption fluxes at the interface. As will be shown in \S\ref{subsec:weak_exchange}, they reduce to a simpler coupled nonlinear system of second-order ODEs:
\begin{subequations} \label{eq:model1}
\begin{align}
         (c_{0} - \alpha c_{0x})_x - \frac{\nu}{\epsilon^2}(\Gamma_0 - c_0) = 0 \quad \text{in} \quad &\mathcal{D}_1, \\  (\beta \Gamma_{0} - \gamma \Gamma_0 \Gamma_{0x} -  \delta \Gamma_{0x})_x - \frac{\nu}{\epsilon^2}(c_0 - \Gamma_0) = 0 \quad \text{in} \quad &\mathcal{D}_1, \\
         (c_{0} - \alpha c_{0x})_x = 0 \quad \text{in} \quad &\mathcal{D}_2.
\end{align}
\end{subequations}
The total flux of surfactant \eqref{eq:dimensionless_surfactant_flux} becomes
\begin{subequations} \label{eq:model2}
\begin{align}
         c_{0} - \alpha c_{0x} + \beta \Gamma_{0} - \gamma \Gamma_0 \Gamma_{0x} -  \delta \Gamma_{0x} &= 1  \quad \text{in} \quad \mathcal{D}_1, \\
         c_{0} - \alpha c_{0x} &= 1 \quad \text{in} \quad \mathcal{D}_2.
\end{align}
\end{subequations}
Continuity of bulk surfactant \eqref{eq:nondimensional_continuity} and bulk surfactant flux \eqref{eq:dimensionless_surfactant_flux} between unit cells, and no flux of interfacial surfactant \eqref{eq:nondimensional_noflux} through contact lines, are together given by
\refstepcounter{equation} 
\label{eq:model3}
\begin{multline} 
        c_0(\phi_x^{-}) = c_0(\phi_x^{+}), \quad c_0(-\phi_x) = c_0(2 - \phi_x), \\ c_0 - \alpha c_{0x} = 1, \quad \beta \Gamma_0 - \gamma \Gamma_0 \Gamma_{0x} -  \delta \Gamma_{0x} = 0 \quad \text{at} \quad x=\pm \phi_x. \tag{\theequation\textit{a--d}}
\end{multline}
Equations \eqref{eq:model1}--\eqref{eq:model3} are the steady leading-order surfactant transport equations, flux constraints and boundary conditions in $\mathcal{D}_1$ and $\mathcal{D}_2$, accounting for bulk advection by the pressure gradient ($c_{0x}$ in \eqref{eq:model1}), surface advection by the pressure gradient ($\beta \Gamma_{0x}$), Marangoni advection ($\gamma (\Gamma_0 \Gamma_{0x})_x$), bulk diffusion $(\alpha c_{0xx})$, surface diffusion $(\delta \Gamma_{0xx})$ and bulk--surface exchange ($\pm\nu(\Gamma_0 - c_0)/\epsilon^2$).
In \S\ref{subsec:strong_exchange}, we derive and discuss further the non-dimensional parameters $\alpha$, $\beta$, $\gamma$ and $\delta$ associated with the strengths of the above physical processes.
In \S\ref{subsec:strong_exchange}, we focus on the limit where bulk--surface exchange of surfactant is strong relative to advection and diffusion, such that $\nu/\epsilon^2 \gg O(1,\,\alpha,\,\beta,\,\delta)$ in \eqref{eq:model1}--\eqref{eq:model3}, and therefore, the $c_0$ and $\Gamma_0$ fields are in equilibrium. 
In \S\ref{subsec:weak_exchange}, we study the limit where bulk--surface exchange of surfactant is comparable to advection and diffusion, such that $\nu/\epsilon^2 \sim O(1,\,\alpha,\,\beta,\,\delta)$ in \eqref{eq:model1}--\eqref{eq:model3}, and therefore, the $c_0$ and $\Gamma_0$ fields are distinct.

\subsection{Strong exchange} \label{subsec:strong_exchange}

\subsubsection{Strong cross-channel diffusion} \label{subsec:strong_diffusion}

We begin by assuming that $\Pen = \Pen_I = \Bi = \Da = \Ma = O(1)$ in the limit $\epsilon \ll 1$.
We call this the ``strong cross-channel diffusion limit''.
In this subsection, we derive the $O(1)$ coefficients $\alpha$, $\beta$, $\gamma$ and $\delta$, such that $\nu/\epsilon^2 \gg O(1)$ for $\epsilon \ll 1$ and $\nu = O(1)$.
We substitute the expansions
\begin{equation} \label{eq:asymptotic_expansion}
	(u,\,v,\,w, \,p,\, c,\, \Gamma) = (u_0,\,v_0,\,w_0, \,p_0,\, c_0, \,\Gamma_0) + \epsilon^2 (u_1,\,v_1,\,w_1, \,p_1,\, c_1, \,\Gamma_1) + ...,
\end{equation}
in the governing equations \eqref{eq:nondimensional_equations}--\eqref{eq:nondimensional_drag} and take the $O(1)$ approximation.
In domains $\mathcal{D}_1$ and $\mathcal{D}_2$,
\refstepcounter{equation} \label{eq:leading_order} 
\begin{equation}
	\boldsymbol{\nabla}_{\perp} \cdot \boldsymbol{u}_{\perp 0}  = 0, \quad \nabla^2_{\perp} \boldsymbol{u}_0 - \boldsymbol{\nabla} p_{0} = \boldsymbol{0}, \quad \Pen^{-1}\nabla^2_{\perp} c_0 - \boldsymbol{u}_{\perp 0} \cdot \boldsymbol{\nabla}_{\perp} c_0 =0.
    \tag{\theequation\textit{a--c}}
\end{equation}
Along the interface $\mathcal{I}$,
\refstepcounter{equation} \label{eq:leading_order_interface_bcs}
\begin{multline}
	\boldsymbol{n}\cdot \boldsymbol{\nabla} u_0 - \Ma \Gamma_{0x} = 0, \quad  \boldsymbol{n}\cdot \boldsymbol{\nabla} w_0 - \Ma \Gamma_{0z} = 0, \quad v_0 = 0, \\ \boldsymbol{n}\cdot \boldsymbol{\nabla} c_0 - Da (c_0 -\Gamma_0)  =0, \quad \Pen^{-1}_I \Gamma_{0zz} + \Bi( c_0 - \Gamma_0) - (w_0 \Gamma_0)_{z} = 0,
    \tag{\theequation\textit{a--e}}
\end{multline}
and on the interface contour $\partial \mathcal{I}$,
\begin{equation} \label{eq:leading_order_interface_boundary bcs}
    w_0 \Gamma_0 - \Pen^{-1}_I \Gamma_{0 z} = 0 \quad  \text{at} \quad z = \pm \phi_z P_z.
\end{equation}
There are inner regions near $x = \pm \phi_x$, within which $u_0 \Gamma_0 - \Pen^{-1}_I \Gamma_{0x} = 0$ is imposed on $\partial \mathcal{I}$, and the boundary conditions $\boldsymbol{u}_0(\phi_x^{-},\,y,\,z) = \boldsymbol{u}_0(\phi_x^{+},\,y,\,z)$ and $\boldsymbol{u}_0(-\phi_x,\,y,\,z) = \boldsymbol{u}_0(2 - \phi_x,\,y,\,z)$ are imposed between $\mathcal{D}_1$ and $\mathcal{D}_2$.
Within the inner regions, the flow and surfactant field are governed by the coupled 3D Stokes and surfactant transport equations.
In the present long-wave theory it is sufficient to impose continuity of $c_0$ between $\mathcal{D}_1$ and $\mathcal{D}_2$, such that
\refstepcounter{equation} 
\label{eq:leading_order_periodicity}
\begin{equation}
    c_0(\phi_x^{-},\,y,\,z) = c_0(\phi_x^{+},\,y,\,z), \quad c_0(-\phi_x,\,y,\,z) = c_0(2 - \phi_x,\,y,\,z), \tag{\theequation\textit{a,\,b}}
\end{equation}
and continuity of volume and surfactant flux between $\mathcal{D}_1$ and $\mathcal{D}_2$, such that
\begin{equation} \label{eq:leading_order_fluid_flux}
    \int_{z=- P_z}^{ P_z} \int_{ y=0}^{2} u_0 \, \text{d}y \, \text{d}z = 1,
\end{equation}
\begin{equation} \label{eq:leading_order_surfactant_flux}
    \int_{z=- P_z}^{P_z} \int_{y=0}^{2} \left(u_0 c_0 - \frac{c_{0x}}{\Pen}\right) \, \text{d}y \, \text{d}z + \frac{2 \Da}{\Bi \Pen }\int_{z=- \phi_z P_z,\, y=0}^{\phi_z P_z ,} \left(u_0 \Gamma_0 - \frac{\Gamma_{0x}}{\Pen_I}\right) \, \text{d}z = 1,
\end{equation}
for all $-\phi_x\leq x\leq 2-\phi_x$.
Along the ridge $\mathcal{R}$ and solid $\mathcal{S}$,
\refstepcounter{equation} \label{eq:leading_order_solid_bcs}
\begin{equation}
	u_{0} = 0, \quad v_0 = 0, \quad w_{0} =0, \quad c_{0y} =0.
    \tag{\theequation\textit{a--d}}
\end{equation}
Transverse periodicity can be rewritten as the symmetry conditions
\refstepcounter{equation} \label{eq:leading_order_transverse_symmetry_bcs}
\begin{equation}  
	u_{0z} = 0, \quad v_{0z} = 0, \quad w_{0} =0, \ \quad c_{0z} =0 \quad \text{at} \quad z = \pm P_z.
    \tag{\theequation\textit{a--d}}
\end{equation}
The drag reduction becomes
\begin{equation}  \label{eq:leading_order_drag}
{DR}_0 = \frac{\Delta p_R - \Delta p_0}{\Delta p_R - \Delta p_U}.
\end{equation}

\begin{figure}
    \centering
    (\textit{a}) \hfill (\textit{b}) \hfill \hfill \hfill \\
    \vspace{-.5cm} \includegraphics[width=.425\textwidth]{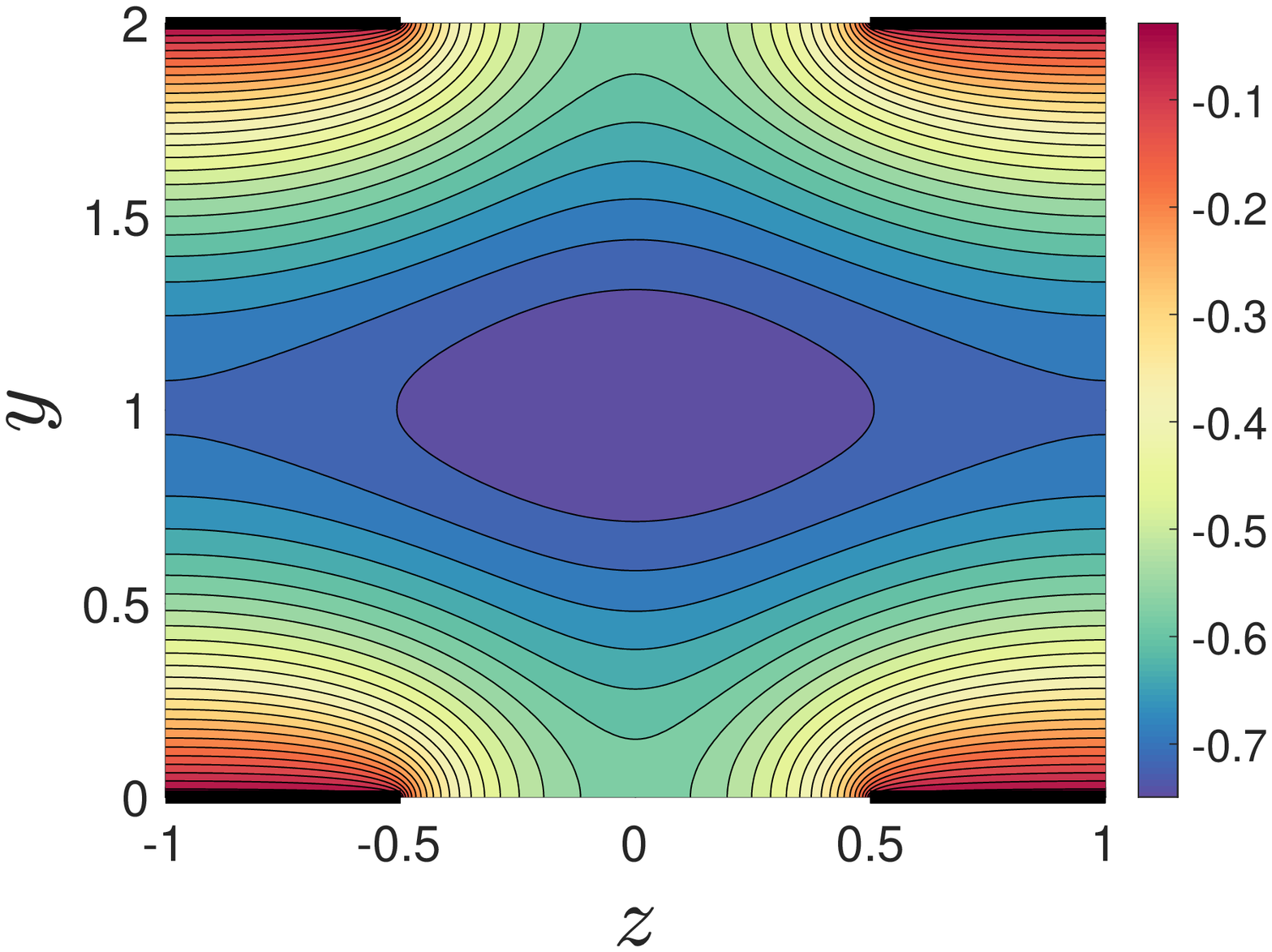} \hspace{.8cm} \includegraphics[width=.425\textwidth]{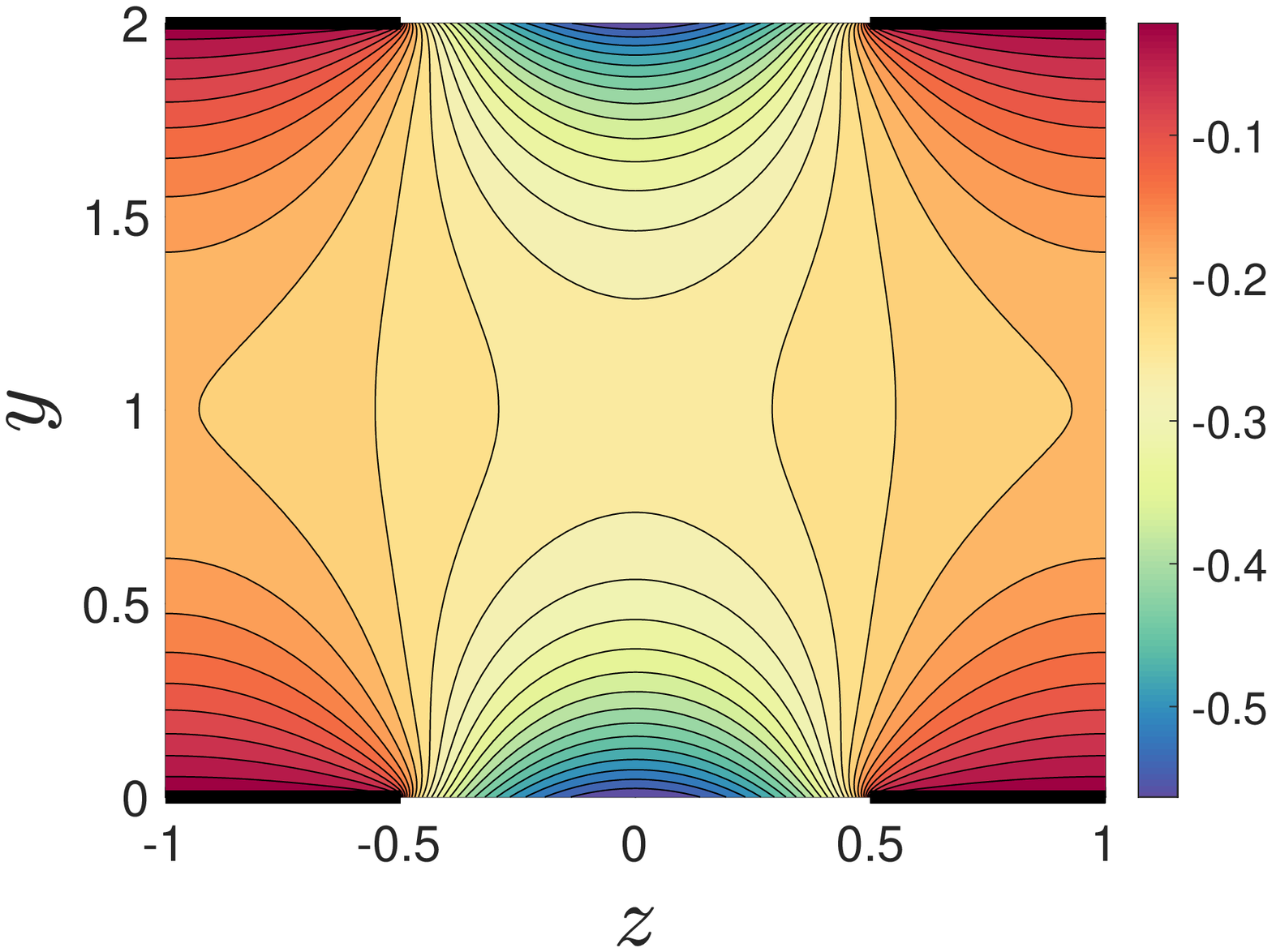}
    \caption{(\textit{a}) Contour plot of $\tilde{U}$ defined by \eqref{eq:tildeu_bvp}, the leading-order contribution to the streamwise flow due to the pressure gradient, $p_{0x}$.
    (\textit{b}) Contour plot of $\bar{U}$ defined by \eqref{eq:baru_bvp}, the leading-order contribution to the streamwise flow due to the surfactant gradient, $\Gamma_{0x}$. 
    Since $p_{0x}<0$, $\tilde{U}$ contributes positively to the leading-order streamwise velocity component $u_0$, whilst $\bar{U}$ contributes negatively, since $\Gamma_{0x}>0$, following \eqref{eq:u_def}.
    The thick black lines represent the solid regions of the SHS with the transverse gas fraction $\phi_z = 0.5$ and width $P_z = 1$.}
    \label{fig:first_order_velocity}
\end{figure}

From \eqref{eq:leading_order}--\eqref{eq:leading_order_transverse_symmetry_bcs}, the leading-order solution simplifies to $\Gamma_0=\Gamma_0(x)$, $c_0 = c_0(x)$ where $c_0=\Gamma_0$, $p_0 = p_0(x)$ and $v_0=w_0=0$. 
The leading order velocities in the cross-section, $v_0$ and $w_0$, vanish due to transverse surfactant gradients decaying exponentially fast in a time-dependent setting, as discussed by \citet{mcnair2022surfactant}. 
Therefore, in our steady problem, the concentration field does not vary in the transverse direction and there are no concentration gradients to generate velocities in the cross-plane.
The streamwise flow is driven by $p_{0x}$, from  (\ref{eq:leading_order}\textit{b}), and $\Gamma_{0x}$, from (\ref{eq:leading_order_interface_bcs}\textit{a}).
Using linear superposition, we can write
\refstepcounter{equation} \label{eq:u_def}
\begin{equation}
	u_0 = \tilde{U} p_{0x} + \Ma \bar{U}\Gamma_{0x} \quad \text{in} \ \mathcal{D}_1 \quad \text{and} \quad u_0 = \breve{U} p_{0x} \quad \text{in} \ \mathcal{D}_2,
	\tag{\theequation\textit{a,\,b}}
\end{equation}
where $\tilde{U}(y,\,z)$ and $\bar{U}(y,\,z)$ are velocity contributions related to the bulk pressure and surfactant gradient, respectively, in $\mathcal{D}_1$, and $\breve{U}(y)$ is the velocity contribution related to the bulk pressure gradient in $\mathcal{D}_2$. 
Hence, we must solve the following boundary-value problems: streamwise flow driven by a pressure gradient over $\mathcal{D}_1$,
\refstepcounter{equation}  \label{eq:tildeu_bvp}
\begin{multline} 
    \nabla^2_{\perp}\tilde{U} = 1, \quad \text{subject to} \quad \tilde{U}_y(0,\,z_s)=0, \quad \tilde{U}(0,\,z_{ns})=0, \quad \tilde{U}_y(2,\,z_s)=0, \\ \tilde{U}(2,\,z_{ns})=0, \quad \tilde{U}_z(y,\,-P_z)=0, \quad \tilde{U}_z(y,\,P_z)=0;
	\tag{\theequation\textit{a--g}}
\end{multline}
streamwise flow driven by a surfactant-induced Marangoni shear stress over $\mathcal{D}_1$,
\refstepcounter{equation} \label{eq:baru_bvp}
\begin{multline} 
    \nabla^2_{\perp}\bar{U} = 0, \quad \text{subject to} \quad \bar{U}_y(0,\,z_s)= 1, \quad \bar{U}(0,\,z_{ns})= 0, \quad \bar{U}_y(2,\,z_s) = -1, \\ \bar{U}(2,\,z_{ns})= 0, \quad \bar{U}_z(y,\,-P_z)=0, \quad \bar{U}_z(y,\,P_z)=0;
	\tag{\theequation\textit{a--g}}
\end{multline}
and streamwise flow driven by a pressure gradient over $\mathcal{D}_2$,
\refstepcounter{equation} \label{eq:hatu_bvp} 
\begin{equation} 
    \breve{U}_{yy} = 1, \quad \text{subject to} \quad \breve{U}(0)= 0, \ \quad \breve{U}(2)=0;
	\tag{\theequation\textit{a--c}}
\end{equation} 
with $z_s\equiv \{z\in [-\phi_z P_z, \,\phi_z P_z]\}$ and $z_{ns} \equiv \{z\in [-P_z,\, -\phi_z P_z]\}\cup\{z\in [\phi_z P_z, \,P_z]\}$.

\begin{figure}
    \centering
    (\textit{a}) \hfill (\textit{b}) \hfill (\textit{c}) \hfill (\textit{d}) \hfill \hfill \hfill \\
    \includegraphics[trim={.0cm .3cm 1.8cm .7cm},clip,width=.24\textwidth]{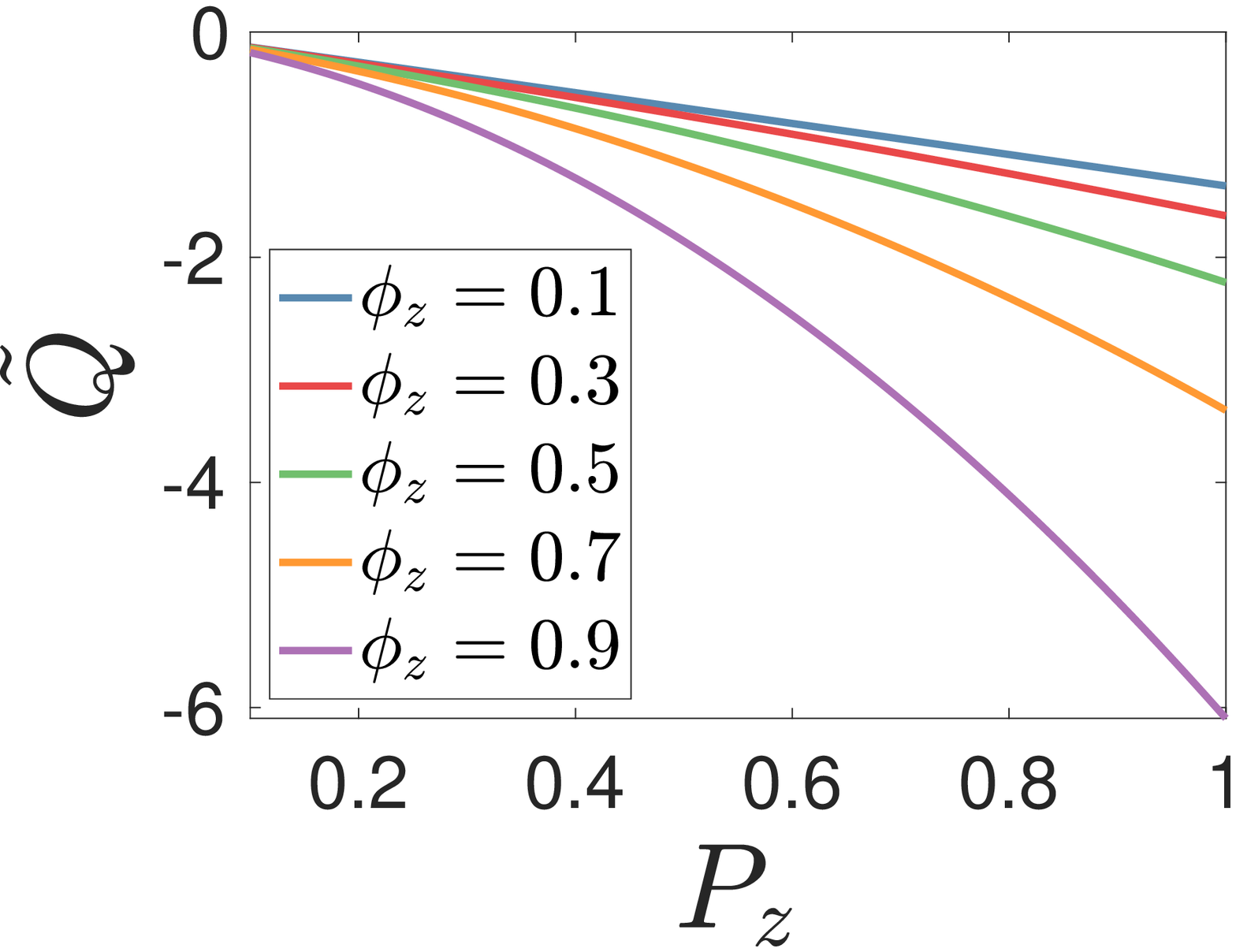} \includegraphics[trim={.0cm .3cm 1.8cm .7cm},clip,width=.24\textwidth]{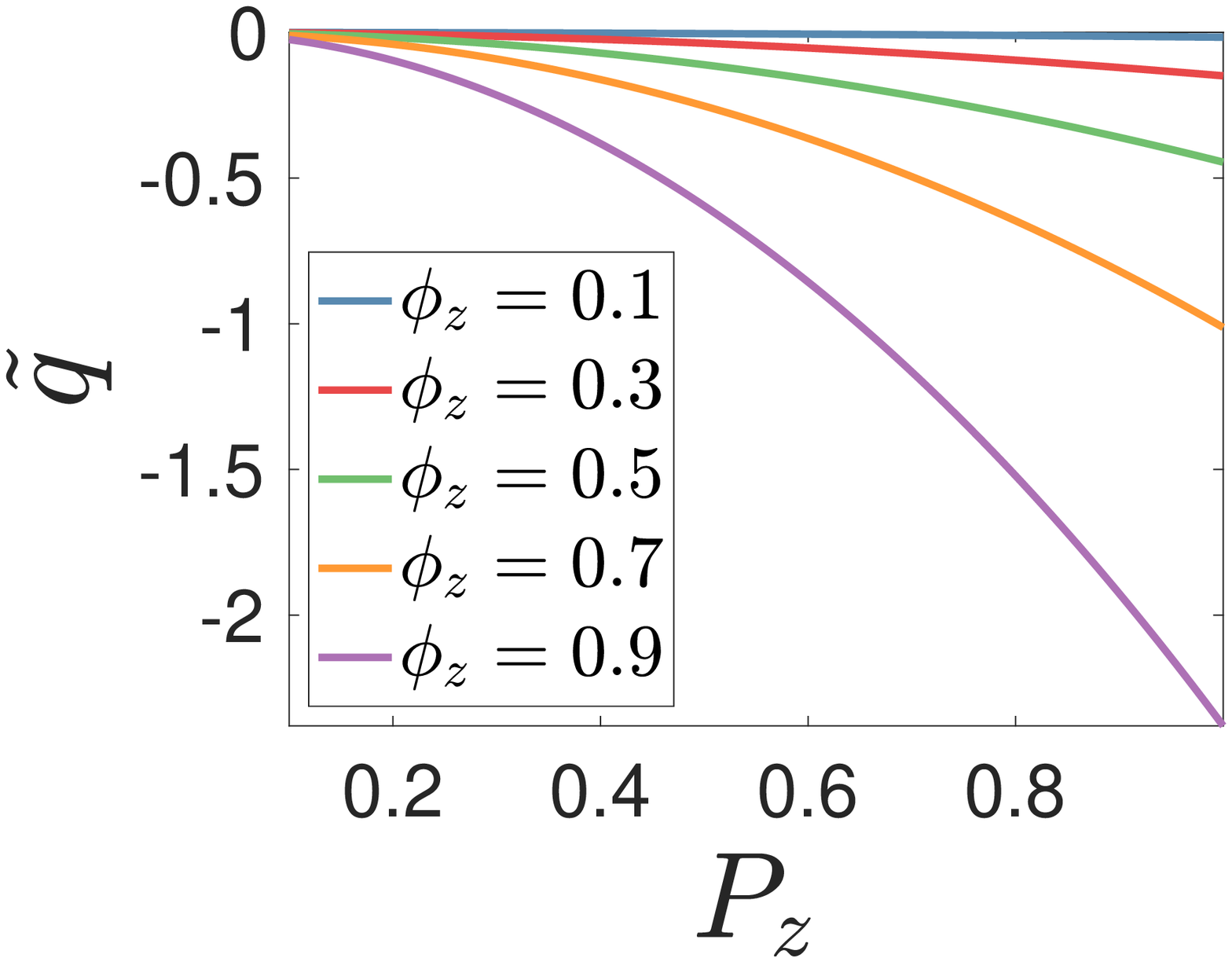}
    \includegraphics[trim={.0cm .3cm 1.8cm .7cm},clip,width=.24\textwidth]{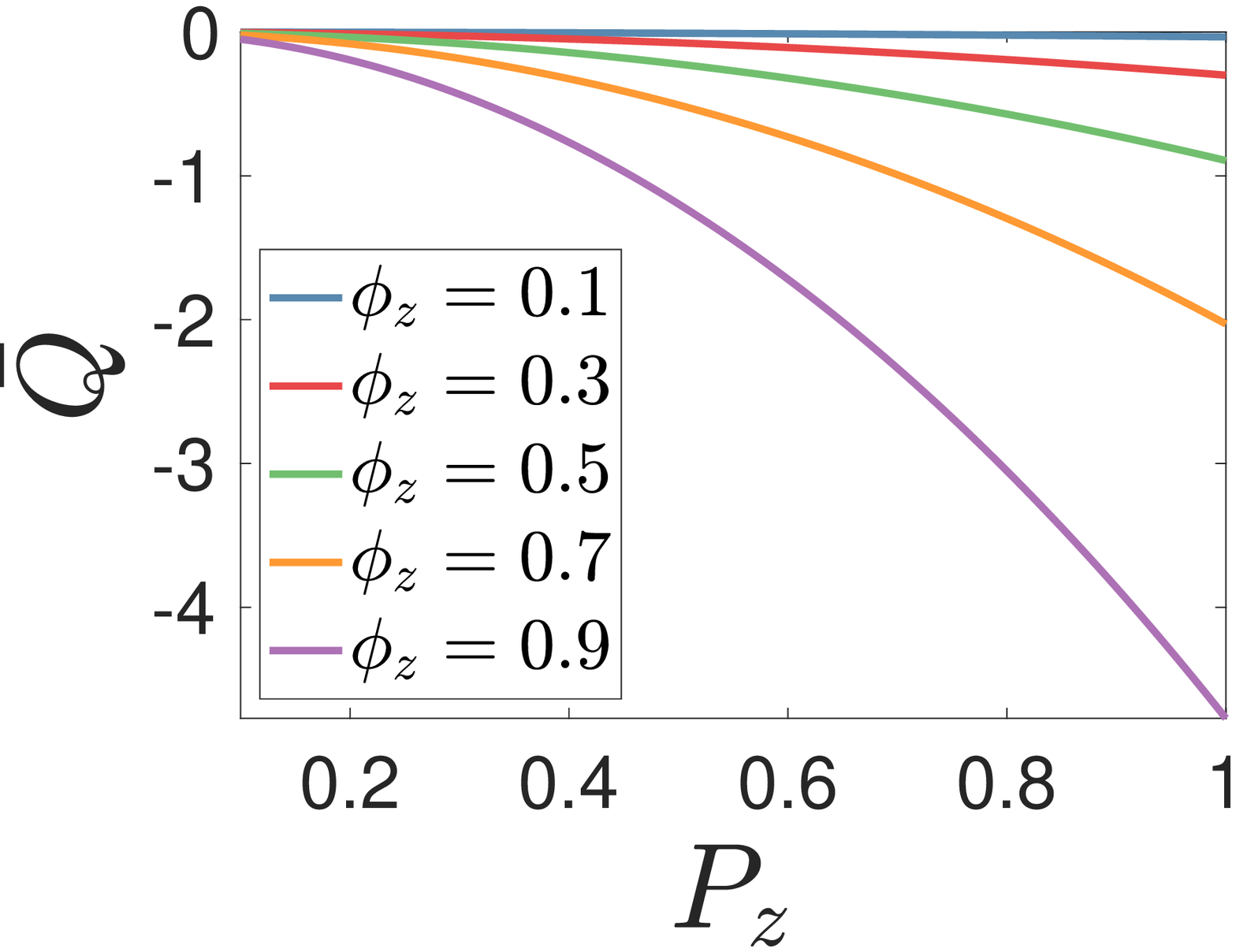} \includegraphics[trim={.0cm .3cm 1.8cm .7cm},clip,width=.24\textwidth]{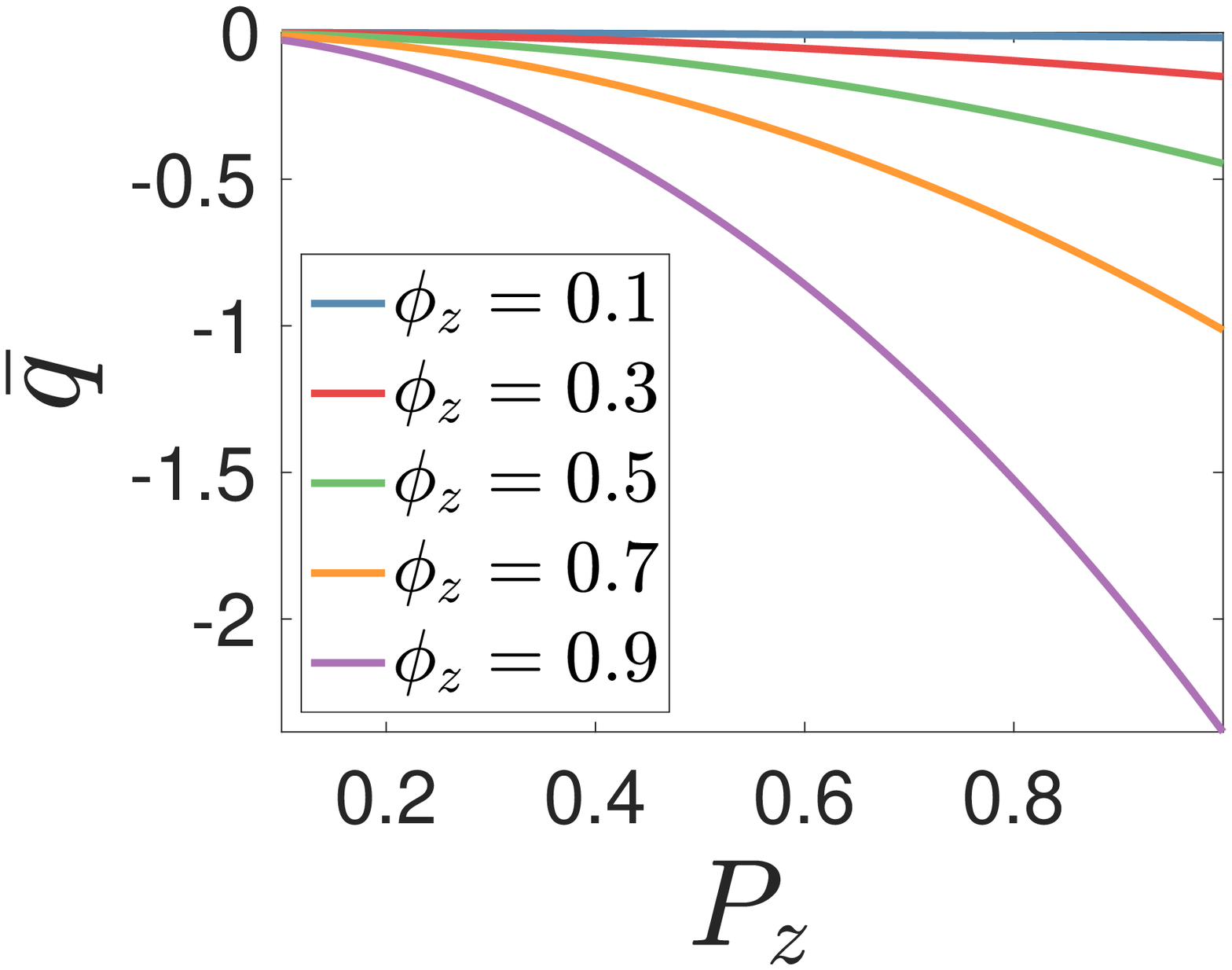}
    \caption{Plot of (\textit{a}) $\tilde{Q}$ defined in (\ref{eq:bulkfluxes}\textit{a}) and (\textit{b}) $\tilde{q}$ defined in (\ref{eq:surfacefluxes}\textit{b}), the contributions to the bulk and surface flux due to the pressure gradient $p_{0x}<0$, for varying $\phi_z$ and $P_z$.
    Plot of (\textit{c}) $\bar{Q}$ defined in (\ref{eq:bulkfluxes}\textit{b}) and (\textit{d}) $\bar{q}$  defined in (\ref{eq:surfacefluxes}\textit{c}), the contributions to the bulk and surface flux due to the surfactant gradient $\Gamma_{0x}>0$, for varying $\phi_z$ and $P_z$.
    Since $p_{0x}<0$, $\tilde{Q}$ and $\tilde{q}$ contribute positively to the leading-order velocity flux, whilst $\bar{Q}$ and $\bar{q}$ contribute negatively, since $\Gamma_{0x}>0$, following (\ref{eq:bulk_flux_constraints}\textit{a}).
    }
    \label{fig:first_order_fluxes}
\end{figure}

Numerical solutions to (\ref{eq:tildeu_bvp}, \ref{eq:baru_bvp}) can be seen in figure \ref{fig:first_order_velocity} for $\phi_z = 0.5$ and $P_z = 1$, computed using the method outlined in Appendix \ref{app:A}. Equation \eqref{eq:hatu_bvp} can be integrated to give $\breve{U} = y(y-2)/2$ and $\breve{Q} = \int_{z=-P_z}^{P_z} \int_{y=0}^2 \breve{U} \, \text{d} y \, \text{d} z = -4 P_z /3$. We define the volume and surface fluxes
\refstepcounter{equation}  \label{eq:bulkfluxes}
\begin{equation}
    \tilde{Q} = \int_{z=- P_z}^{P_z} \int_{y=0}^{2} \tilde{U} \, \text{d} y \, \text{d} z, \quad \bar{Q} =  \int_{z=- P_z}^{P_z} \int_{y=0}^{2} \bar{U} \,  \text{d} y \, \text{d} z,
	\tag{\theequation\textit{a,\,b}}
\end{equation}
\refstepcounter{equation}  
\label{eq:surfacefluxes}
\begin{equation}
q = \int_{z=-\phi_z P_z, \, y=0}^{\phi_z P_z} u_0 \, \text{d} z, \quad \tilde{q} = \int_{z=-\phi_z P_z, \, y=0}^{\phi_z P_z} \tilde{U} \, \text{d} z, \quad
\bar{q} = \int_{z=-\phi_z P_z, \, y=0}^{\phi_z P_z} \bar{U} \, \text{d} z . \tag{\theequation\textit{a--c}}
\end{equation}
Bulk ($\tilde{Q}$, $\bar{Q}$) and surface fluxes ($\tilde{q}$, $\bar{q}$) are plotted as functions of $\phi_z$ and $P_z$ in figure \ref{fig:first_order_fluxes}. 

Next, we substitute the expansions \eqref{eq:asymptotic_expansion} into the governing equations \eqref{eq:nondimensional_equations}--\eqref{eq:nondimensional_periodicity} to evaluate the $O(\epsilon^2)$ equations. 
The $O(\epsilon^2)$ system is required to calculate the first-order cross-channel flow field $(v_1, \, w_1)$, pressure field $p_1$, bulk $c_1$ and surface concentration field $\Gamma_1$ driven by streamwise gradients of the leading-order quantities ($\boldsymbol{u}_0=(u_0(x,y,z),\, 0,\, 0)$, $c_0(x)$ and $\Gamma_0(x)$), and to close the leading-order problem. 
In $\mathcal{D}_1$ and $\mathcal{D}_2$, 
\refstepcounter{equation} \label{eq:first_order}
\begin{equation}
	\boldsymbol{\nabla}_{\perp} \cdot \boldsymbol{u}_{\perp 1} = -u_{0x}, \quad 
	\nabla^2_{\perp} \boldsymbol{u}_{\perp 1} - \boldsymbol{\nabla}_{\perp} p_{1} = \boldsymbol{0}, \quad 
	\Pen^{-1}\nabla^2_{\perp} c_1 = - \Pen^{-1} c_{0xx} + u_0 c_{0x}.
	\tag{\theequation\textit{a--c}}
\end{equation}
Along $\mathcal{I}$,
\refstepcounter{equation} \label{eq:first_order_interface_bcs}
\begin{multline} 
	\boldsymbol{n}\cdot \boldsymbol{\nabla} w_1 - \Ma \Gamma_{1z} = 0, \quad
	v_1 = 0, \quad \boldsymbol{n}\cdot \boldsymbol{\nabla} c_1 - Da (c_1 -\Gamma_1)  =0, \\ \Pen^{-1}_I \Gamma_{1zz} + \Bi( c_1 - \Gamma_1) - \Gamma_0 w_{1z} = - \Pen_I^{-1} \Gamma_{0xx} + (u_0 \Gamma_0)_x,
	\tag{\theequation\textit{a--d}}
\end{multline}
and on $\partial \mathcal{I}$,
\begin{equation} \label{eq:first_order_interface_boundary bcs}
    w_1 \Gamma_0 - \Pen^{-1}_I \Gamma_{1 z} = 0 \quad \text{at} \quad z = \pm \phi_z P_z.
\end{equation}
Along $\mathcal{R}$ and $\mathcal{S}$,
\refstepcounter{equation} \label{eq:first_order_solid_bcs}
\begin{equation}
	v_1 = 0, \quad w_{1} =0, \quad c_{1y} =0.
    \tag{\theequation\textit{a--c}}
\end{equation}
Transverse periodicity can be rewritten as the symmetry conditions
\refstepcounter{equation} \label{eq:first_order_transverse_symmetry_bcs}
\begin{equation}  
	v_{1z} = 0, \quad w_{1} =0, \ \quad c_{1z} =0 \quad \text{at} \quad z = \pm P_z.
    \tag{\theequation\textit{a--c}}
\end{equation}

Solvability conditions imposed on \eqref{eq:first_order}--\eqref{eq:first_order_transverse_symmetry_bcs} constrain $u_0$, $c_0$ and $\Gamma_0$ at $O(1)$. 
The forcing on the right-hand sides of (\ref{eq:first_order}, \ref{eq:first_order_interface_bcs}) must be orthogonal to each vector in the null space of the linear operator that is adjoint to the left-hand sides of (\ref{eq:first_order}, \ref{eq:first_order_interface_bcs}).
Conveniently, these conditions are provided by the conservation arguments that result in the leading-order velocity and surfactant flux conditions (\ref{eq:leading_order_fluid_flux}, \ref{eq:leading_order_surfactant_flux}).
Substituting the leading-order streamwise velocity \eqref{eq:u_def} into the leading-order velocity flux condition \eqref{eq:leading_order_fluid_flux}, we obtain
\refstepcounter{equation} \label{eq:bulk_flux_constraints} 
\begin{equation} 
    \tilde{Q} p_{0x} + \Ma \bar{Q} \Gamma_{0x} =1, \quad q = \tilde{q} p_{0x} + \Ma \bar{q} \Gamma_{0x} \quad \text{in} \quad \mathcal{D}_1, \quad \breve{Q} p_{0x} = 1 \quad \text{in} \quad \mathcal{D}_2.
	\tag{\theequation\textit{a--c}}
\end{equation}
Substituting \eqref{eq:bulk_flux_constraints} into the leading-order surfactant flux condition \eqref{eq:leading_order_surfactant_flux} and remembering that $c_0(x)=\Gamma_0(x)$, we obtain the strong cross-channel diffusion problem
\begin{subequations} \label{eq:strong_diff}
\begin{align}
      \left(\beta + 1\right)c_0 - \gamma c_0 c_{0x} - \left(\alpha + \delta \right)c_{0x} &= 1 \quad \text{in} \quad \mathcal{D}_1, \\
      c_0 - \alpha c_{0x}  &= 1 \quad \text{in} \quad \mathcal{D}_2,
\end{align}
\end{subequations}
subject to continuity of bulk surfactant (\ref{eq:model3}\textit{a,\,b}).
In \eqref{eq:strong_diff}, we have introduced
\begin{subequations} \label{eq:coefficients}
\begin{align}
   \alpha &= \frac{4 P_z}{\Pen} = \frac{4 \hat{D} \hat{P}_z}{\hat{H}^2 \hat{U}} \quad \text{(bulk diffusion)}, \label{eq:coefficients1} \\ 
   \beta &= \frac{2 \Da \tilde{q}}{\Bi\Pen \tilde{Q}} = \frac{2 \hat{K}_a \tilde{q}}{\hat{H} \hat{K}_d \tilde{Q}} \quad \text{(partition coefficient)}, \label{eq:coefficients2} 
   \\  \gamma &=  \frac{2 \Ma \Da}{\Bi\Pen}\left(\frac{\tilde{q}\bar{Q}}{\tilde{Q}} - \bar{q}\right) = \frac{2 \hat{K}_a \hat{A} \hat{G}}{\hat{H}\hat{K}_d \hat{\mu} \hat{U}}\left(\frac{\tilde{q}\bar{Q}}{\tilde{Q}} - \bar{q}\right) \quad \text{(surfactant strength)}, \label{eq:coefficients3} \\
   \delta &= \frac{4 \phi_z P_z \Da}{\Bi \Pen \Pen_I} = \frac{4 \hat{D}_I \hat{K}_a \phi_z \hat{P}_z}{\hat{H}^3 \hat{K}_d \hat{U}} \quad \text{(surface diffusion)}. \label{eq:coefficients4}
\end{align}
\end{subequations}
The bulk diffusion coefficient $\alpha>0$ is a rescaled (by $4 \hat{P}_z/\hat{H}$) inverse P\'{e}clet number $1/\Pen = \hat{D}/(\hat{H} \hat{U})$, nominally assumed $O(1)$. The  coefficient $\alpha$ characterizes the ratio of streamwise diffusion to streamwise advection in the bulk for a given normalised channel width, assumed $O(1)$.  
The partition coefficient $\beta>0$ is the rescaled (by $2 \tilde{q}/\tilde{Q}$) normalised surfactant depletion depth $L_d = \hat{K}_a/(\hat{H} \hat{K}_d)$. 
The coefficient $\beta$ in (\ref{eq:strong_diff}\textit{a}) determines the portion of the surfactant flux transported by the pressure-driven flow along the interface in comparison to the surfactant flux transported by the pressure-driven flow in the bulk. 
For $\beta\gg 1$ ($\beta\ll 1$), the  surfactant advection flux is strongest at the interface (bulk).
The surfactant strength coefficient $\gamma>0$ is the rescaled (by $2(\tilde{q}\bar{Q}/\tilde{Q} - \bar{q})$) product of $L_d$ and the Marangoni number $\Ma = \hat{A} \hat{G} / (\hat{\mu} \hat{U})$, the ratio of surface tension changes due to interfacial surfactant to viscous forces. 
The coefficient $\gamma$ characterizes the nonlinear impact of streamwise surfactant-induced Marangoni stresses on the streamwise surfactant flux at the interface and in the bulk.
The surface diffusion parameter $\delta>0$ is a rescaled (by $4 \phi_z P_z$) product of $L_d$ and the inverse surface P\'{e}clet number $1/\Pen_I = \hat{D}_I/(\hat{H} \hat{U})$, nominally assumed $O(1)$. The coefficient $\delta$ measures the ratio of interfacial streamwise surface diffusion to streamwise advection for a given normalised interface width.

In summary, the leading-order problem \eqref{eq:leading_order}--\eqref{eq:leading_order_transverse_symmetry_bcs} has been reduced to the 5-parameter problem \eqref{eq:strong_diff} subject to (\ref{eq:model3}\textit{a,\,b}), a special case (where $\nu/\epsilon^2 \gg O(1,\,\alpha,\,\beta,\,\delta)$) of the 6-parameter problem \eqref{eq:model1}--\eqref{eq:model3}, requiring the solution of a first-order ODE to determine $c_0$ and hence ${DR}_0$ (explained in \S\ref{subsec:drag_reduction} below).
Then, the leading-order streamwise velocity \eqref{eq:u_def} and first-order problem \eqref{eq:first_order}--\eqref{eq:first_order_transverse_symmetry_bcs} can be used to construct the 3D flow field.

\subsubsection{Moderate cross-channel diffusion} \label{subsec:weak_diffusion}

In general, when cross-channel diffusion is weak and cross-channel gradients in the concentration field are comparable to the cross-channel average, a numerical technique must be employed to evaluate $c$.
We call this the ``generalised limit''.
In Appendix \ref{app:B}, we identify the regions of parameter space where cross-channel gradients first emerge by perturbing around the cross-channel-averaged concentration.
Writing $c_0 = \langle c_0 \rangle (x) + c_0'(x,\,y,\,z)$, we assume that $\|c_0'\|\equiv\sqrt{\langle c_0'^2\rangle} \ll \langle c_0 \rangle$ for all $x$, seeking to characterise the leading-order effects of shear dispersion in this problem \citep{taylor1953dispersion}.
We expect shear-dispersion effects to arise in surfactant-contaminated SHS channels because the pressure-driven flow advects surfactant at different speeds across the channel. 
The advected surfactant is then mixed by cross-channel diffusion, which in turn results in longitudinal dispersion of the surfactant distribution.
We call this the ``moderate cross-channel diffusion limit''.
The validity of this assumption is evaluated \textit{a posteriori} in Appendix \ref{app:D}.

To develop this limit, we assume that $1/\Pen = 1/\Pen_I = \Bi =  O(\epsilon^{2})$ and $\Da= \Ma = O(1)$ for $\epsilon \ll 1$, substituting the asymptotic expansion \eqref{eq:asymptotic_expansion} into the governing equations \eqref{eq:nondimensional_equations}--\eqref{eq:transport_eqs} (as in \S\ref{subsec:strong_diffusion}).
We then use the first-order system to eliminate $c_0'$ and the leading-order surfactant flux constraint to derive the moderate cross-channel diffusion problem, which is effectively a modification to the strong cross-channel diffusion problem \eqref{eq:strong_diff} accounting for shear dispersion,
\begin{subequations} \label{eq:weak_diffusion}
    \begin{align}
    \left(\beta + 1\right)c_0 - \gamma c_0 c_{0x} - \frac{\epsilon^2}{\alpha}\left(s_1 c_{0x} + s_2\frac{\gamma}{ \beta} c_{0x}^2 + s_3\frac{\gamma^2}{ \beta^2} c_{0x}^3\right) &= 1 \quad \text{in} \quad \mathcal{D}_1, \\
    c_0 - \frac{\epsilon^2}{\alpha} \left(s_4 c_{0x}\right) &= 1 \quad \text{in} \quad \mathcal{D}_2,
    \end{align}
\end{subequations}
subject to continuity of bulk surfactant (\ref{eq:model3}\textit{a,\,b}). 
We have defined the $O(1)$ coefficients
\refstepcounter{equation} \label{eq:s}
\begin{multline}
    s_1 =  - \frac{16 P_z^2 \langle \tilde{U} \tilde{C} \rangle}{\tilde{Q}^2}, \quad
    s_2 = - \frac{16 P_z^2 \tilde{q} (\tilde{Q} ( \langle \bar{U}\tilde{C}\rangle + \langle \tilde{U}\bar{C}\rangle)- 2 \bar{Q} \langle \tilde{U} \tilde{C} \rangle) }{\tilde{Q}^2 (\tilde{q}\bar{Q}-\bar{q}\tilde{Q})}, \\ 
    s_3 = - \frac{16 P_z^2 \tilde{q}^2 (\bar{Q}^2\langle \tilde{U}\tilde{C}\rangle - \bar{Q}\tilde{Q}(\langle \bar{U}\tilde{C}\rangle + \langle \tilde{U}\bar{C}\rangle) + \tilde{Q}^2\langle \bar{U}\bar{C}\rangle)}{\tilde{Q}^2 (\tilde{q}\bar{Q}-\bar{q}\tilde{Q})^2}, \quad
    s_4 = - \frac{16 P_z^2 \langle \breve{U} \breve{C}\rangle}{\breve{Q}^2}, \tag{\theequation\textit{a--d}}
\end{multline}
where $\tilde{C}$, $\bar{C}$ and $\breve{C}$ are solutions of the boundary-value problems \eqref{eq:atildec1}--\eqref{eq:ahatc}, respectively (see Appendix~\ref{app:B}), which depend on geometrical factors and $\Da$.
Here, ($\tilde{U}, \, \bar{U}$) are given by (\ref{eq:tildeu_bvp}, \ref{eq:baru_bvp}), ($\tilde{Q}, \, \bar{Q})$ by \eqref{eq:bulkfluxes}, and ($\tilde{q}, \, \bar{q}$) by \eqref{eq:surfacefluxes}. The coefficients $s_1>0$, $s_2$ and $s_3$ are plotted for different $\phi_z$, for $P_z=1$ and as a function of  $Da$ in figure~\ref{fig:s_i} (Appendix~\ref{app:B}). The coefficient $s_4=3/35$, since  $\breve{U} = y(y-2)/2$, $\breve{C} = 1/5 - y^3/6 + y^4/24$ and $\breve{Q} = - 4 P_z /3$ in $\mathcal{D}_2$.
Comparing the strong \eqref{eq:strong_diff} and moderate cross-channel diffusion problems \eqref{eq:weak_diffusion}, streamwise diffusion terms proportional to $\alpha$ and $\delta$ in \eqref{eq:strong_diff} have been replaced by shear dispersion terms proportional to $\epsilon^2/\alpha$ in  \eqref{eq:weak_diffusion}, some of which are nonlinear due to Marangoni effects.

\subsubsection{Composite equation} \label{subsubsec:composite_equations}

The ODEs in the strong cross-channel diffusion limit \eqref{eq:strong_diff} can be combined with the ODEs in the moderate cross-channel diffusion limit \eqref{eq:weak_diffusion} to construct the composite approximation
\begin{subequations} \label{eq:composite_equation}
\begin{align}
    (\beta+1) c_0 - \gamma c_0 c_{0x} - (\alpha+\delta) c_{0x} - \frac{\epsilon^2}{\alpha}\left(s_1 c_{0x} + s_2\frac{ \gamma}{ \beta} c_{0x}^2 + s_3\frac{ \gamma^2}{ \beta^2} c_{0x}^3\right) &= 1 \ \text{in} \ \mathcal{D}_1, \\
    c_0 - \alpha c_{0x} - \frac{\epsilon^2}{\alpha} \left(s_4 c_{0x}\right)&= 1 \ \text{in} \ \mathcal{D}_2,
\end{align}
\end{subequations}
subject to continuity of bulk surfactant (\ref{eq:model3}\textit{a,\,b}).
The dependence of $\alpha$, $\beta$, $\gamma$ and $\delta$ on $\Ma$, $\Pen$, $\Pen_I$, $\Bi$, $\Da$, $\phi_z$ and $P_z$ is given in (\ref{eq:coefficients}).
The dependence of $s_1$, $s_2$, $s_3$ and $s_4$ on $\Da$, $\phi_z$ and $P_z$ is given in \eqref{eq:s} and discussed in Appendix \ref{app:B}.
Together, the solution to \eqref{eq:composite_equation} depends on nine dimensionless parameters.
The numerical method used to solve (\ref{eq:model3}\textit{a}, \ref{eq:model3}\textit{b}, \ref{eq:composite_equation}) is outlined in Appendix \ref{subsec:transport}.
Incorporation of shear dispersion effects in \eqref{eq:composite_equation}, through perturbation of the cross-channel-averaged concentration ($c_0 = \langle c_0 \rangle(x) + c_0'(x,\, y,\, z)$), enables the strong-exchange model to explore regimes where cross-channel concentration gradients first become significant ($\|c_0'\| = O( \langle c_0 \rangle)$ for some $x$), enabling us to estimate approximately the boundary of validity of the 1D model \eqref{eq:model1}--\eqref{eq:model3} (Appendix \ref{app:D}).
When bulk diffusion is strong enough for shear dispersion to become negligible, \eqref{eq:composite_equation} reduces to \eqref{eq:strong_diff}, or \eqref{eq:model2} with $c_0 = \Gamma_0$.


\subsection{Moderate exchange} \label{subsec:weak_exchange}

In \S\ref{subsec:strong_exchange}, we assumed that $\Da= O(1)$ in order to have strong exchange at the interface, such that bulk and interfacial surfactants were in equilibrium at leading-order, i.e. $c_0=\Gamma_0$.
To study the case where $c_0$ and $\Gamma_0$ are not in equilibrium, we rescale $\Bi = \epsilon^2 \mathscr{B}$ and $\Da = \epsilon^2 \mathscr{D}$ with $\mathscr{B} = \mathscr{D} = O(1)$, whilst retaining $\Pen = \Pen_I = \Ma = O(1)$. 
Substituting the expansion \eqref{eq:asymptotic_expansion} into the governing equations \eqref{eq:nondimensional_equations}--\eqref{eq:transport_eqs}, then the leading-order strong cross-channel diffusion equations \eqref{eq:leading_order}, (\ref{eq:leading_order_interface_bcs}\textit{a--c}), \eqref{eq:leading_order_interface_boundary bcs}--\eqref{eq:leading_order_fluid_flux} and \eqref{eq:leading_order_solid_bcs}--\eqref{eq:leading_order_transverse_symmetry_bcs} are recovered.
Along the interface $\mathcal{I}$
\refstepcounter{equation} \label{eq:weak_ex_leading_order_interface_bcs}
\begin{equation} 
	c_{0y} =0, \quad \Pen^{-1}_I \Gamma_{0zz} - (w_0 \Gamma_0)_{z} = 0.
    \tag{\theequation\textit{a,\,b}}
\end{equation}
The total flux of surfactant \eqref{eq:dimensionless_surfactant_flux} is given by
\begin{equation} \label{eq:weak_ex_leading_order_surfactant_flux}
    \int_{z=- P_z}^{P_z} \int_{y=0}^{2} \left(u_0 c_0 - \frac{c_{0x}}{\Pen}\right) \, \text{d}y \, \text{d}z + \frac{2 \mathscr{D}}{\mathscr{B} \Pen }\int_{z=- \phi_z P_z, \, y=0}^{\phi_z P_z} \left(u_0 \Gamma_0 - \frac{\Gamma_{0x}}{\Pen_I}\right) \, \text{d}z = 1.
\end{equation}
The bulk and surface fluxes \eqref{eq:transport_eqs} are related to adsorption--desorption fluxes at the interface via
\begin{subequations} \label{eq:bulk_and_surfface_we}
\begin{align}
    \frac{\text{d}}{\text{d} x}\left(\int_{z=- P_z}^{P_z} \int_{ y=0}^{2}\left(u_0 c_0 - \frac{c_{0x}}{\Pen}\right) \, \text{d} y \, \text{d} z\right) &= \frac{2 \mathscr{D}}{\Pen} \int_{z=- \phi_z P_z, \, y=0}^{\phi_z P_z} (\Gamma_0 - c_0) \, \text{d} z , \\
    \frac{\text{d}}{\text{d} x}\left(\int_{z=-\phi_z P_z,\, y=0}^{\phi_z P_z} \left(u_0 \Gamma_0 - \frac{\Gamma_{0x}}{\Pen_I}\right) \, \text{d}z\right) &= - \mathscr{B} \int_{z=- \phi_z P_z,\, y=0}^{\phi_z P_z} (\Gamma_0 - c_0) \, \text{d} z.
\end{align}
\end{subequations}
The cross-channel velocity field again decays rapidly in time to $v_0 = w_0 = 0$. 
However, the bulk concentration $c_0 = c_0(x)$ and surface concentration $\Gamma_0 = \Gamma_0(x)$ are no longer equal at leading-order. 
The streamwise velocity field is given by $u_0 = \tilde{U} p_{0x} + \Ma \bar{U}\Gamma_{0x}$ in $\mathcal{D}_1$ and $u_0 = \breve{U} p_{0x}$ in $\mathcal{D}_2$, where $\tilde{U}$, $\bar{U}$ and $\breve{U}$ are given by \eqref{eq:tildeu_bvp}--\eqref{eq:hatu_bvp}.

At $O(\epsilon^2)$, the first-order strong cross-channel diffusion equations \eqref{eq:first_order}, (\ref{eq:first_order_interface_bcs}\textit{a,\,b}) and \eqref{eq:first_order_interface_boundary bcs}--\eqref{eq:first_order_transverse_symmetry_bcs} are recovered.
On $\mathcal{I}$,
\refstepcounter{equation} \label{eq:weak_ex_first_order_interface_bcs}
\begin{equation}
	c_{1y} = \mathscr{D}(c_0 - \Gamma_0), \quad \Pen^{-1}_I \Gamma_{1zz} - \Gamma_0 w_{1z} = - \Pen^{-1}_I \Gamma_{0xx}  + (u_0 \Gamma_0)_x - \mathscr{B}(c_0 - \Gamma_0).
	\tag{\theequation\textit{a,\,b}}
\end{equation}
This system is required to evaluate the first-order cross-channel flow, $(v_1,\,w_1)$, driven by streamwise gradients of $u_0$, $c_0$ and $\Gamma_0$, and to close the leading-order problem.

Substituting the streamwise velocity \eqref{eq:u_def} into the velocity flux condition \eqref{eq:leading_order_fluid_flux}, we recover $\tilde{Q} p_{0x} + \Ma \bar{Q} \Gamma_{0x} =1$ in $\mathcal{D}_1$ and $\breve{Q} p_{0x} = 1 $ in $\mathcal{D}_2$, where $\tilde{Q}$ is given by (\ref{eq:bulkfluxes}\textit{a}), $\bar{Q}$ is given by (\ref{eq:bulkfluxes}\textit{b}) and $\breve{Q}=- 4 P_z /3$.  
Substituting the streamwise velocity \eqref{eq:u_def} into \eqref{eq:bulk_and_surfface_we} we derive the steady leading-order surfactant transport equations \eqref{eq:model1}.
Substituting \eqref{eq:u_def} into the surfactant transport equations \eqref{eq:weak_ex_leading_order_surfactant_flux} (or adding (\ref{eq:model1}\textit{a}) and (\ref{eq:model1}\textit{b}) and integrating) we recover integral constraints on the surfactant transport equations \eqref{eq:model1}, which are given by surfactant flux constraints \eqref{eq:model2}.
The coefficients $\alpha$, $\beta$, $\gamma$ and $\delta$ are defined in \eqref{eq:coefficients} and
\begin{equation} \label{eq:coefficients_weak_exchange}
   \nu = \frac{4 \phi_z P_z \Da}{\Pen} = \frac{4 \hat{K}_a \phi_z \hat{P}_z}{\hat{H} \hat{U}} \quad \text{(surface exchange)}.
\end{equation}
The surface exchange coefficient $\nu$ is a rescaled (by $4 \phi_z \hat{P}_z$) product of the inverse bulk P\'{e}clet and \Damkohler numbers, comparing the rate of adsorption of surfactant contained in the layer of fluid $\hat{H}$ with the rate of advection.
At the domain boundaries, we impose continuity of bulk surfactant (\ref{eq:model3}\textit{a,\,b}), continuity of bulk surfactant flux between $\mathcal{D}_1$ and $\mathcal{D}_2$ (\ref{eq:model3}\textit{c}) and no flux of interfacial surfactant (\ref{eq:model3}\textit{d}).
For $\nu/\epsilon^2  \gg O(1,\,\alpha,\,\beta,\,\delta)$ then $c_0 = \Gamma_0$ at leading-order, recovering the transport equations \eqref{eq:strong_diff} in the strong cross-channel diffusion and strong exchange limit.
The numerical method used to solve the 1D surfactant transport equations \eqref{eq:model1}--\eqref{eq:model3} is outlined in Appendix \ref{subsec:transport}.


\subsection{Drag reduction} \label{subsec:drag_reduction}

As mentioned in \S\ref{sec:formulation}, we study the transition from an immobilised interface where ${DR}_0=0$, to a shear-free interface where ${DR}_0=1$.
When the interface is shear-free, $\tilde{Q} p_{0x} = 1$ in $\mathcal{D}_1$ and $\breve{Q} p_{0x} = 1$ in $\mathcal{D}_2$, such that integrating $-p_{0x}$ across the period gives
\begin{equation}
    \Delta p_U = -\int_{x = -\phi_x}^{\phi_x} p_{0x} \, \text{d} x - \int_{x = \phi_x}^{2 - \phi_x} p_{0x} \, \text{d} x = -\frac{2\phi_x}{\tilde{Q}} - \frac{2(1-\phi_x)}{\breve{Q}}.
\end{equation} 
In the limit where the interface is immobilized, $\breve{Q} p_{0x} = 1$ in $\mathcal{D}_1$ and $\mathcal{D}_2$, giving $\Delta p_R = - 2/\breve{Q}$.
Note that when $q=0$, the surface velocity flux condition (\ref{eq:bulk_flux_constraints}\textit{b}) implies that $\tilde{q} p_{0x} + \Ma \bar{q}\Gamma_{0x}=0$. 
We can substitute this into the bulk velocity flux condition (\ref{eq:bulk_flux_constraints}\textit{a}), to find $p_{0x}(\tilde{Q}\bar{q} - \bar{Q}\tilde{q})=\bar{q}$ in $\mathcal{D}_1$.
However, we also have $p_{0x} = 1/\breve{Q}$ in $\mathcal{D}_1$ for $q=0$.
Combining these two expressions for $p_{0x}$ in $\mathcal{D}_1$, gives $\tilde{Q}\bar{q} - \bar{Q}\tilde{q} = \bar{q}\breve{Q}$ and $\bar{q}(\tilde{Q} - \breve{Q}) = \bar{Q}\tilde{q}$.
It follows that $\breve{Q}\bar{Q}\tilde{q}/((\tilde{q}\bar{Q} - \bar{q}\tilde{Q})(\breve{Q} - \tilde{Q})) = 1$, which is valid for any $q$.
Between these no-slip and shear-free limits, $\tilde{Q} p_{0x} + \Ma \bar{Q} \Gamma_{0x} =1$ in $\mathcal{D}_1$ and $\breve{Q} p_{0x} = 1$ in $\mathcal{D}_2$, such that integrating $-p_{0x}$ across the period gives
\begin{equation} \label{eq:leading_order_drop}
    \Delta p_0 = -\int_{x = -\phi_x}^{\phi_x} p_{0x} \, \text{d} x - \int_{x = \phi_x}^{2 - \phi_x} p_{0x} \, \text{d} x = -\frac{2\phi_x}{\tilde{Q}} - \frac{2(1-\phi_x)}{\breve{Q}}  + \frac{\Ma \bar{Q} \Delta \Gamma_0}{\tilde{Q}},
\end{equation}
where we have defined $\Delta \Gamma_0 = \Gamma_0(\phi_x) - \Gamma_0(-\phi_x) > 0$ ($\Delta \Gamma_0 = \Delta c_0$ in the strong exchange limit).
Using the definition of $(\beta,\, \gamma)$ in (\ref{eq:coefficients2}, \ref{eq:coefficients3}) and substituting $\Delta p_U$, $\Delta p_R$ and $\Delta p_0$ into the leading-order drag reduction \eqref{eq:leading_order_drag}, we have
\begin{equation} \label{eq:dr_def}
        {DR}_0 = 1 - \frac{\gamma \Delta \Gamma_0}{2\phi_x \beta}.
\end{equation}
In \eqref{eq:dr_def}, the first term gives the drag reduction when the interface is shear-free and the second term measures the impact of surfactant.
In order to derive the expression in \eqref{eq:dr_def}, we used the fact that $\breve{Q}\bar{Q}\tilde{q}/((\tilde{q}\bar{Q} - \bar{q}\tilde{Q})(\breve{Q} - \tilde{Q})) = 1$.

Drag reduction is  one possible measure of the performance of surfactant-contaminated SHSs. 
An alternative measure, commonly used for laminar flows over SHSs, is the effective slip length $\lambda_e$ \citep{landel2020theory, temprano2023single}, defined as the uniform slip length applied to the top and bottom boundaries of an equivalent channel of the same height as the SHS channel and which has the same flow rate as the SHS channel under the same pressure gradient \citep{lauga2003effective}.
To evaluate $\lambda_e$, we integrate the leading-order streamwise momentum equation (\ref{eq:leading_order}\textit{b}) for an equivalent channel with the mixed boundary conditions (\ref{eq:leading_order_interface_bcs}, \ref{eq:leading_order_solid_bcs}) replaced by $\lambda_e u_{0y} - u_0 = 0$ on $\mathcal{I}$, $\mathcal{R}$ and $\mathcal{S}$. 
We obtain $u_0 = \check{U} \Delta p_{0x}$ where $\check{U} = y(y-2)/2 - \lambda_e$ and $p_{0x}$ is the same pressure gradient as in the SHS channel. 
The flux is $\check{Q} = \int_{z = - P_z}^{P_z} \int_{y = 0}^{2} \check{U} p_{0x} \text{d}y  \text{d}z = (\breve{Q} - 2 P_z \lambda_e)p_{0x}$, or by integrating over one period $\check{Q} = (\breve{Q} - 2 P_z \lambda_e) \Delta p_{0} / 2$.
Equating the flux of the equivalent channel with the flux of the SHS channel, $\tilde{Q}p_{0x} + \Ma \bar{Q}\Gamma_{0x} = 1$, we find 
\begin{equation} \label{eq:lambda_e}
    \lambda_e = \frac{{DR}_0(\Delta p_R - \Delta p_U)}{P_z \Delta p_R(\Delta p_U {DR}_0  + \Delta p_R (1-{DR}_0))},
\end{equation}
which can be used to convert results from ${DR}_0$ to $\lambda_e$.

\section{Results} \label{sec:results} 

In \S\ref{subsec:Strong exchange}, we investigate the leading-order drag reduction (${DR}_0$), bulk surfactant concentration ($c_0$), interfacial surfactant concentration ($\Gamma_0$) and flow field in the strong-exchange problem, (\ref{eq:composite_equation}, \ref{eq:model3}\textit{a}, \ref{eq:model3}\textit{b}), by varying the bulk diffusion ($\alpha$), partition coefficient ($\beta$), surfactant strength ($\gamma$) and surface diffusion ($\delta$).
In \S\ref{subsec:Weak exchange}, we address the moderate-exchange problem, \eqref{eq:model1}--\eqref{eq:model3}, by varying $\alpha$, $\beta$, $\gamma$, $\delta$ and the exchange strength ($\nu$).
In both strong- and moderate-exchange problems, we identify three primary areas of the parameter space. 
The Marangoni-dominated (M) region, where the interfacial surfactant gradient is sufficient to immobilise the liquid--gas interface (low drag reduction); the advection-dominated region (A), where interfacial surfactant has been swept to the downstream stagnation point and the liquid--gas interface is mostly shear free (high drag reduction); and the diffusion-dominated (D) region, where the interfacial surfactant gradient has been attenuated by diffusion and the liquid--gas interface is mostly shear free (high drag reduction).
Throughout \S\ref{sec:results}, the gas fraction and transverse period width are maintained at $\phi_x = \phi_z = 0.5$ and $P_z = 1$ for simplicity. 
However, general asymptotic solutions are derived for any $\phi_x$, $\phi_z$ and $P_z$.

\subsection{Strong exchange} \label{subsec:Strong exchange}

\subsubsection{Drag reduction} \label{Weak partition coefficient}

\begin{figure}
\begin{minipage}{0.67\textwidth}
\centering
(\textit{a}) \hfill \hfill \hfill \\
\includegraphics[trim={0 1cm 0 1.5cm},clip,width=\linewidth]{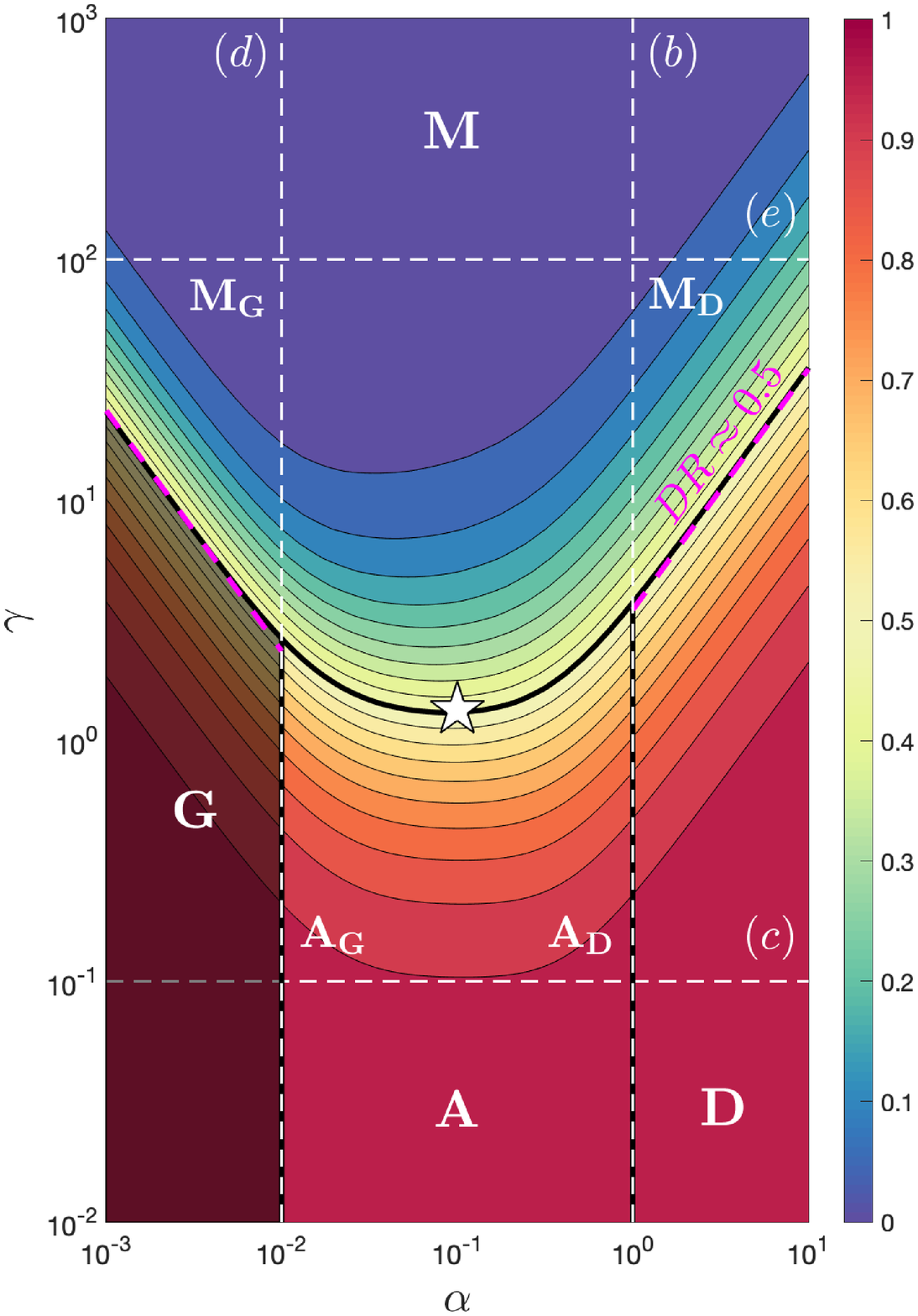}
\end{minipage}
\begin{minipage}{0.29\textwidth}
\centering
(\textit{b}) \hfill \hfill \hfill \\
\includegraphics[width=\linewidth]{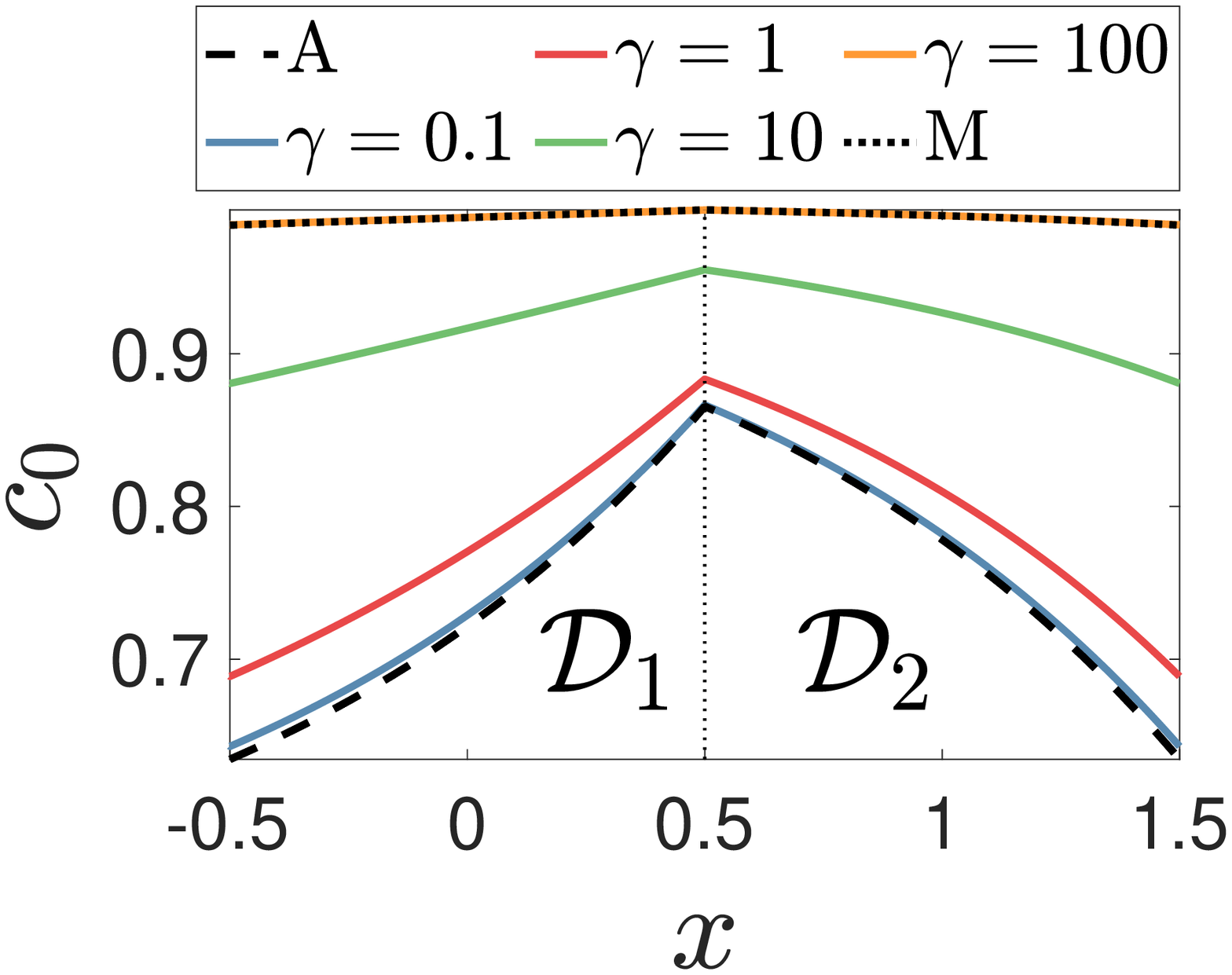} \\
\vspace{-.5cm}
(\textit{c}) \hfill \hfill \hfill \\
\includegraphics[width=\linewidth]{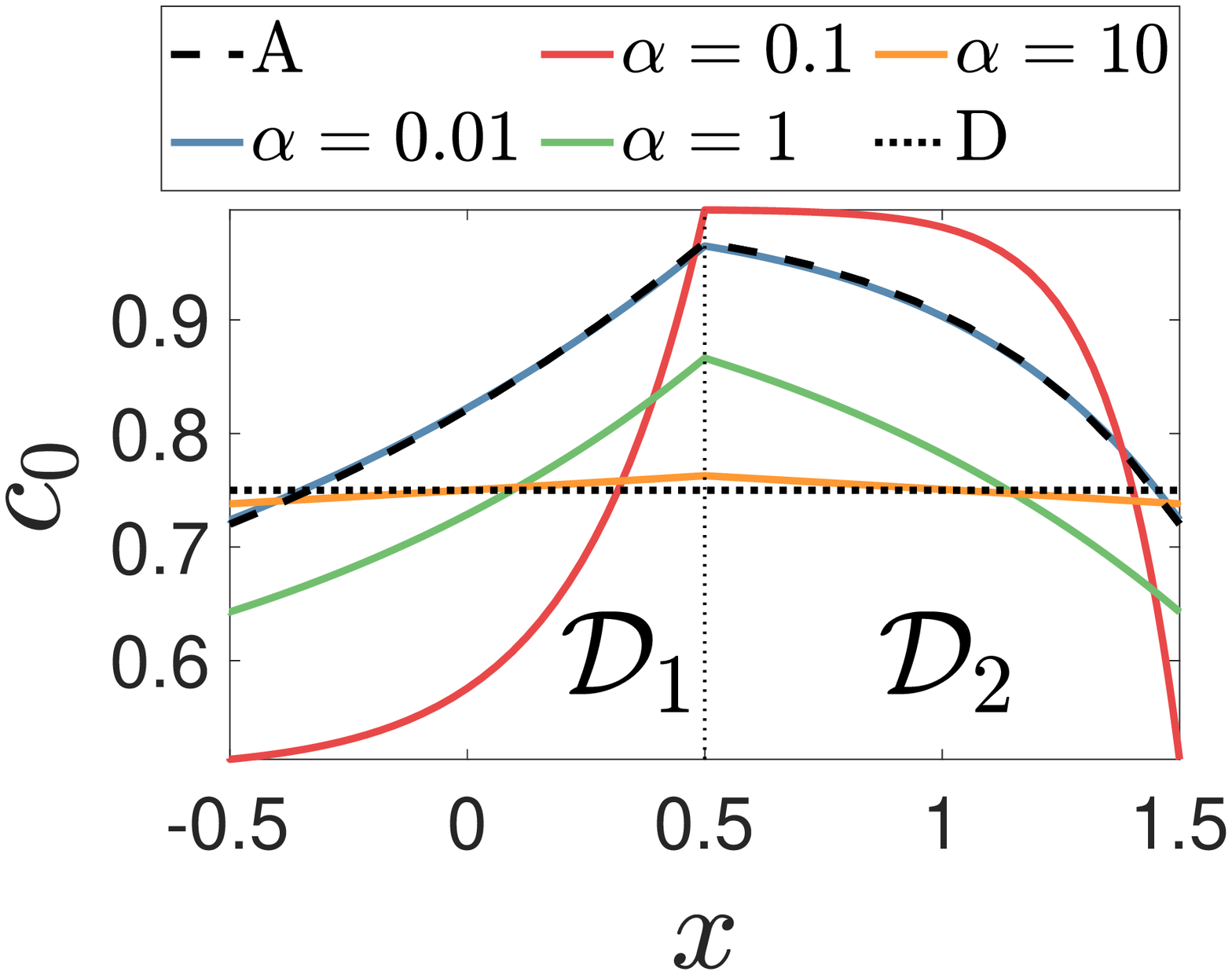} \\
\vspace{-.5cm}
(\textit{d}) \hfill \hfill \hfill \\
\includegraphics[width=\linewidth]{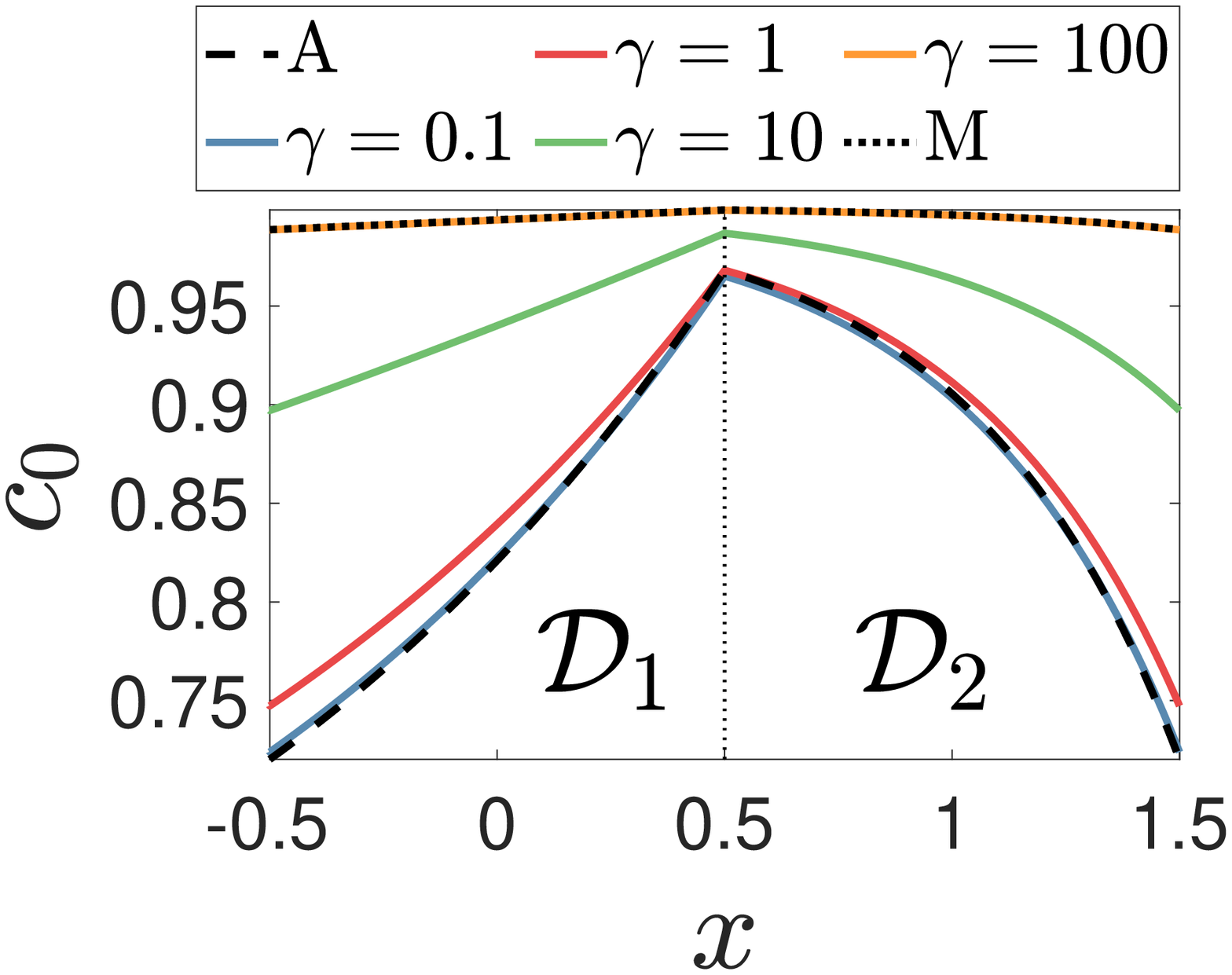} \\
\vspace{-.5cm}
(\textit{e}) \hfill \hfill \hfill \\
\includegraphics[width=\linewidth]{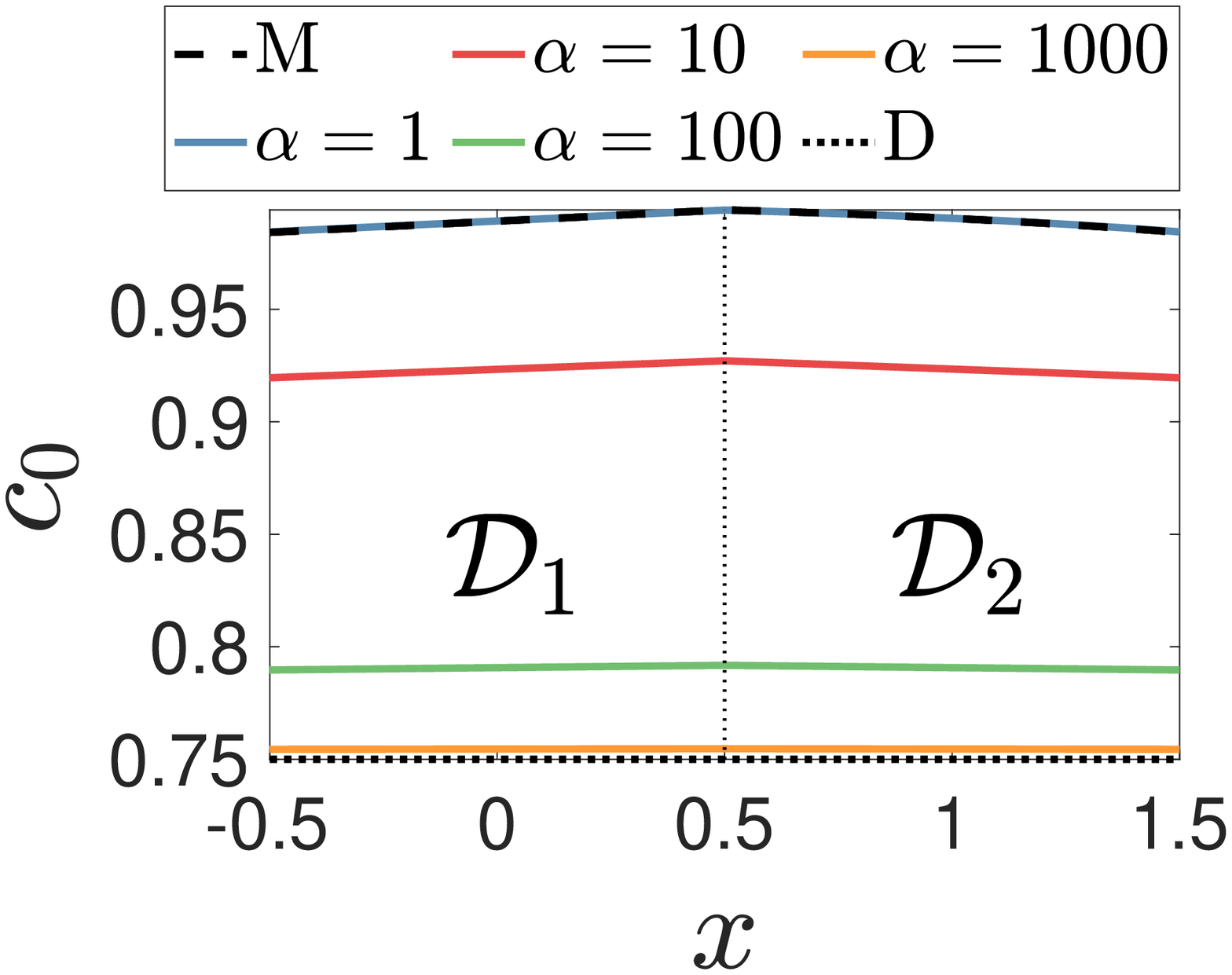} \\
\end{minipage}
\caption{
The leading-order drag reduction (${DR}_0$) and surfactant distribution ($c_0=\Gamma_0$) in the strong-exchange problem, for $\beta = 1$, $\phi_x = 0.5$, $\epsilon = 0.1$, $\phi_z = 0.5$, $\Da = 1$ and $P_z = 0.5$, computed using \eqref{eq:composite_equation}.
(\textit{a}) Contours of ${DR}_0$, where ${DR}_0 = 0$ exhibits a no-slip SHS and ${DR}_0 = 1$ exhibits a shear-free plastron. 
The Marangoni (M), advection (A) and diffusion-dominated (D) regions are separated by black lines and ${DR}_0$ is approximated by (\ref{eq:drs_m}), (\ref{eq:drs_a}) and (\ref{eq:drs_d}) for the M, A and D regions, respectively.
The generalised (G) region shaded in brown is discussed in Appendix \ref{app:D}.
The dashed magenta lines describing when ${DR}_0 \approx0.5$ are given by (\ref{eq:smboundary}, \ref{eq:mdboundary}).
(\textit{b}--\textit{e}) Plots of $c_0$ for varying surfactant strength ($\gamma$) and bulk diffusion ($\alpha$), where the asymptotic curves are: (\textit{b}) $\alpha = \delta = 1$, A: $--$ \eqref{eq:zeta_ll_1}, M: $\cdot\cdot\cdot$ \eqref{eq:zeta_gg_1} with $\gamma = 100$, (\textit{c}) $\gamma = 0.1$, A: $--$ \eqref{eq:zeta_ll_1} with $\alpha = \delta = 0.01$, D: $\cdot\cdot\cdot$ \eqref{eq:alpha_gg_1} with $\alpha = \delta = 10$, (\textit{d}) $\alpha = \delta = 0.01$, A: $--$ \eqref{eq:zeta_ll_1}, M: $\cdot\cdot\cdot$ \eqref{eq:zeta_gg_1} with $\gamma = 100$, and (\textit{e}) $\gamma = 100$, M: $--$ \eqref{eq:zeta_gg_1} with $\alpha = \delta = 1$, D: $\cdot\cdot\cdot$ \eqref{eq:alpha_gg_1} with $\alpha = \delta = 1000$.
The star identifies the point of the $(\alpha, \, \gamma)$-plane where we examine the flow field in \S\ref{subsubsec:ff_strong_exchange}.}
\label{fig:map_1}
\end{figure}

Figure \ref{fig:map_1}(\textit{a}) shows how ${DR}_0$ varies with the bulk diffusion ($\alpha$) and surfactant strength ($\gamma$), for $\beta = 1$.
In \S\ref{subsec:Strong exchange}, we set $\alpha = \delta$ for simplicity. 
However, general asymptotic solutions are derived for any $\delta$.
The governing equations simplify in different regions of the parameter space, where subsets of the terms in (\ref{eq:composite_equation}) are dominant.
These distinct physical balances are reflected by limits or transitions of ${DR}_0$, as well as variations in the concentration profiles.
The three primary areas of the parameter space, regions M, D and A, are separated by black lines in figure \ref{fig:map_1}(\textit{a}).
They are analysed asymptotically in Appendix \ref{app:C}. 
Another more general region (G) exists where cross-channel gradients in concentration can be comparable to the cross-channel average (shaded in figure \ref{fig:map_1}\textit{a}). As region G lies beyond the model predictions, its boundaries delimit the domain of validity of the model (Appendix \ref{app:D}). 
To contextualise the changes in ${DR}_0$, we examine how $c_0$ and $\Gamma_0$ vary across one period in figure \ref{fig:map_1}(\textit{b}--\textit{e}).
Recall that $\mathcal{D}_1$ and $\mathcal{D}_2$ represent the subdomains over the plastron and solid ridge, respectively, ($\mathcal{D}_1$ and $\mathcal{D}_2$ are separated by the vertical dotted line at $x = \phi_x=0.5$), and $c_0 = \Gamma_0$ in $\mathcal{D}_1$. 

To start our journey around the parameter space in figure \ref{fig:map_1}(\textit{a}), we address region M, in which Marangoni effects are sufficiently strong to render the interface almost immobile.
That is, the surfactant is transported along the liquid--gas interface by the flow and accumulates at stagnation points, where the adverse Marangoni stress generated is  large enough to reduce the streamwise velocity at the liquid--gas interface to negligible values along the whole length of the plastron.
As shown in Appendix \ref{subsec:a_strong_surfactant_strength}, ${DR}_0$ is close to zero in  region M and takes the value
\begin{equation} \label{eq:drs_m}
{DR}_0 \approx \frac{1}{\gamma}\left(\alpha + \delta + \frac{\epsilon^2 s}{\alpha} +  \frac{\phi_x(E+1)}{\left(E-1\right)} \right)
\quad \text{for} \quad \gamma \gg \max\left(1,\,\alpha,\,\beta,\,\delta,\,\frac{\epsilon^2}{\alpha}\right),
\end{equation}
where $E \equiv \exp(2\alpha(1-\phi_x)/(\alpha^2 + \epsilon^2 s_4))$ and $s \equiv s_1 + s_2 + s_3$ (note that $s > 0$ for all $\phi_z$, $P_z$ and $\Da$ examined in Appendix \ref{app:B}).
The leading-order drag reduction can be simplified within two sub-regions: $\text{M}_\text{D}$, in which Marangoni effects start to compete with bulk diffusion, and $\text{M}_\text{G}$, in which Marangoni effects start to compete with shear dispersion. 
In the $\text{M}_\text{D}$ case, \eqref{eq:drs_m} reduces to ${DR}_0 \approx \alpha / (\gamma(1 - \phi_x)) + \delta / \gamma$ for $\gamma \gg (\alpha,\, \delta) \gg \max(1,\,\beta,\,\epsilon^2/\alpha)$, 
demonstrating how bulk diffusion (noting that $\alpha/\gamma \propto \hat{\mu} \hat{D}\hat{P}_z/(\hat{A}\hat{G} \hat{H}\hat{L}_d)$) and surface diffusion (noting that $\delta/\gamma \propto \hat{\mu} \hat{D}_I \phi_z \hat{P}_z / (\hat{A}\hat{G}\hat{H}^2)$) weaken the immobilizing effects of surfactant. 
In the $\text{M}_\text{G}$ case, \eqref{eq:drs_m} reduces to ${DR}_0 \approx \epsilon^2 (s_4 \phi_x + s (1 - \phi_x)) / (\gamma\alpha(1 - \phi_x))$ for 
$\gamma \gg \epsilon^2 / \alpha \gg \max(1,\,\alpha,\,\beta,\, \delta)$,
demonstrating how shear dispersion starts to mobilise the interface (noting that $\epsilon^2/(\alpha\gamma) \propto \hat{\mu} \hat{U}^2\hat{H}^5/(\hat{D}\hat{A}\hat{G}\hat{L}_d\hat{P}_z\hat{P}_x^2)$ shows a quadratic dependence on the bulk flow speed).

Representative concentration profiles in region M are shown in figure \ref{fig:map_1}(\textit{b,\,d,\,e}).
Here, the leading-order concentration solution in $\mathcal{D}_1$ (Appendix \ref{subsec:a_strong_surfactant_strength}) is linear with a shallow gradient,
\begin{equation} \label{eq:zeta_gg_1}
c_{0} \approx 1 + \frac{\beta}{\gamma}\left(x - \frac{\phi_x (E + 1)}{E - 1}\right) \quad \text{for} \quad \gamma \gg \max\left(1,\,\alpha,\,\beta,\,\delta,\,\frac{\epsilon^2}{\alpha}\right).
\end{equation}
The gradient of $c_0$ is controlled by both advection and Marangoni effects (noting that $\beta/\gamma\propto 1/Ma= \hat{\mu}\hat{U}/(\hat{A}\hat{G})$, which shows that an advective flow is needed to set up a Marangoni gradient), providing a surface shear stress sufficient to immobilise the liquid--gas interface.
From \eqref{eq:zeta_gg_1}, the magnitude of the concentration in $\mathcal{D}_1$  is weakly regulated by fluxes driven by bulk diffusion or shear dispersion in domain $\mathcal{D}_2$, via the factor $\alpha+s_4 \epsilon^2/\alpha$  in $E$ that appears due to continuity of concentration across $\mathcal{D}_1$ and $\mathcal{D}_2$.

The M region transitions into the G region across the GM boundary, where the concentration field starts to develop appreciable cross-channel gradients.
In Appendix \ref{app:D}, we show that a 3D perturbation ($c_0'$) to the cross-channel-averaged surfactant field ($\langle c_0 \rangle$) becomes comparable to $\langle c_0 \rangle$ when Marangoni effects compete with shear dispersion, with $\gamma = O(\epsilon^2/\alpha)$ for $\beta=O(1)$.
Similar to streamwise diffusion, shear dispersion increases  drag reduction when approaching the GM boundary, as shown in figure \ref{fig:map_1}(\textit{a}).
We conjecture that at 50\% drag reduction, shear dispersion becomes large enough to induce an appreciable cross-channel gradient, defining a possible boundary between region M and G. However, 3D numerical simulations would be required to test this hypothesis and determine precisely the GM boundary, beyond which our asymptotic results \eqref{eq:drs_m} and \eqref{eq:zeta_gg_1} in region M are no longer valid.
Therefore, our asymptotic approximation below to determine the GM boundary based on estimating ${DR}_0 = 0.5$ should be used qualitatively rather than quantitatively.
At the GM boundary, we find (Appendix \ref{subsec:a_weak_diff_and_surf})
\begin{equation} \label{eq:smboundary}
    {DR}_0 \approx 0.5 \quad \text{when} \quad \gamma = \frac{\epsilon^2(2 s_{1}(1 - \phi_x) + s_4 \phi_x (2 + \beta))}{2 \alpha \left(1- \phi_x\right)} \quad \text{for} \quad \alpha \ll 1,
\end{equation}
which gives the leftmost dashed magenta line in figure \ref{fig:map_1}(\textit{a}).
This approximation agrees with the numerical solution of the 1D strong-exchange problem in the limit $\alpha \rightarrow 0$.
As the immobilizing effect of surfactant decreases across the GM boundary, larger streamwise velocities in the cross-plane imply that $c_0$ decreases from $c_0\approx 1$ to smaller values along the channel (see figure~\ref{fig:map_1}\textit{d}). 
This is due to the velocity and surfactant flux being fixed.

The M region transitions into the D region through the DM boundary for $\gamma=O(\alpha)$ or $\gamma=O(\delta)$, which can be defined (see Appendix \ref{subsec:a_strong_diff_and_surf}) by
\begin{equation} \label{eq:mdboundary}
    {DR}_0 \approx 0.5 \quad \text{when} \quad \gamma = \delta + \frac{\alpha(2 + \beta \phi_x)}{2(1 - \phi_x)} \quad \text{for} \quad \min(\alpha,\,\delta) \gg 1,
\end{equation}
which gives the rightmost dashed magenta line in figure \ref{fig:map_1}(\textit{a}).
The asymptotic approximation \eqref{eq:mdboundary} agrees with the numerical solution of \eqref{eq:composite_equation} (thick black line in figure \ref{fig:map_1}\textit{a}) for $\min(\alpha,\,\delta) \gg 1$.
Similar to the GM transition, the immobilizing effect of surfactant decreases as we move from M to D by increasing the strength of bulk diffusion. 
By increasing the strength of bulk diffusion, bulk and interfacial surfactant gradients are attenuated because bulk--surface exchange is strong; this allows the streamwise velocity  and drag reduction to increase. 
Since the velocity and surfactant flux are fixed, the increase in velocity leads to a decrease in $c_0$, as shown in figure \ref{fig:map_1}(\textit{e}).

In summary, the factors promoting drag reduction from a state of interfacial immobilisation (in region M) are bulk diffusion (moving into region D), shear dispersion (moving into region G) or reduced surfactant strength (moving into region G, D or A, of which more details are given below).
As we move into region D (or G), the surfactant in the bulk of the channel spreads out via diffusion (or shear dispersion), transporting surfactant from areas of high concentration to low concentration. 
Then, because exchange between the bulk and interfacial surfactant is strong, the concentration fields rapidly equilibrate; the interfacial surfactant gradient is also attenuated by these diffusive processes and the drag reduction $DR_0$ increases.

We next turn our attention to region D, where diffusion is sufficiently strong for surfactant to be distributed almost uniformly along the bulk and interface.
This implies that there are almost no Marangoni stresses to increase the drag and the liquid--gas interface is almost shear-free.
As shown in Appendix \ref{subsec:a_strong_diffusion}, ${DR}_0$ is close to unity in region D and is given by
\begin{equation} \label{eq:drs_d}
{DR}_0 \approx 1 - \frac{(1 - \phi_x)\gamma}{(1 +\phi_x \beta)\alpha + (1 - \phi_x)\delta} \quad \text{for} \quad \min(\alpha,\,\delta) \gg \max(1,\,\gamma),
\end{equation}
for $\beta=O(1)$.
Equation \eqref{eq:drs_d} shows how drag-promoting Marangoni effects are weakened by strong diffusion in the bulk ($\alpha$) or at the interface ($\delta$).
Figure \ref{fig:map_1}(\textit{c,\,e}) shows the surfactant profiles in region D in the limit of strong diffusion, where $\alpha$ or $\delta$ are large compared to bulk advection and Marangoni effects.
The leading-order solution in $\mathcal{D}_1$ is uniform along $x$, such that
\begin{equation}
\label{eq:alpha_gg_1}
    c_{0} \approx \frac{\alpha + \delta(1-\phi_x)}{\alpha(\beta\phi_x + 1) + \delta(1-\phi_x)} \quad \text{for} \quad \min(\alpha,\,\delta) \gg \max(1 ,\,\gamma),
\end{equation}
demonstrating how diffusion eliminates gradients of $c_0$ throughout $\mathcal{D}_1$ and $\mathcal{D}_2$ (the weak gradients contributing to ${DR}_0$ in \eqref{eq:drs_d} appear at higher order and are detailed in Appendix \ref{subsec:a_strong_diffusion}).
Therefore, drag reduction in region D is impeded by decreasing the bulk and surface diffusion (if $\gamma\gg 1$) or increasing the surfactant strength (i.e. moving into region M).

\begin{figure}
\centering
(\textit{a}) \hfill (\textit{b}) \hfill \hfill \hfill \\
\vspace{-.5cm} 
\includegraphics[trim={0 0.75cm 0 1cm}, clip,width=.425\textwidth]{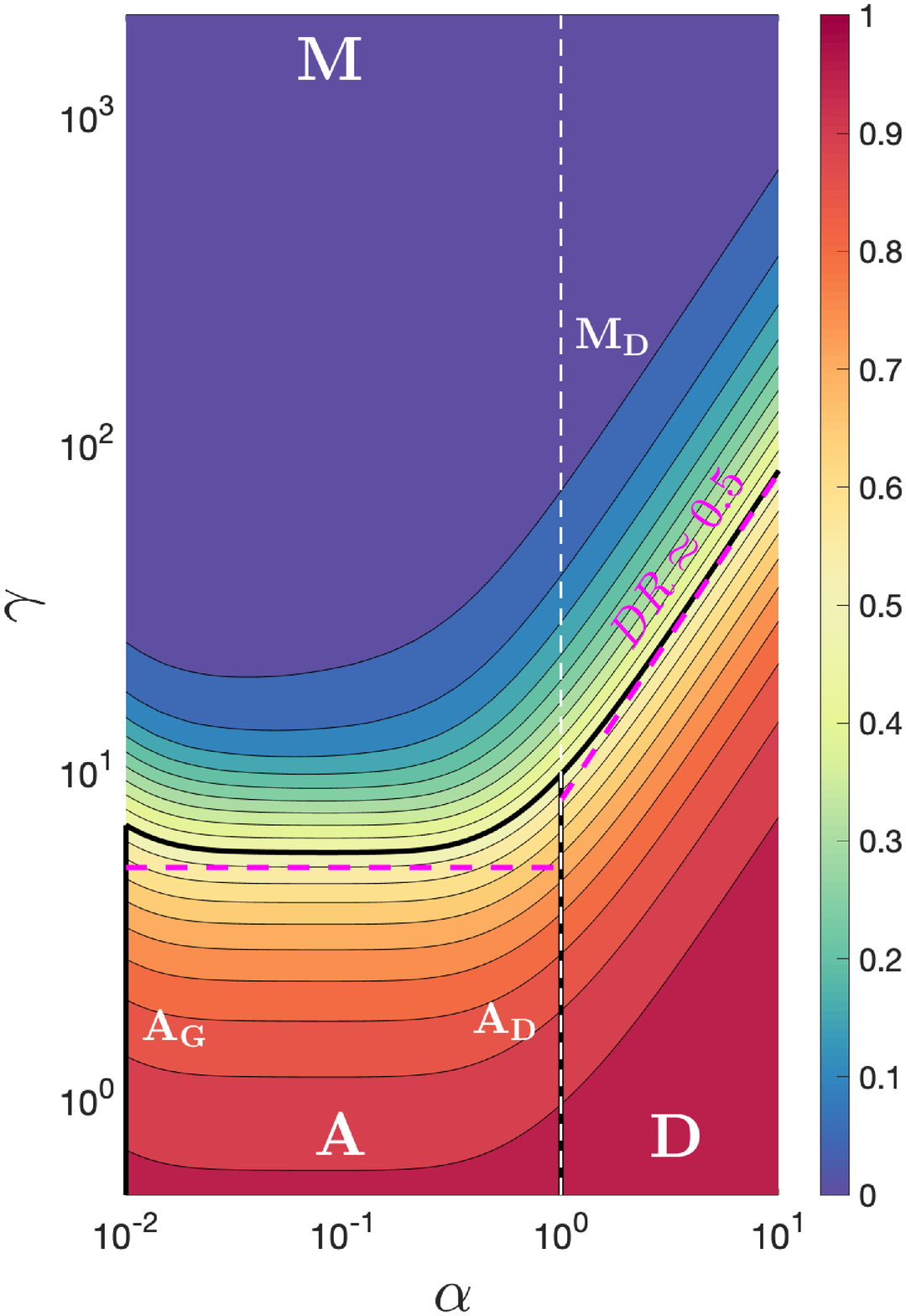} \hspace{.8cm}
\includegraphics[trim={0 0.75cm 0 1cm}, clip,width=.425\textwidth]{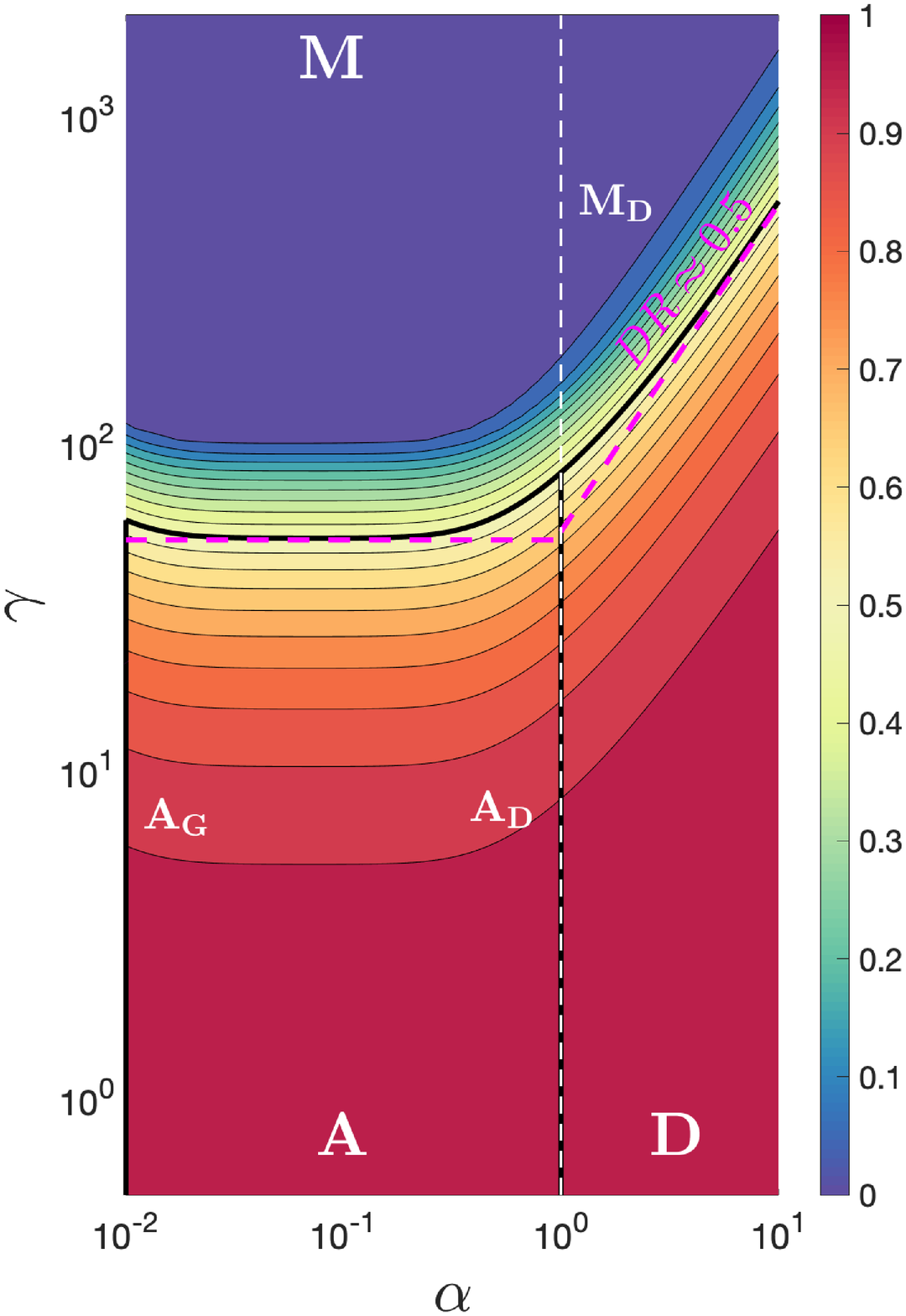} \\
(\textit{c}) \hfill (\textit{d}) \hfill \hfill \hfill \\
\vspace{-.4cm} 
\includegraphics[trim={0 0cm 0 0cm}, clip,width=.425\textwidth]{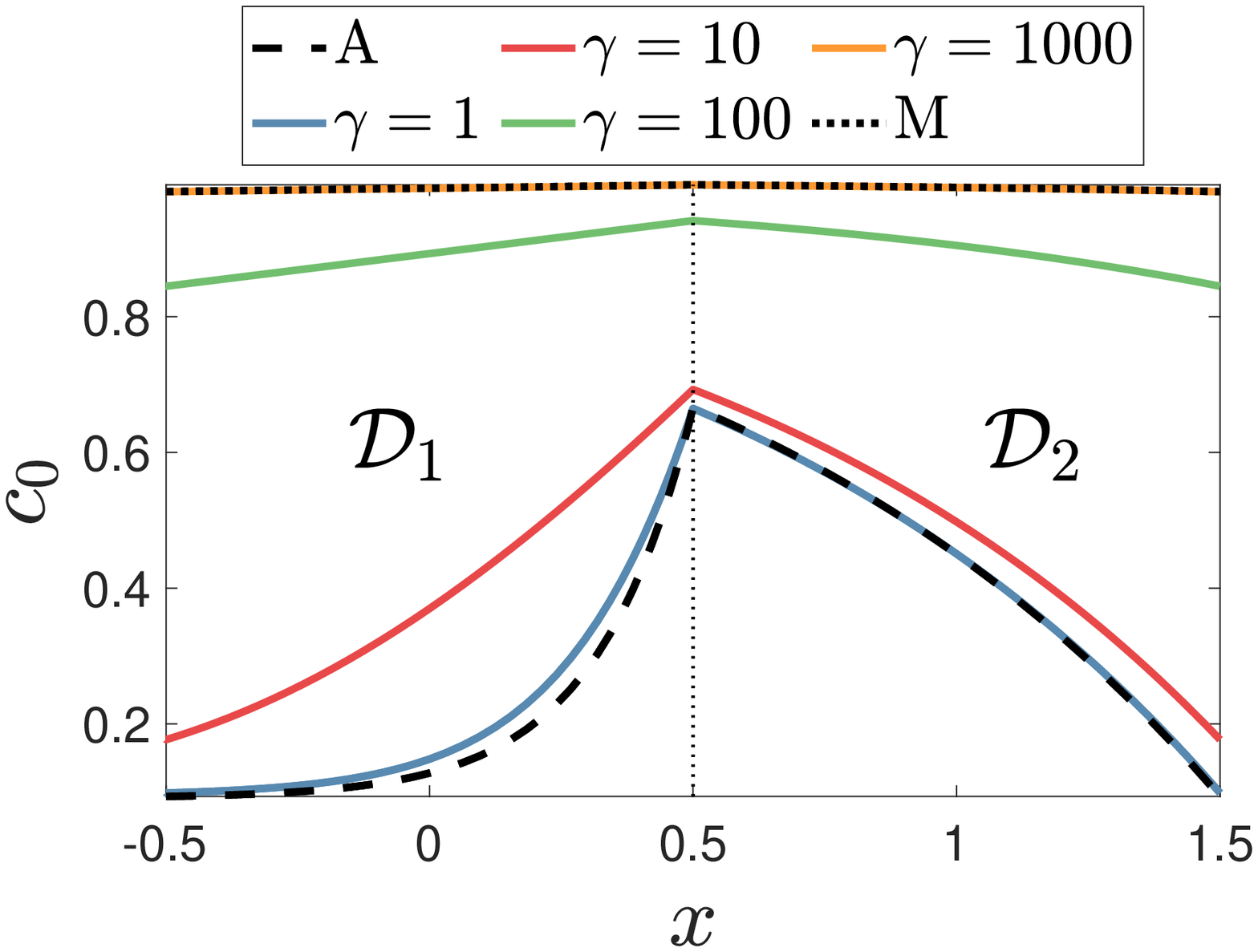} \hspace{.8cm}
\includegraphics[trim={0 0cm 0 0cm}, clip,width=.425\textwidth]{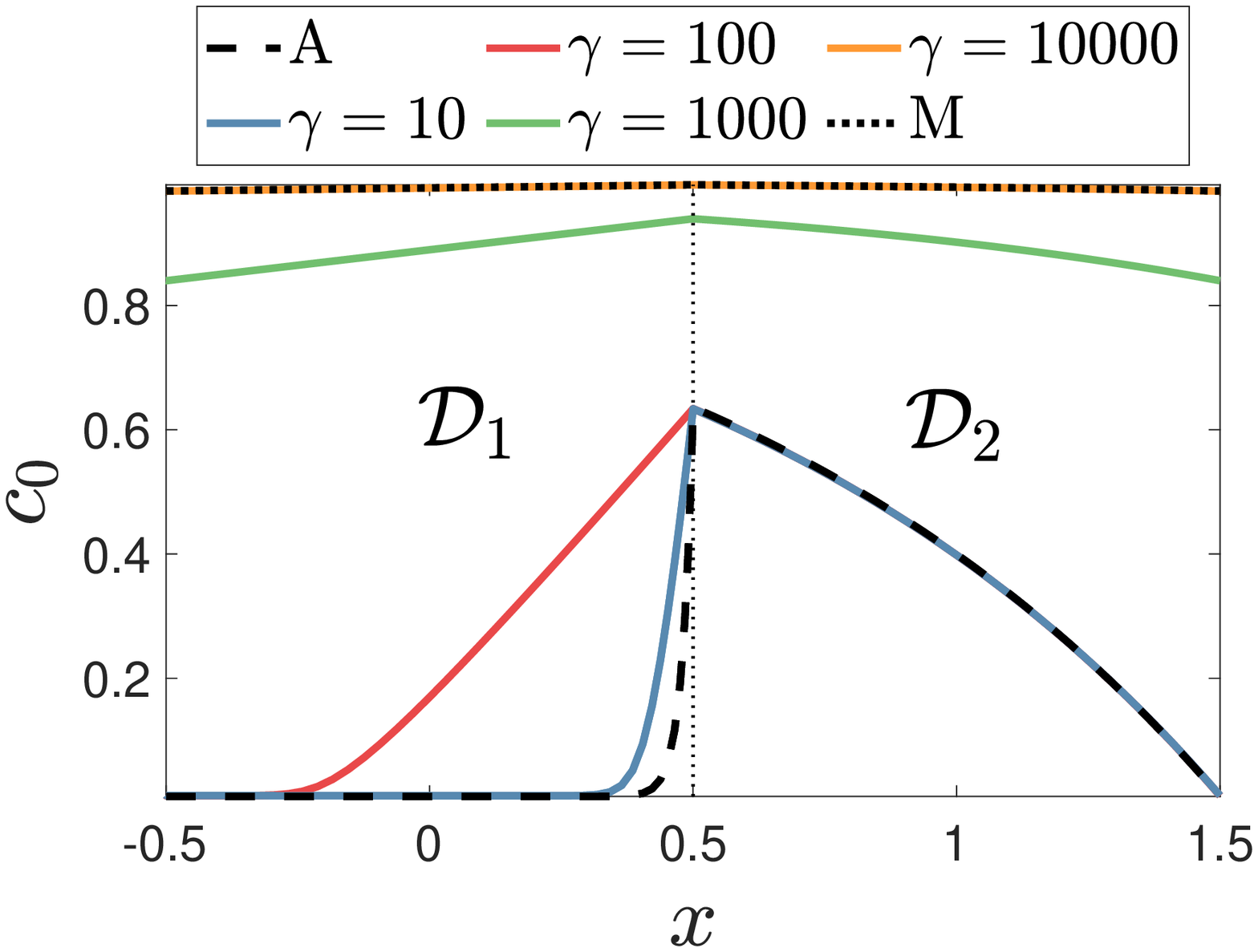}
\caption{
The leading-order drag reduction (${DR}_0$) and surfactant distribution ($c_0=\Gamma_0$) in the strong-exchange problem, for $\phi_x = 0.5$, $\epsilon = 0.1$, $\phi_z = 0.5$, $\Da=1$ and $P_z = 0.5$, computed using \eqref{eq:composite_equation}.
Contours of ${DR}_0$ for (\textit{a}) $\beta = 10$ and (\textit{b}) $\beta = 100$, where ${DR}_0 = 0$ exhibits a no-slip SHS and ${DR}_0 = 1$ exhibits a shear-free plastron at the SHS.
The Marangoni (M), advection (A) and diffusion-dominated (D) regions are separated by black lines and ${DR}_0$ is approximated by (\ref{eq:drs_m}), (\ref{eq:drs_a}) and (\ref{eq:drs_d}) for the M, A and D regions, respectively.
The dashed magenta lines describing when ${DR}_0 \approx0.5$ are given by (\ref{eq:largebetaflat}) and (\ref{eq:mdboundary}) for the AM and DM boundaries, respectively.
Plots of $c_0$ for varying surfactant strength ($\gamma$) and $\alpha = \delta = 1$, where (\textit{c}) $\beta = 10$, A: $--$ \eqref{eq:zeta_ll_1}, M: $\cdot\cdot\cdot$  \eqref{eq:zeta_gg_1} with $\gamma = 1000$, and (\textit{d}) $\beta = 100$, A: $--$ \eqref{eq:zeta_ll_1}, M: $\cdot\cdot\cdot$  \eqref{eq:zeta_gg_1} with $\gamma = 10000$.}
\label{fig:beta}
\end{figure}

We finally consider region A, in which surfactants can adsorb onto the interface but generate weak Marangoni stresses, allowing the streamwise flow to advect the interfacial surfactant towards the downstream stagnation point.
In this advection-dominated region, we show in Appendix \ref{subsec:a_weak_surfactant_strength} that ${DR}_0$ is once again close to the shear-free value
\begin{equation} \label{eq:drs_a}
{DR}_0 \approx  1 -\frac{\gamma}{2\phi_x\left(\beta +1\right)} \quad \text{for} \quad \epsilon^2 \ll \alpha \ll 1, \quad \gamma \ll \min(1,\,\beta). \\
\end{equation}
Region A can also be divided into two sub-regions: $\text{A}_\text{G}$ which balances advection and shear dispersion, and $\text{A}_\text{D}$ which balances advection and diffusion; these balances give the same ${DR}_0$ as in \eqref{eq:drs_a}.
The corresponding concentration profiles in region A are shown in figure \ref{fig:map_1}(\textit{b,\,c,\,d}). 
The leading-order solution in $\mathcal{D}_1$ exhibits a surfactant gradient that increases monotonically towards the downstream end of the plastron
\begin{equation} \label{eq:zeta_ll_1}
    c_0 \approx \frac{1}{\beta +1}  + \frac{\beta}{(\beta+1)}\exp\left(\frac{(1+\beta)(x-\phi_x)}{\alpha +\delta + \epsilon^2 s_1 / \alpha }\right) \ \text{for} \ \epsilon^2 \ll \alpha \ll 1,\ \gamma \ll \min(1,\,\beta).
\end{equation}
The downstream boundary layer in \eqref{eq:zeta_ll_1} can be short compared to the plastron length, such that the surface shear stress is negligible almost everywhere on the interface and the bulk flow experiences a largely shear-free boundary. 

Region A transitions into region D across the AD boundary, where advection and diffusion balance. 
The concentration and drag reduction at the AD boundary can be obtained in closed form following the method shown in Appendix \ref{subsec:a_weak_surfactant_strength}. 
However, in figure 4(\textit{a}), we see that crossing the AD boundary (shown by the vertical black line at $\alpha=1$ and $\gamma \ll 1$) does not affect ${DR}_0$ to leading-order.
When $\gamma \ll \min(1,\,\beta)$ and the surfactant strength is weak, either the streamwise velocity at the surface is large enough to advect most of the interfacial surfactant to the downstream stagnation point ($\alpha \ll 1$) or diffusion is strong enough to attenuate any interfacial surfactant gradient that forms ($\alpha \gg 1$).
As the interface is mostly shear-free in both A and D, we do not pursue this limit. 
Bulk diffusion dominates over bulk advection for $\alpha \gg O(1)$, so we employ $\alpha = 1$ to illustrate the AD boundary in figure \ref{fig:map_1}(\textit{a}) (via the thick vertical black line on the right of the A region). 
When bulk diffusion is weak, shear dispersion dominates over bulk advection for $\epsilon^2 / \alpha \gg 1$ (Appendix \ref{app:D}), so we use $\alpha = \epsilon^2$ to illustrate the AG boundary  (the thick vertical black line on the left of the A region in figure \ref{fig:map_1}\textit{a}). 
Region A transitions into region M across the AM boundary, such that Marangoni effects dominate advection for $\gamma \gg \max(1,\,\beta)$.

\begin{figure}
\centering
(\textit{a}) \hfill (\textit{b}) \hfill \hfill \hfill \\
\vspace{-.5cm} \includegraphics[width=.425\textwidth]{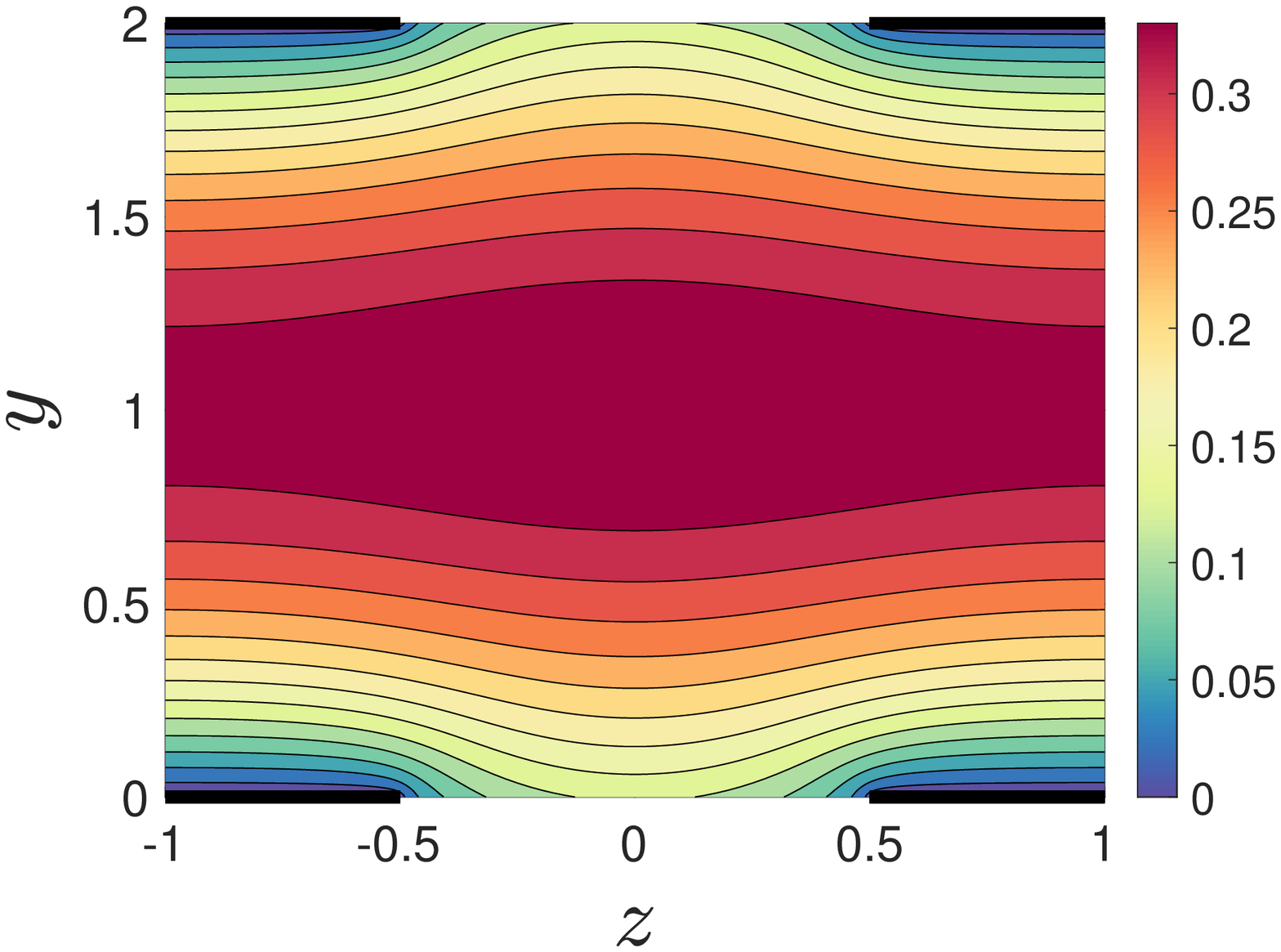} \hspace{.8cm} \includegraphics[width=.425\textwidth]{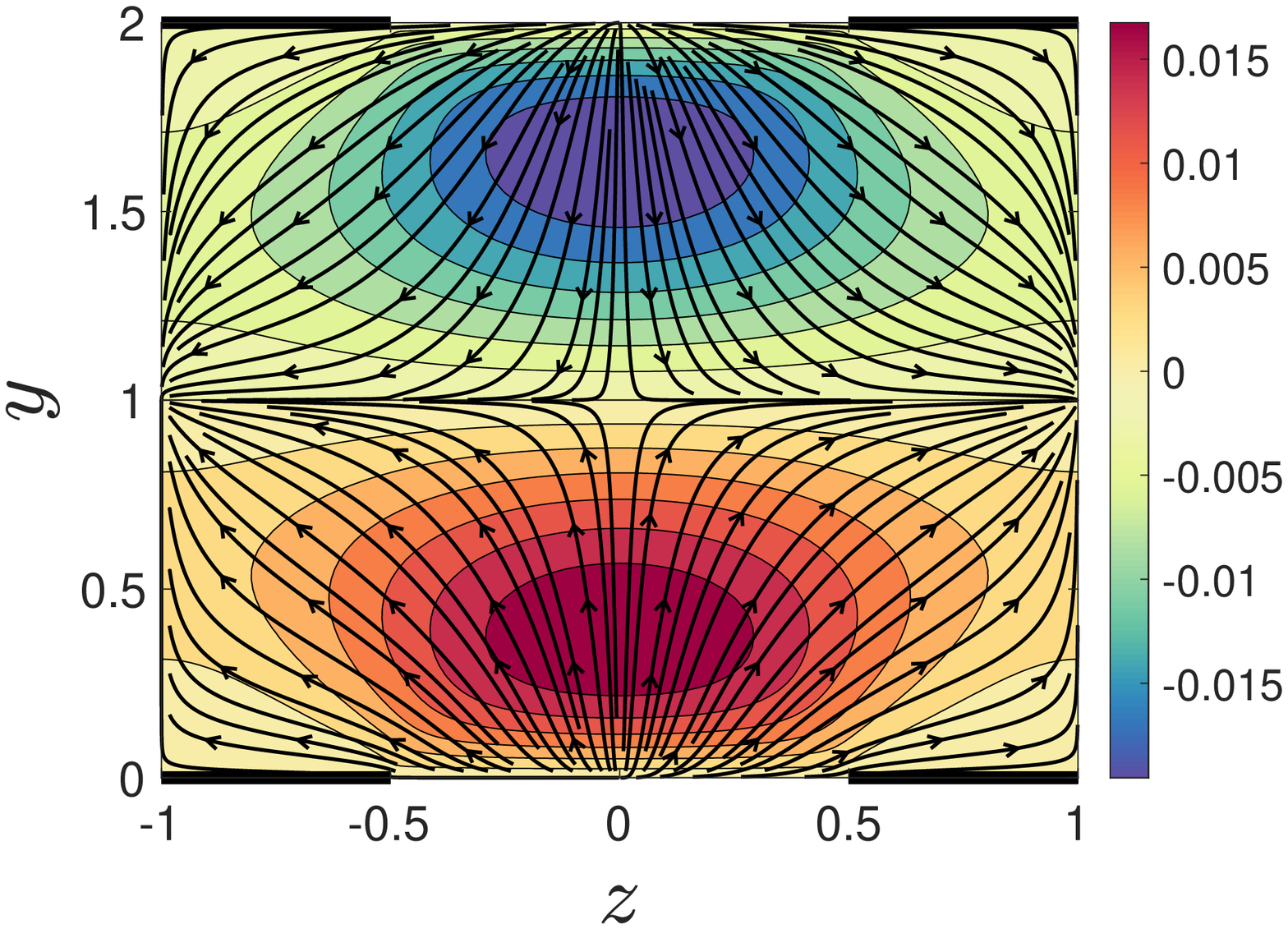} \\
(\textit{c}) \hfill (\textit{d}) \hfill \hfill \hfill \\
\vspace{-.5cm} \includegraphics[width=.425\textwidth]{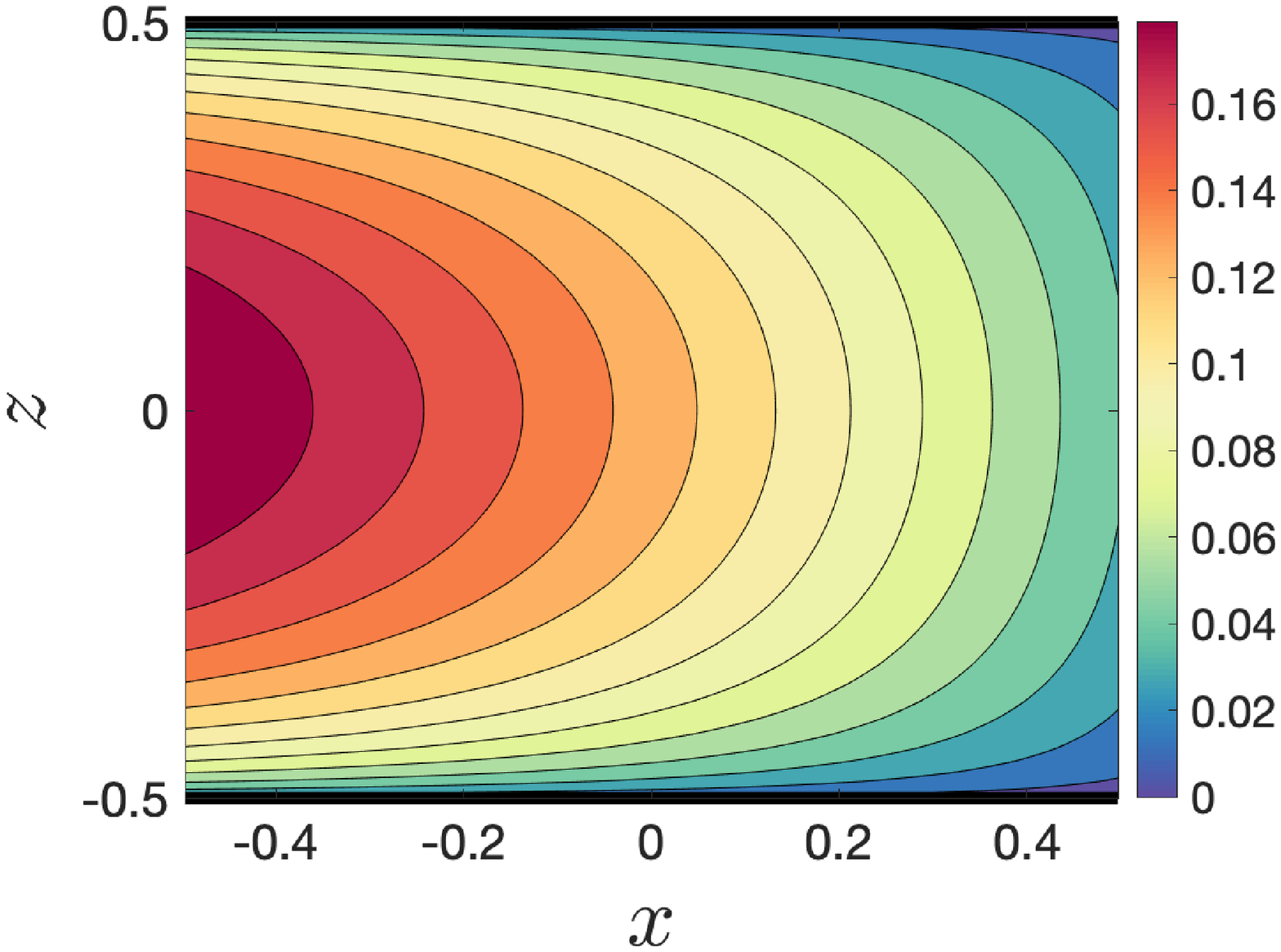} \hspace{.8cm} \includegraphics[width=.425\textwidth]{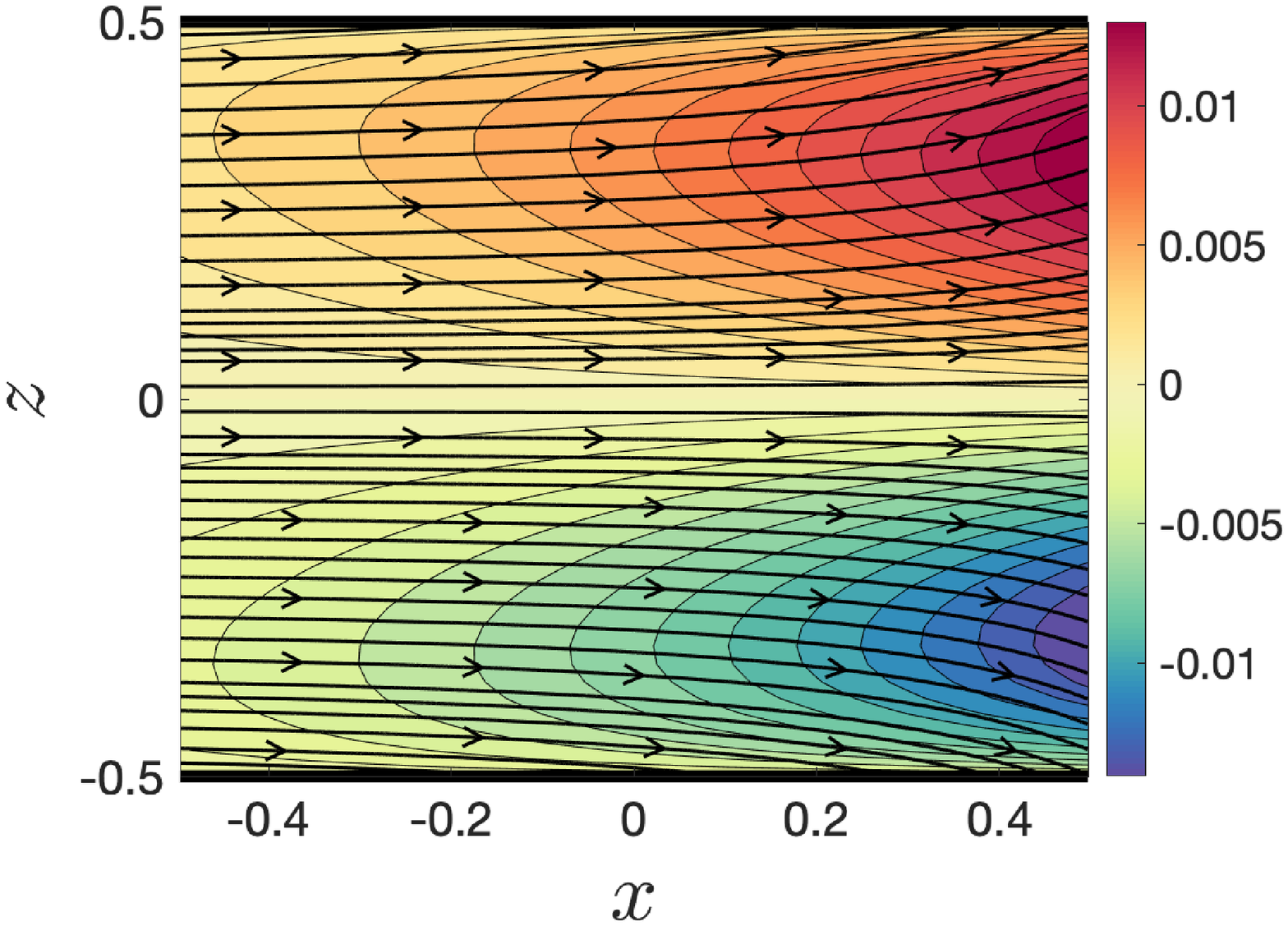}
\caption{
Contour maps of the flow field in the strong-exchange problem, for $\alpha = 0.1$, $\beta = 1$, $\gamma = 1.3$, $\delta = 0.1$, $\phi_x = 0.5$, $\phi_z = 0.5$ and $P_z = 0.5$, corresponding to $50$\% normalized drag reduction given by the star in figure \ref{fig:map_1}(\textit{a}).
(\textit{a}) Leading-order streamwise velocity $u_0$ and (\textit{b}) leading-order wall-normal velocity $v_1$ with $(v_1,\,w_1)$ streamlines at the centre of the plastron, $x=0$. 
(\textit{c}) Leading-order streamwise velocity $u_0$ and (\textit{d}) leading-order transverse velocity $w_1$ with $(u_0, \,w_1)$ streamlines at the interfaces, $y=0$ or $2$.
The thick black lines in  (\textit{a--d}) represent the solid regions of the SHS.
}
\label{fig:ff_1}
\end{figure}

We next investigate the dependence of the drag reduction on the partition coefficient ($\beta$) by evaluating ${DR}_0$ for $\beta=10$ in figure \ref{fig:beta}(\textit{a}) and $\beta = 100$ in figure \ref{fig:beta}(\textit{b}).
As $\beta$ grows, the advective flux of surfactant at the liquid-gas interface increases, sweeping more of the interfacial surfactant towards the downstream stagnation point. This means that a larger portion of the upstream part of the interface is shear-free, whilst the surfactant concentration remains unchanging at the downstream end of the interface. Overall, the net (nonlinear) effect with increasing $\beta$ is a reduction in drag.
The limit $\beta \gg 1$ also corresponds to the near-insoluble surfactant limit, with nearly all surfactant being transported along the interface in region $\mathcal{D}_1$.
A number of the asymptotic approximations given in \eqref{eq:drs_m}--\eqref{eq:zeta_ll_1} simplify for $\beta \gg 1$: \eqref{eq:smboundary} reduces to ${DR}_0\approx 0.5$ when $\gamma = \epsilon^2 s_4 \beta\phi_x/(2\alpha(1-\phi_x))$ for 
$\alpha \ll 1 \ll \beta$,
such that shear dispersion in $\mathcal{D}_2$ determines ${DR}_0$; \eqref{eq:mdboundary} reduces to ${DR}_0\approx 0.5$ when $\gamma = \alpha\beta\phi_x/(2(1-\phi_x))$ 
for $\alpha \gg \beta \gg 1$ and $\alpha = O(\delta)$,
with bulk diffusion determining ${DR}_0$.

Another feature that emerges from figure \ref{fig:beta}(\textit{a,\,b}) is the flattening of the central contours of ${DR}_0$ at the AM boundary as $\beta$ increases for $\beta \gg 1$, due to  advection at the interface becoming stronger.
To extract this feature asymptotically, in Appendix \ref{subsec:a_strong_diff_strong_surfactant_strength} we assume $\beta \gg 1 \gg \max(\alpha,\, \delta,\, \epsilon^2/\alpha)$ and $\gamma = O(\beta)$ at the AM boundary, showing that ${DR}_0 \approx 1 - \gamma/(2 \beta \phi_x)$ for $\gamma / \beta \leq 2\phi_x$ and $\epsilon^2\ll \alpha\ll1$ (the case where $\gamma / \beta > 2\phi_x$ is also considered in Appendix \ref{subsec:a_strong_diff_strong_surfactant_strength} and gives $DR_0\approx -(1/\beta+\gamma \ln(1-2\beta\phi_x/\gamma)/(2\phi_x\beta^2))$ provided $1 - 2\phi_x \beta / \gamma \gg \exp{(-\beta)}$). 
This demonstrates how a large surface advective flux weakens the immobilizing effects of surfactant (noting that $\gamma/\beta \propto \hat{A}\hat{G}/\hat{\mu}\hat{U}$). 
Using this expression for ${DR}_0$ at the AM boundary, we predict that
\begin{equation} \label{eq:largebetaflat}
    {DR}_0 \approx 0.5 \quad \text{when} \quad \gamma = \phi_x \beta \quad \text{for} \quad \beta \gg 1 \gg \max\left(\alpha,\, \delta,\, \frac{\epsilon^2}{\alpha}\right),
\end{equation}
which gives the central dashed magenta lines in figure \ref{fig:beta}(\textit{a,\,b}) (agreement with the 1D numerical solution of the strong-exchange problem improves for increasing $\beta$).
At the AM boundary, the surfactant concentration at the interface exhibits a stagnant cap \citep{he1991size}, or a piecewise-linear distribution, if $\gamma/\beta \leq 2\phi_x$, where the upstream interface is almost shear-free and the downstream interface is effectively no-slip.
In Appendix \ref{subsec:a_strong_diff_strong_surfactant_strength}, we show that 
\begin{equation} \label{eq:stag_cap}
c_0 \approx
\begin{cases}
    \displaystyle 0 \ \text{for} \ -\phi_x \leq x \leq x_0, \\
    \displaystyle \frac{\beta}{\gamma}(x - \phi_x) +1 \ \text{for} \ x_0 \leq x \leq \phi_x,
\end{cases}
\text{for} \ \min(\beta,\, \gamma) \gg 1 \gg \max\left(\alpha,\, \delta,\, \frac{\epsilon^2}{\alpha}\right),
\end{equation}
where $x_0 = \phi_x - \gamma/\beta$ (see, e.g., the red curve for $\gamma=100$ in figure \ref{fig:beta}\textit{d}). If   $\gamma/\beta > 2\phi_x$, then the interfacial concentration distribution at the AM boundary is linear, $c_0\approx \beta/\gamma (x-\phi_x)+1$ at leading-order.
When the surface advection overcomes Marangoni effects, for $\beta \gg \gamma$, surfactants can accumulate at the downstream stagnation point. 
The blue curve for $\gamma = 10$ in figure \ref{fig:beta}\textit{d} demonstrates that $c_0 \approx 0$ throughout most of $\mathcal{D}_1$: the large surfactant advective flux at the interface contracts the downstream boundary layer, making the interface almost shear-free.
In summary, stronger nonlinearity in the form of stagnant cap profiles in the interfacial concentration  can appear  with increasing $\beta$ near the transition from region A to M. 
This results in a sharp decline in ${DR}_0$ towards zero, which can be seen as the AM transition becomes less smooth, i.e. the vertical distance between ${DR}_0=0.5$ and ${DR}_0 = 0.05$ contours decreases from figure \ref{fig:beta}(\textit{a}) to (\textit{b}) as the surface advection increases from $\beta=10$ to $\beta=100$. 
However, we note that  the location of the transition is captured by the simple expression \eqref{eq:largebetaflat} when $\gamma/\beta \leq 2\phi_x$.

\subsubsection{Flow field} \label{subsubsec:ff_strong_exchange}

Whenever there is partial drag reduction, $0 \leq {DR}_0 < 1$, the surfactant has a non-uniform concentration in $\mathcal{D}_1$, reducing the streamwise velocity at the plastron and generating a secondary flow in the bulk. 
To illustrate, videos of the leading-order streamwise velocity $u_0$, wall-normal velocity $v_1$, transverse velocity $w_1$ and the streamwise gradient of the surfactant distribution $c_{0x}$ are given in supplementary movie 1, evaluated at a point in the parameter space where ${DR}_0 \approx 0.5$, shown by the star in figure \ref{fig:map_1}(\textit{a}).
The corresponding flow field is shown in figure \ref{fig:ff_1}.
Figure \ref{fig:ff_1}(\textit{a}) shows $u_0$ at the centre of domain $\mathcal{D}_1$, where $x=0$.
The streamwise velocity, built from the components shown in figure \ref{fig:first_order_velocity}, attains a maximum value at the channel centre, decaying towards either SHS ($y = 0,\,2$) to satisfy no-slip at the solid wall ($-P_z \leq z \leq \phi_z$ and $\phi_z\leq z \leq P_z$), whilst allowing for slip over the plastron ($-\phi_z \leq z \leq \phi_z$). 
The slip velocity at $\mathcal{I}$ reduces as one progresses through $\mathcal{D}_1$, in accordance with the rise in $c_{0x}$ for increasing $x$; see figure \ref{fig:ff_1}(\textit{c}) and supplementary movie 1.
The associated secondary cross-channel velocity field $(v_1,w_1)$ shown in figure \ref{fig:ff_1}(\textit{b}) (recall that the leading-order components are $(v_0,w_0)=(0,0)$) advects particles from $\mathcal{I}$ towards the centre of the channel ($y=1$ and $z=\pm P_z$).
The absolute maximum of $v_1$ occurs above the centre of the plastron ($z=0$), whereas the absolute maximum of $w_1$ occurs above the transverse contact lines ($z=\pm \phi_z$).
Supplementary movie 1 and figure \ref{fig:ff_1}$(d)$ show how $v_1$ and $w_1$ grow in magnitude as one progresses through $\mathcal{D}_1$ along the positive $x$-direction as the interface is immobilized.
As mentioned in \S\ref{subsec:strong_diffusion}, the present long-wave model does not capture rapid adjustments of the flow field near the contact lines at $x= \pm \phi_x$, where domains $\mathcal{D}_1$ and $\mathcal{D}_2$ meet.

\subsection{Moderate exchange} \label{subsec:Weak exchange}

We now turn our attention to the moderate-exchange problem \eqref{eq:model1}--\eqref{eq:model3}, in which the bulk and interfacial concentration fields decouple for $\nu/\epsilon^2 \sim O(1, \, \alpha, \, \beta,\, \delta)$. 
At the plastron, surfactant adsorbs onto the interface at the upstream end ($c_0>\Gamma_0$) and desorbs at the downstream end ($\Gamma_0>c_0$), as regulated by the bulk--surface exchange parameter $\nu$ in \eqref{eq:coefficients_weak_exchange}. 
The moderate-exchange model does not take into account shear-dispersion effects, however, we ensure that the bulk diffusion strength is larger than the threshold identified in the strong-exchange problem (\S\ref{Weak partition coefficient}), $\alpha \gg \epsilon^2$, where cross-channel concentration gradients first become significant.

\subsubsection{Drag reduction} \label{Without surface diffusion}

\begin{figure}
\begin{minipage}{0.67\textwidth}
\centering
(\textit{a}) \hfill \hfill \hfill \\
\includegraphics[trim={0 1cm 0 1.5cm},clip,width=\linewidth]{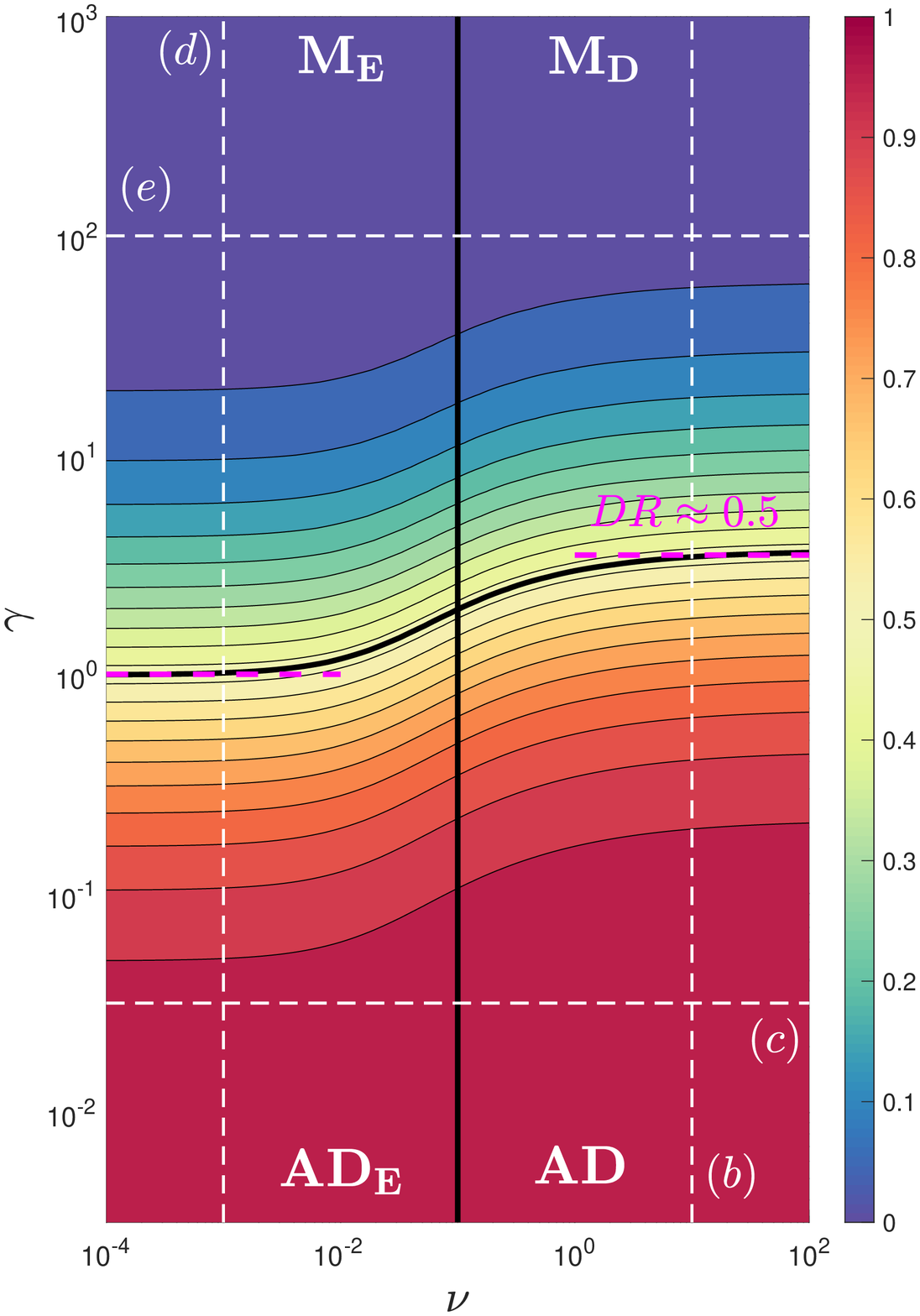}
\end{minipage}
\begin{minipage}{0.29\textwidth}
\centering
(\textit{b}) \hfill \hfill \hfill \\
\includegraphics[width=\linewidth]{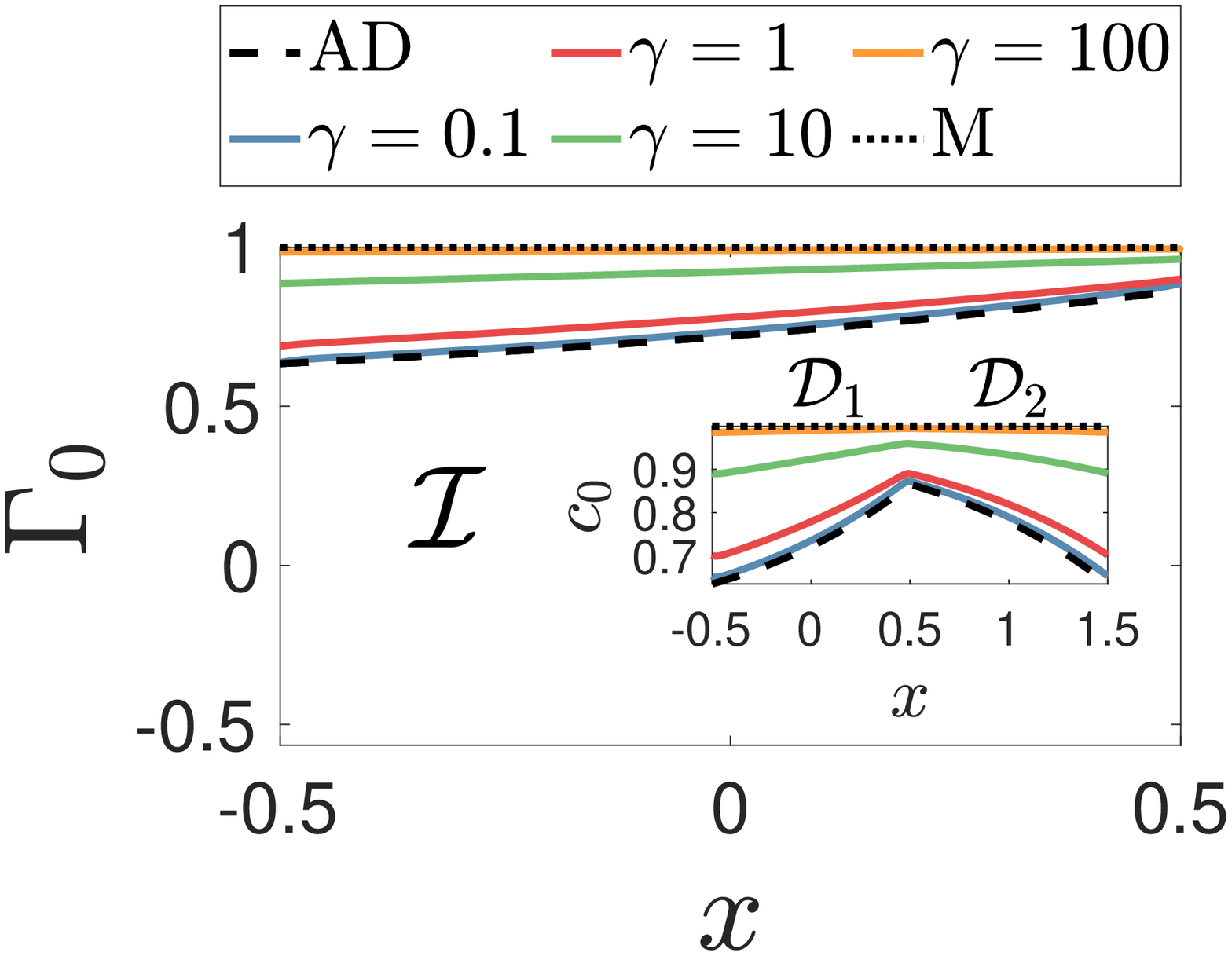} \\
\vspace{-.5cm}
(\textit{c}) \hfill \hfill \hfill \\
\includegraphics[width=\linewidth]{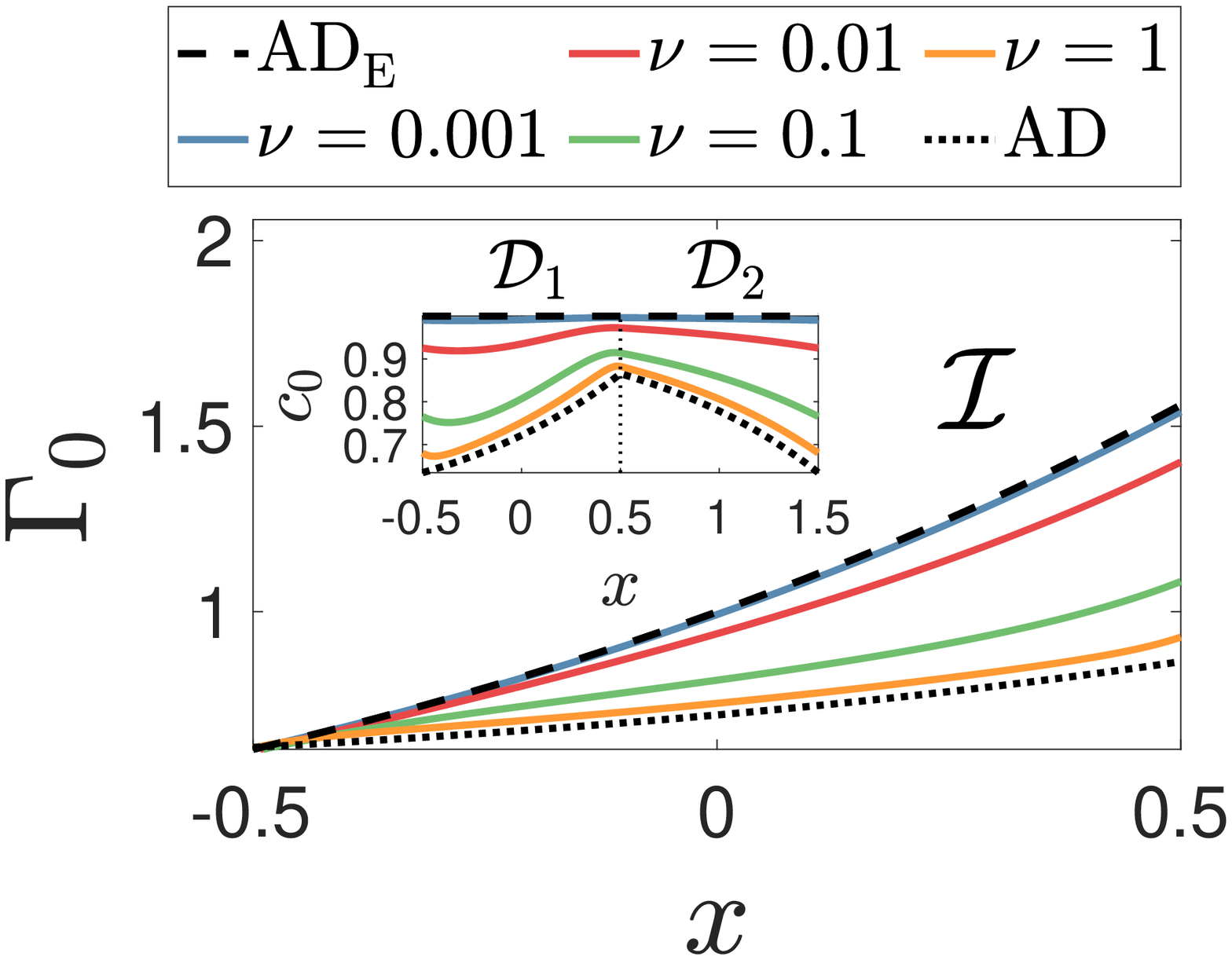} \\
\vspace{-.5cm}
(\textit{d}) \hfill \hfill \hfill \\
\includegraphics[width=\linewidth]{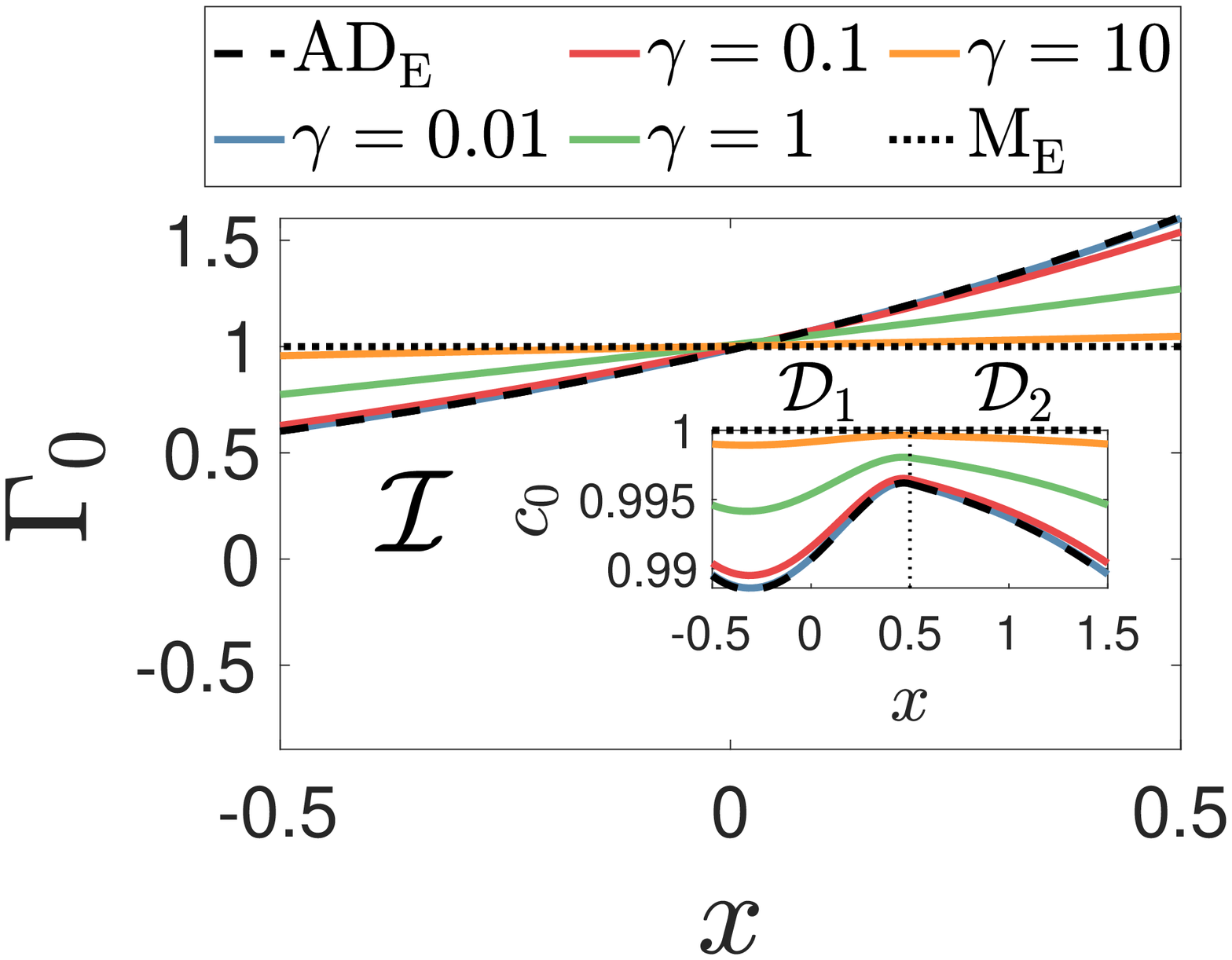} \\
\vspace{-.5cm}
(\textit{e}) \hfill \hfill \hfill \\
\includegraphics[width=\linewidth]{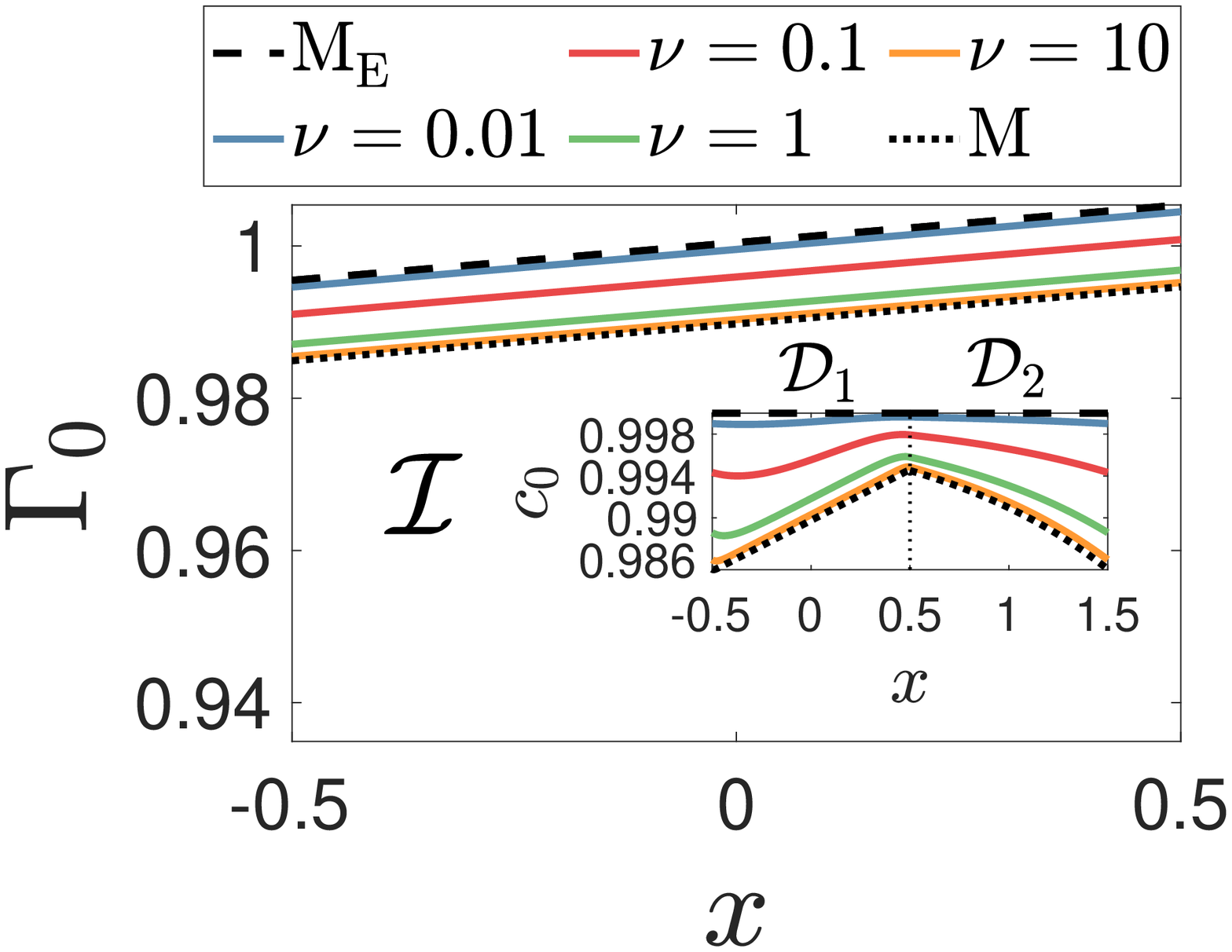} \\
\end{minipage}
\caption{
The leading-order drag reduction (${DR}_0$), bulk surfactant ($c_0$) and interfacial surfactant distribution ($\Gamma_0$) in the moderate-exchange problem, for $\alpha = 1$, $\beta = 1$, $\delta = 1$, $\phi_x = 0.5$, $\epsilon = 0.1$, $\phi_z = 0.5$ and $P_z = 0.5$, computed using \eqref{eq:model1}--\eqref{eq:model3}.
(\textit{a}) Contours of ${DR}_0$, where ${DR}_0 = 0$ exhibits a no-slip SHS and ${DR}_0 = 1$ exhibits a shear-free plastron.
The Marangoni (M), advection-diffusion (AD), Marangoni-exchange ($\text{M}_\text{E}$) and advection-diffusion-exchange ($\text{AD}_\text{E}$) regions are separated by black lines and ${DR}_0$ is approximated by (\ref{eq:drs_m}), (\ref{eq:drs_a}), (\ref{eq:drs_m_we}) and (\ref{eq:we_dr_d}) in the M, AD, $\text{M}_\text{E}$ and $\text{AD}_\text{E}$ regions, respectively.
The dashed magenta lines describing when ${DR}_0\approx0.5$ are given by (\ref{eq:mdboundary}, \ref{eq:we_50_1}).
Plots of $\Gamma_0$ and $c_0$ for varying surfactant strength ($\gamma$) and exchange strength ($\nu$), where the asymptotic curves are: (\textit{b}) $\nu = 10$, AD: $--$ \eqref{eq:zeta_gg_1} with $\gamma = 100$, M: $\cdot\cdot\cdot$ \eqref{eq:zeta_ll_1}, (\textit{c}) $\gamma = 0.03$, $\text{AD}_\text{E}$: $--$ \eqref{eq:we_d}, AD: $\cdot\cdot\cdot$ \eqref{eq:zeta_ll_1}, (\textit{d}) $\nu = 0.001$, $\text{AD}_\text{E}$: $--$ \eqref{eq:we_d}, $\text{M}_\text{E}$: $\cdot\cdot\cdot$ \eqref{eq:we_m} with $\gamma = 10$, and (\textit{e}) $\gamma = 100$, $\text{M}_\text{E}$: $--$ \eqref{eq:we_m}, M: $\cdot\cdot\cdot$ \eqref{eq:zeta_gg_1}.
}
\label{fig:map_3}
\end{figure}

Figure \ref{fig:map_3}(\textit{a}) shows how ${DR}_0$ varies with surface exchange strength ($\nu$) and surfactant strength ($\gamma$), for $\alpha = 1$, $\beta = 1$ and $\delta =1$.
Similar to the strong-exchange problem in \S\ref{Weak partition coefficient}, the moderate-exchange governing equations \eqref{eq:model1}--\eqref{eq:model2} simplify in different regions of the parameter space, representing distinct physical balances.
The analysis in the moderate-exchange problem is simplified by the fact that for $\nu/\epsilon^2 \gg \max(1,\,\alpha,\,\beta,\,\delta)$ the moderate-exchange problem transitions to the strong-exchange problem, which is discussed in detail in \S\ref{Weak partition coefficient}.
In the strong-exchange limit, $\nu/\epsilon^2 \gg \max(1,\,\alpha,\,\beta,\,\delta)$, we reach the AD boundary and sub-region $\text{M}_\text{D}$ identified in figure \ref{fig:map_1}(\textit{a}). Accordingly, the regimes AD and $\text{M}_\text{D}$ are also identified in figure \ref{fig:map_3}(\textit{a}) where bulk--surface exchange is strong.
New regimes $\text{M}_\text{E}$ and $\text{AD}_\text{E}$ appear for weak exchange, $\nu/\epsilon^2 \ll \max(1,\,\alpha,\,\beta,\,\delta)$, and are analysed asymptotically in Appendix \ref{app:E}.
The four primary areas of the parameter space, namely $\text{M}_\text{E}$, $\text{AD}_\text{E}$, M and AD, are separated by black lines in figure \ref{fig:map_3}(\textit{a}).
Corresponding surface concentration distributions $\Gamma_0(x)$ (where, unlike the strong-exchange problem, $c_0 \neq \Gamma_0$ in $\mathcal{D}_1$) are illustrated in figure \ref{fig:map_3}(\textit{b--e}).
The primary feature of figure \ref{fig:map_3}(\textit{a}) is that the drag-reduction transition from large $DR_0$ to small $DR_0$ shifts only modestly, despite the exchange strength $\nu$ varying across many orders of magnitude, for reasons that we explain below.

We begin our exploration of the parameter space in the moderate-exchange problem (figure \ref{fig:map_3}) with  region $\text{M}_\text{E}$, where Marangoni effects are strong and surfactant exchange between the bulk and interface is weak.
In Appendix \ref{subsec:a_we_strong_marangoni} we show that ${DR}_0$ remains close to the immobilised value
\begin{equation} \label{eq:drs_m_we}
{DR}_0 \approx \frac{\delta}{\gamma} \quad \text{for} \quad \gamma \gg \max(1,\,\alpha,\,\beta,\,\delta), \quad  \frac{\nu}{\epsilon^2} \ll \min(1,\,\alpha,\,\beta,\,\delta).
\end{equation} 
Here, the immobilizing effects of surfactant are weakened by strong interfacial diffusion ($\delta/\gamma \propto \hat{\mu} \hat{D}_I \phi_z \hat{P}_z / (\hat{A}\hat{G}\hat{H}^2)$). 
The corresponding concentration fields $c_0$ and $\Gamma_0$ are shown in figure \ref{fig:map_3}(\textit{d,\,e}).
The leading-order solution on $\mathcal{I}$ in $\text{M}_\text{E}$ is given by
\begin{equation} \label{eq:we_m}
    \Gamma_{0} \approx 1 + \frac{\beta x}{\gamma} \quad \text{for} \quad \gamma \gg \max(1,\,\alpha,\,\beta,\,\delta), \quad  \frac{\nu}{\epsilon^2} \ll \min(1,\,\alpha,\,\beta,\,\delta).
\end{equation}
The gradient of $\Gamma_0$ in region $\text{M}_\text{E}$ is controlled by the relative strength of surface advection and Marangoni effects, $\beta/\gamma$, which is sufficiently small to immobilise the liquid--gas interface. 
Furthermore, there is very little adsorption and desorption at the interface and the bulk and interfacial surfactant concentrations are close to their background values at leading order (i.e. $c_0 \approx 1$ and $\Gamma_0 \approx 1$).
In contrast to \eqref{eq:zeta_gg_1} in M and the strong-exchange problem of \S\ref{Weak partition coefficient}, where $c_0=\Gamma_0$ and $\Gamma_0 < 1$ for all $x$, in $\text{M}_{\text{E}}$ and the weak-exchange problem, $c_0 \approx 1$ and $\Gamma_0 > 1$ for $x > 0$ (see, e.g., the dashed and blue curves in figure \ref{fig:map_3}\textit{e}), in order to satisfy the net flux condition \eqref{eq:dimensionless_net_flux} at leading-order. 
Furthermore, $c_0$ has turning points for $-\phi_x < x < \phi_x$ where the adsorption--desorption fluxes attain a local maximum and then decrease towards the contact lines, as shown by the green and orange curves in the inset in figure \ref{fig:map_3}(\textit{d}).

Region $\text{M}_\text{E}$ gives way to region $\text{M}_\text{D}$ as the exchange strength increases.
It is notable that ${DR}_0 \approx \delta/\gamma$ (see (\ref{eq:drs_m_we})) is regulated mainly by surface diffusion in region $\text{M}_\text{E}$, whilst both bulk and surface diffusion control $DR_0$ in region $\text{M}_\text{D}$, where ${DR}_0 \approx \alpha/ (\gamma(1 - \phi_x)) + \delta/\gamma$ (see \S\ref{Weak partition coefficient}). This is due to the absence of the strong coupling between $\Gamma_0$ and $c_0$ in region $\text{M}_\text{E}$, which imposed $\Gamma_0=c_0$ in region $\text{M}_\text{D}$. 
In figure \ref{fig:map_3}(\textit{a}), we find that $\nu = 10 \epsilon^2$ best captures the centre of this transition (this can also be tested by varying $\epsilon$).
As exchange weakens and the bulk and interfacial surfactant move out of equilibrium, variations in $\Gamma_0$ increase, thus making the interface more susceptible to Marangoni effects that increase the drag. 

\begin{figure}
\centering
(\textit{a}) \hfill (\textit{b}) \hfill \hfill \hfill \\
\vspace{-.5cm} 
\includegraphics[trim={0 0.75cm 0 1cm}, clip,width=.425\textwidth]{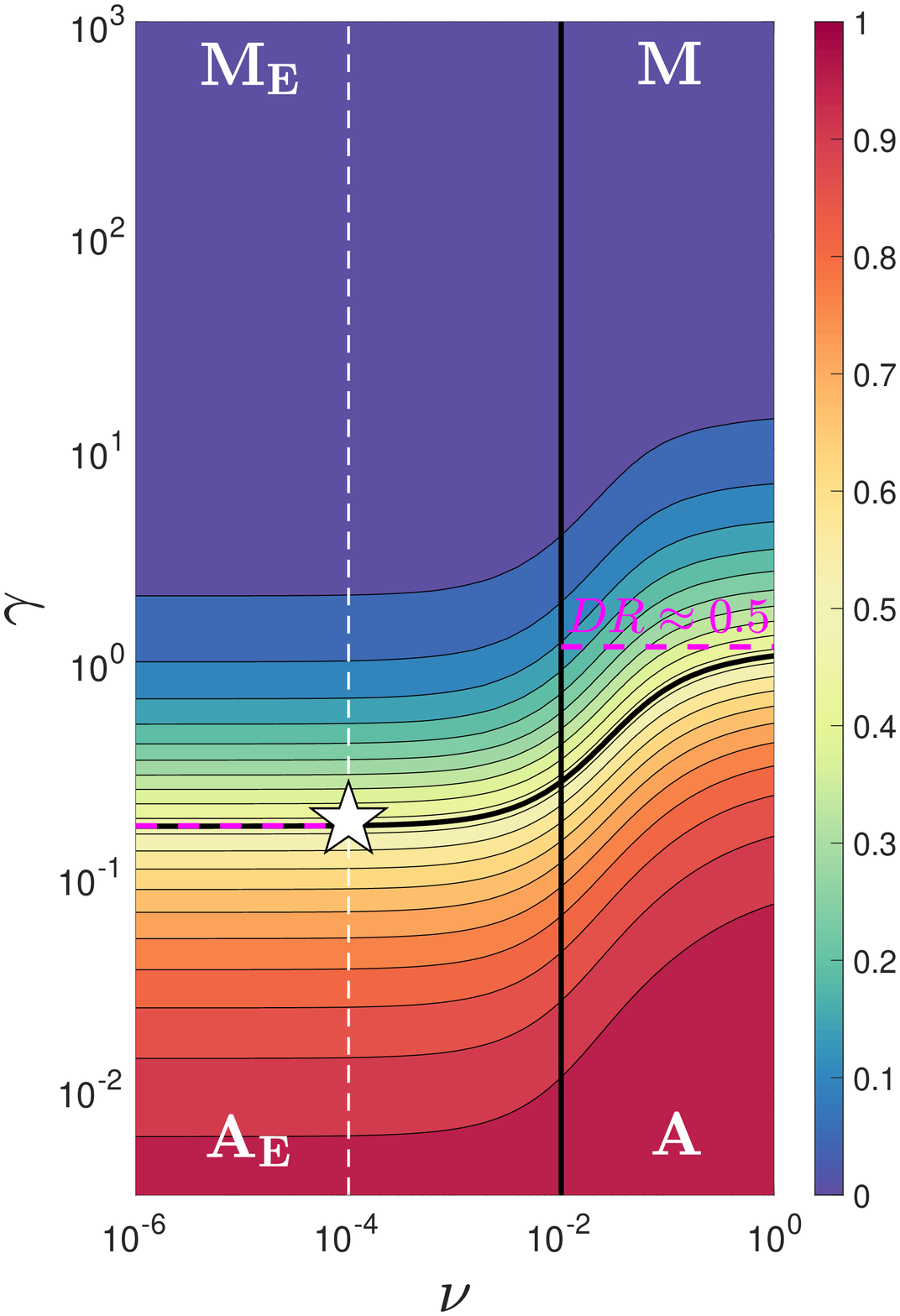} \hspace{.8cm}
\includegraphics[trim={0 0.75cm 0 1cm}, clip,width=.425\textwidth]{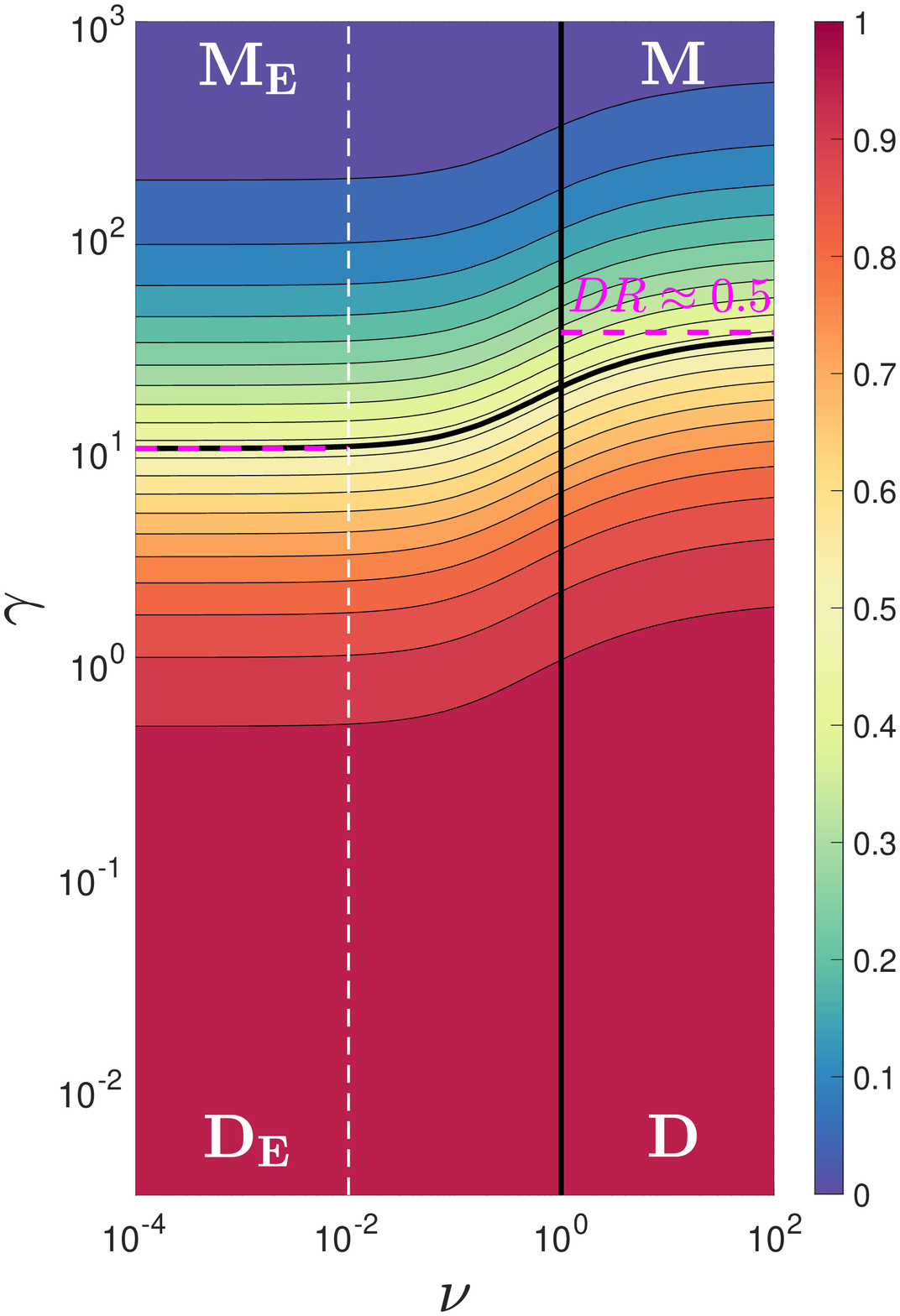} \\
(\textit{c}) \hfill (\textit{d}) \hfill \hfill \hfill \\
\vspace{-.4cm} 
\includegraphics[trim={0 0cm 0 0cm}, clip,width=.425\textwidth]{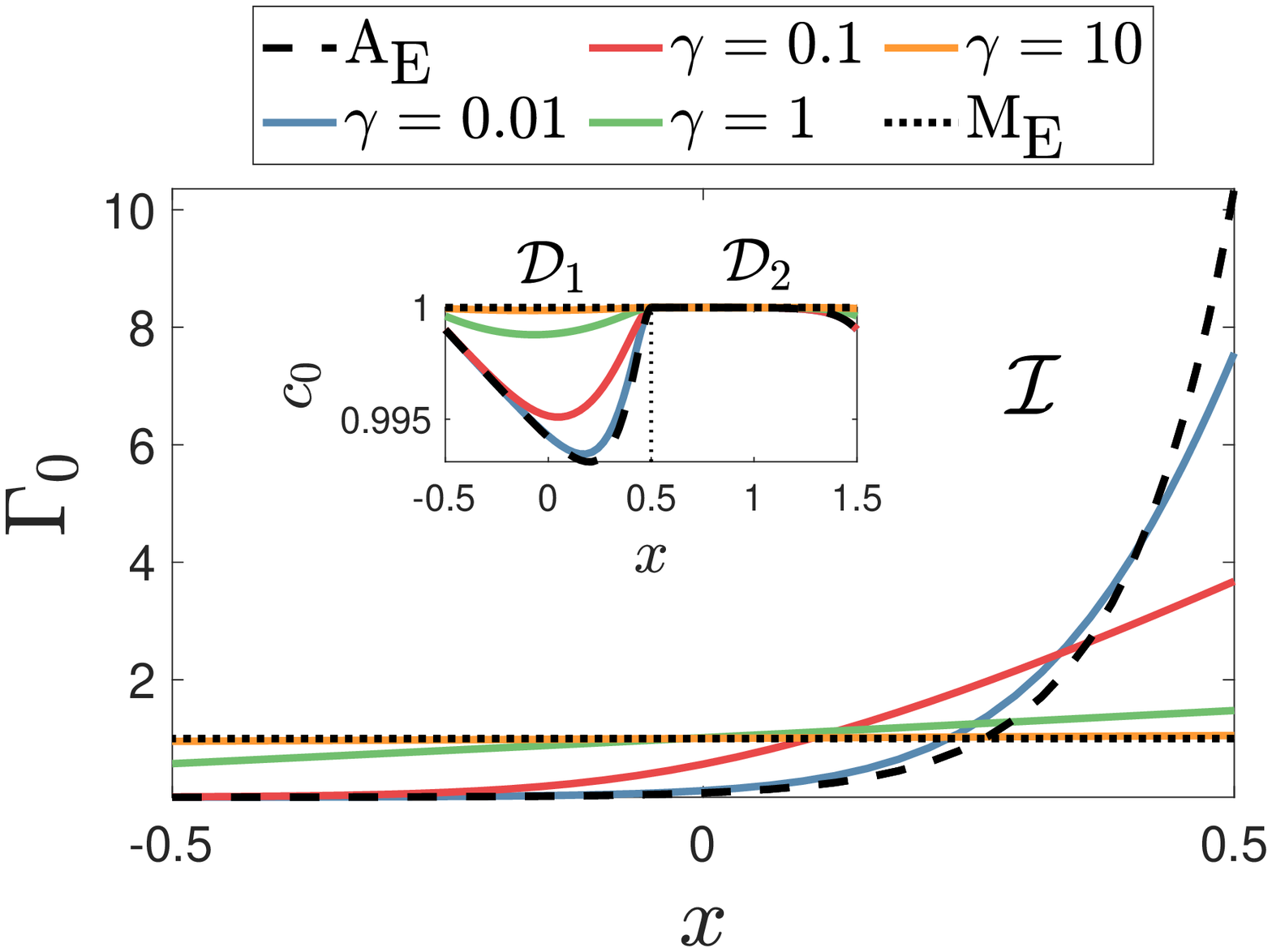} \hspace{.8cm}
\includegraphics[trim={0 0cm 0 0cm}, clip,width=.425\textwidth]{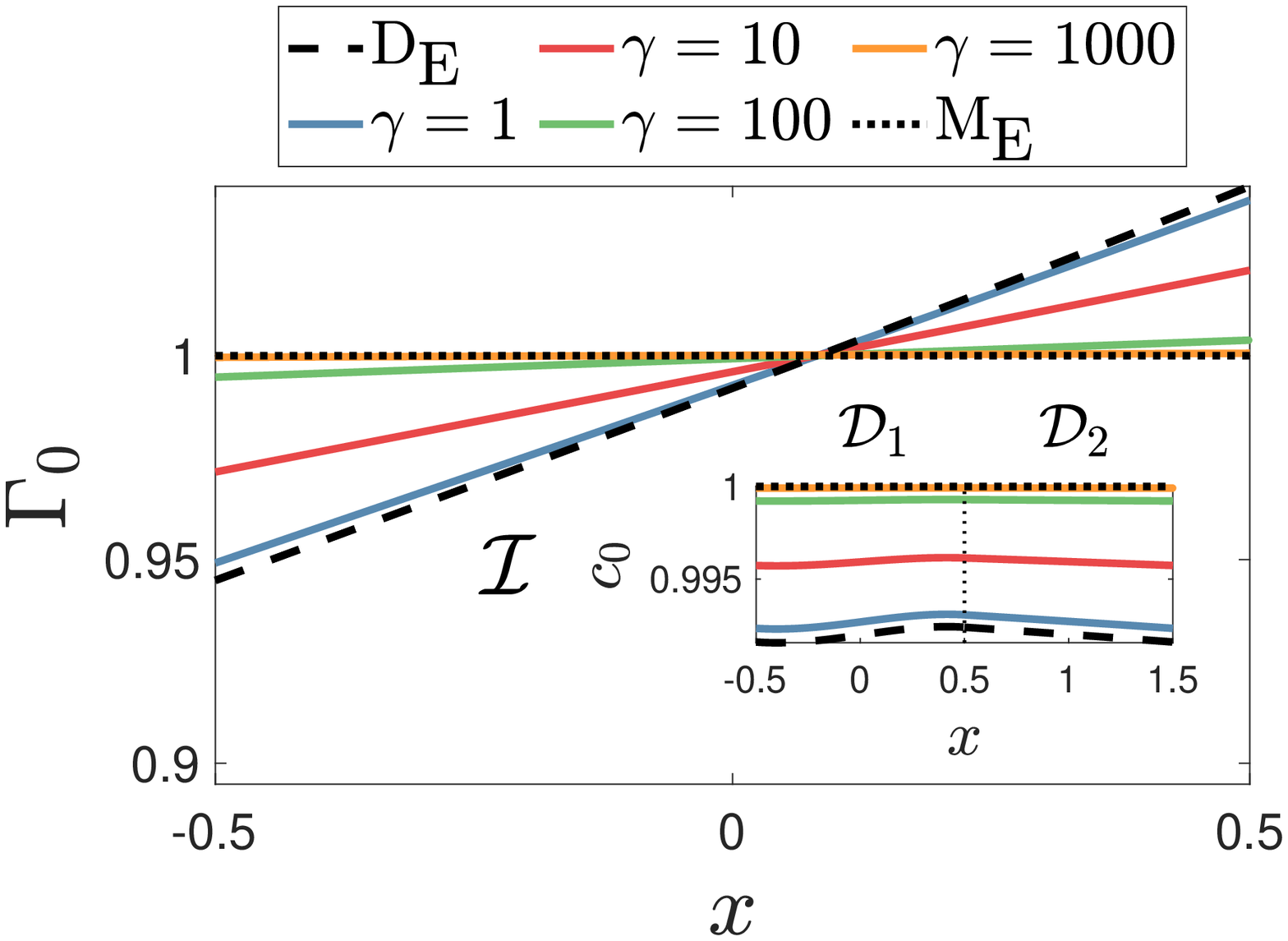}
\caption{
The leading-order drag reduction (${DR}_0$), bulk surfactant ($c_0$) and interfacial surfactant distribution ($\Gamma_0$) in the moderate-exchange problem, for $\beta = 1$, $\phi_x = 0.5$, $\epsilon = 0.1$, $\phi_z = 0.5$ and $P_z = 0.5$, computed using \eqref{eq:model1}--\eqref{eq:model3}.
Contours of ${DR}_0$ for (\textit{a}) $\alpha = \delta = 0.1$ and (\textit{b}) $\alpha = \delta = 10$, where ${DR}_0 = 0$ exhibits a no-slip SHS and ${DR}_0 = 1$ exhibits a shear-free plastron at the SHS. 
The Marangoni (M), advection (A), diffusion (D), Marangoni-exchange ($\text{M}_\text{E}$), advection-exchange ($\text{A}_\text{E}$), diffusion-exchange ($\text{D}_\text{E}$) regions are separated by black lines and ${DR}_0$ is approximated by (\ref{eq:drs_m}), (\ref{eq:drs_a}), (\ref{eq:drs_d}), (\ref{eq:drs_m_we}), (\ref{eq:we_dr_d}) and (\ref{eq:we_dr_d}) in M, A, D, $\text{M}_\text{E}$, $\text{A}_\text{E}$ and $\text{D}_\text{E}$, respectively.
The dashed magenta lines describing when ${DR}_0 \approx0.5$ are given by (\ref{eq:mdboundary}) in (\textit{a,b}) (right curves), (\ref{eq:we_50_1}) in (\textit{b}) and (\ref{eq:we_50_2}) (\textit{a}).
Plots of $\Gamma_0$ and $c_0$ for varying surfactant strength ($\gamma$), where (\textit{c}) $\alpha = \delta = 0.1$ and $\nu = 10^{-4}$, $\text{A}_\text{E}$: $--$ \eqref{eq:we_d}, $\text{M}_\text{E}$: $\cdot\cdot\cdot$ \eqref{eq:we_m} with $\gamma = 10$, and (\textit{d}) $\alpha = \delta = 10$ and $\nu = 0.01$, $\text{D}_\text{E}$: $--$ \eqref{eq:we_dd}, $\text{M}_\text{E}$: $\cdot\cdot\cdot$ \eqref{eq:we_m} with $\gamma = 1000$.
The star identifies the point of the $(\nu, \, \gamma)$-plane where we examine the flow field in \S\ref{subsubsec:ff_weak_exchange}.}
\label{fig:diff}
\end{figure}

Depending on the strength of diffusion (via the parameters $\alpha$ and $\delta$, which vary orthogonally to the $(\nu,\gamma)$-plane in figure~\ref{fig:map_3}\textit{a}), the lower half of the $(\nu,\,\gamma)$-plane is composed of either a sub-region of A, D, or the transition region AD between A and D. In figure~\ref{fig:map_3}(\textit{a}), since $\alpha=\delta=1$, this corresponds to the transition region AD.
In contrast, in figure \ref{fig:diff}(\textit{a}) we plot ${DR}_0$ for $\alpha = \delta=0.1$  and $\alpha = \delta=10$ in figure \ref{fig:diff}(\textit{b}).
The $\text{M}_\text{E}$ region transitions into the diffusion-exchange-dominated ($\text{D}_\text{E}$) region across the $\text{M}_\text{E}\text{D}_\text{E}$ boundary for $\min(\alpha, \,\delta) \gg O(1)$ (figure \ref{fig:diff}(\textit{b})), whereas the $\text{M}_\text{E}$ region transitions into the advection-exchange-dominated ($\text{A}_\text{E}$) region across the $\text{M}_\text{E}\text{A}_\text{E}$ boundary for $\max(\alpha, \,\delta) \ll O(1)$ (figure \ref{fig:diff}(\textit{a})). 
At the $\text{M}_\text{E}\text{D}_\text{E}$ boundary, the asymptotic analysis in Appendix \ref{subsec:a_we_strong_marangoni_diffusion} shows that
\begin{equation} \label{eq:we_50_1}
    {DR}_0 \approx 0.5 \quad \text{when} \quad \gamma = \delta \quad \text{for} \quad \min(\alpha,\, \delta) \gg 1, \quad \frac{\nu}{\epsilon^2} \ll 1,
\end{equation}
which gives the leftmost dashed line in figure \ref{fig:diff}(\textit{b}).
This asymptote agrees with the numerical solution presented in figure \ref{fig:diff}(\textit{b}) for $\delta \gg 1$ and $\nu \rightarrow 0$.
The range of validity for the asymptote given in \eqref{eq:we_50_1} is extended to $\delta = O(1)$ to include the $\text{M}_\text{E}\text{AD}_\text{E}$ boundary, for which we give partial justification in Appendix \ref{subsec:a_we_strong_marangoni_o1diff}.
At the $\text{M}_\text{E}\text{A}_\text{E}$ boundary, Appendix \ref{subsec:a_we_strong_marangoni_weak_diffusion} demonstrates that
\begin{equation} \label{eq:we_50_2}
{DR}_0 \approx 0.5 \quad \text{when} \quad \gamma = \frac{\beta \phi_x}{4} + \frac{\delta}{2} \quad \text{for} \quad \max(\alpha, \, \delta) \ll 1, \quad \frac{\nu}{\epsilon^2} \ll 1,
\end{equation}
which gives the leftmost dashed line in figure \ref{fig:diff}(\textit{a}).
This asymptote approximates the numerical solution presented in figure \ref{fig:diff}(\textit{a}) where $\delta = 0.1$ as $\nu \rightarrow 0$. 
Agreement improves for smaller $\delta$.
It is notable that the threshold \eqref{eq:we_50_2} differs from the strong-exchange limit \eqref{eq:largebetaflat} by a factor of 4.

We next analyse the drag reduction in both the $\text{A}_\text{E}$ and $\text{D}_\text{E}$ regions, where advection or diffusion induce a near uniform distribution of surfactant at the interface. 
In Appendix \ref{subsec:a_we_sd_d} we show that ${DR}_0$ in regions $\text{A}_\text{E}$ and $\text{D}_\text{E}$ is close to the shear-free value
\begin{equation} \label{eq:we_dr_d}
{DR}_0 \approx 1 - \frac{\gamma}{\delta} \quad \text{for} \quad  \gamma \ll \min(1,\,\alpha,\,\beta,\,\delta), \quad  \frac{\nu}{\epsilon^2} \ll \min(1,\,\alpha,\,\beta,\,\delta).
\end{equation} 
Equation \eqref{eq:we_dr_d} shows how drag-promoting Marangoni effects are weakened by strong diffusion at the interface, when surfactant exchange between the bulk and interface is weak. 
Figure \ref{fig:diff}(\textit{c,\,d}) shows the surfactant profiles in $\text{A}_\text{E}$ and $\text{D}_\text{E}$, the leading-order solution on $\mathcal{I}$ being given by 
\begin{equation} \label{eq:we_d}
    \Gamma_{0} \approx \frac{2 \phi_x \beta \exp\left(\frac{\beta \left(\phi_x +x\right)}{\delta }\right)}{\delta \left(\exp\left(\frac{2 \beta \phi_x }{\delta }\right)-1\right)} \quad \text{for} \quad  \gamma \ll \min(1,\,\alpha,\,\beta,\,\delta), \quad  \frac{\nu}{\epsilon^2} \ll \min(1,\,\alpha,\,\beta,\,\delta).
\end{equation}
From \eqref{eq:we_d}, weak surface diffusion, $\delta \ll O(1)$, decreases (increases) $\Gamma_0$ at the upstream (downstream) end of the interface. 
A downstream boundary layer forms to satisfy the mass balance condition \eqref{eq:dimensionless_net_flux}, condensing the Marangoni effect to a small region near $x=\phi_x$. 
The streamwise velocity flows over a almost shear-free boundary at the upstream end of the interface. 
When surface diffusion is strong, $\delta \gg O(1)$, \eqref{eq:we_d} reduces to 
\begin{equation} \label{eq:we_dd}
\Gamma_0 = 1 + \frac{\beta x}{\delta} \quad \text{for} \quad  \gamma \ll \min(1,\,\alpha,\,\beta,\,\delta), \quad  \frac{\nu}{\epsilon^2} \ll \min(1,\,\alpha,\,\beta,\,\delta).
\end{equation}
Noting that $\beta/\delta \propto \hat{H}^2 \hat{U}/(\hat{D}_I \phi_z\hat{P}_z)$, \eqref{eq:we_dd} demonstrates how the gradient of $\Gamma_0$ is controlled by the ratio of advection to diffusion at the interface, which allows surface diffusion to regulate ${DR}_0$ in \eqref{eq:we_dr_d}.
In figure \ref{fig:diff}(\textit{a,\,b}), the horizontal boundary of the $\text{M}_\text{E}$ region (shown with a black solid line) moves downwards when surface diffusion decreases ($\delta=0.1$ in figure \ref{fig:diff}\textit{a}) and upwards when it increases ($\delta=10$ in figure \ref{fig:diff}\textit{b}).
Figure \ref{fig:diff}(\textit{c,\,d}) demonstrates how increasing surface diffusion attenuates gradients of surfactant at the interface, increasing the drag reduction from sub-region $\text{A}_\text{E}$ to $\text{D}_\text{E}$ in figure \ref{fig:diff}(\textit{a,\,b}).

\subsubsection{Flow field}
\label{subsubsec:ff_weak_exchange}

\begin{figure}
\centering
(\textit{a}) \hfill (\textit{b}) \hfill \hfill \hfill \\
\vspace{-.5cm} \includegraphics[width=.425\textwidth]{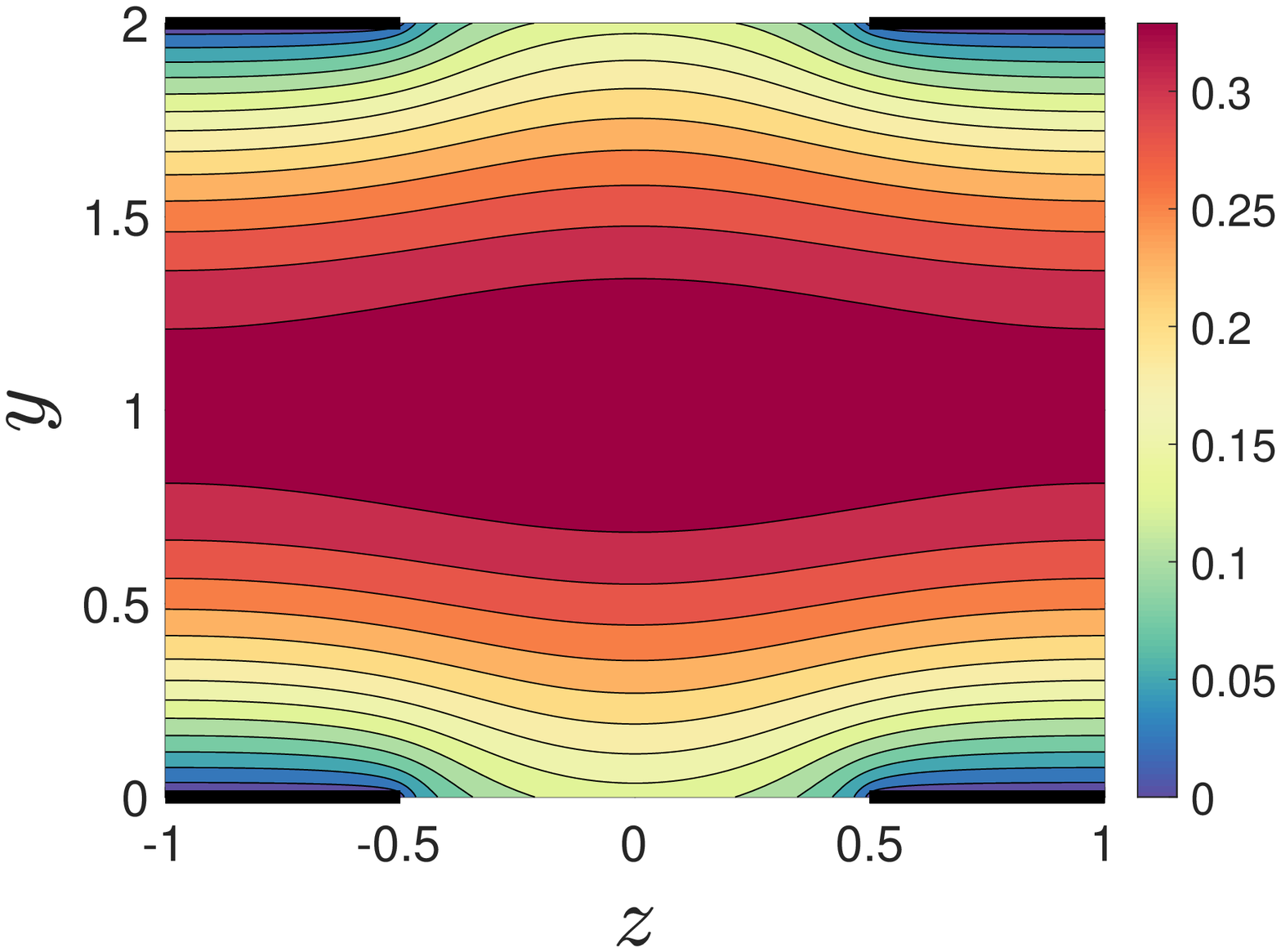} \hspace{.8cm} \includegraphics[width=.425\textwidth]{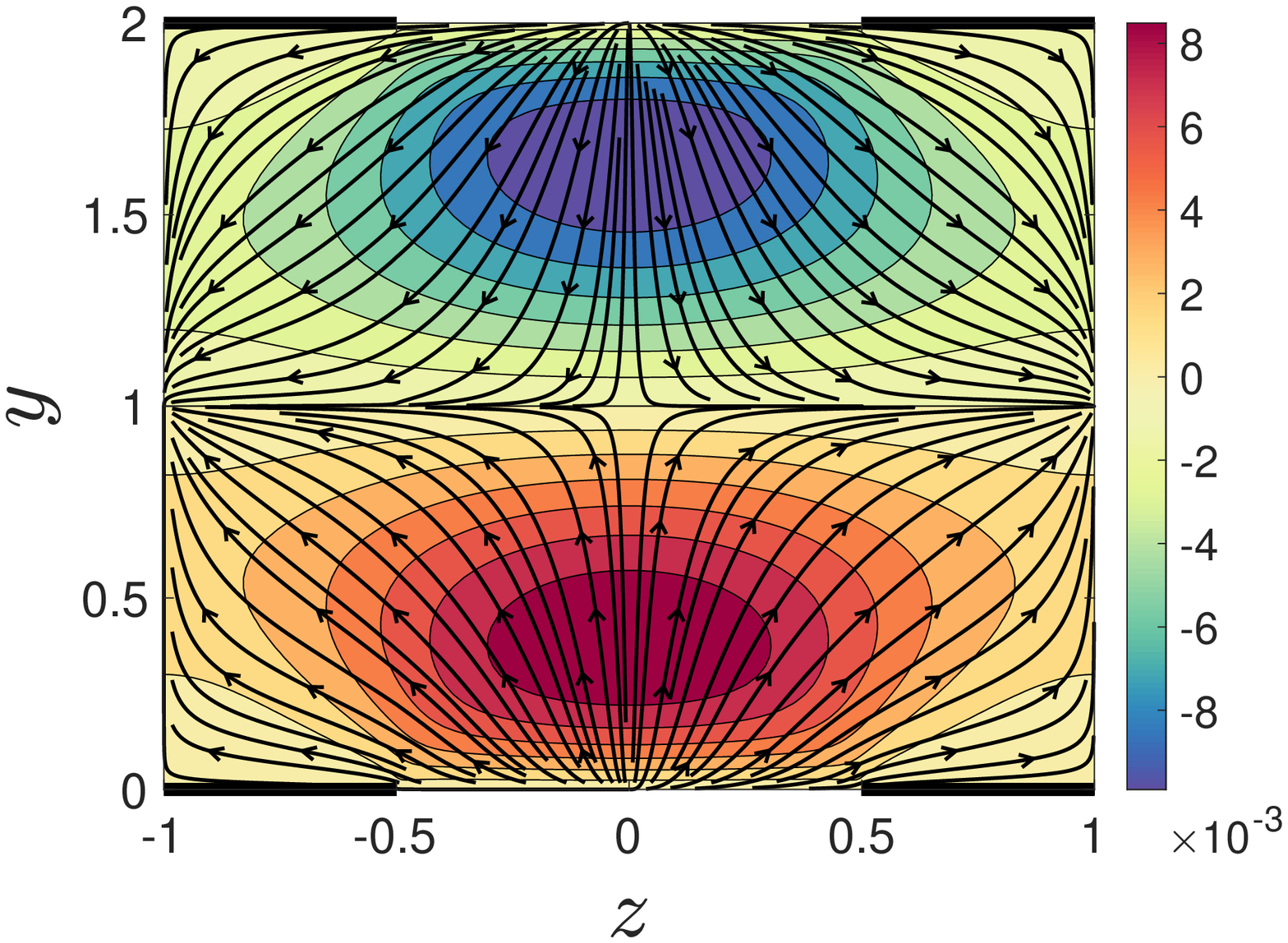} \\
(\textit{c}) \hfill (\textit{d}) \hfill \hfill \hfill \\
\vspace{-.5cm} \includegraphics[width=.425\textwidth]{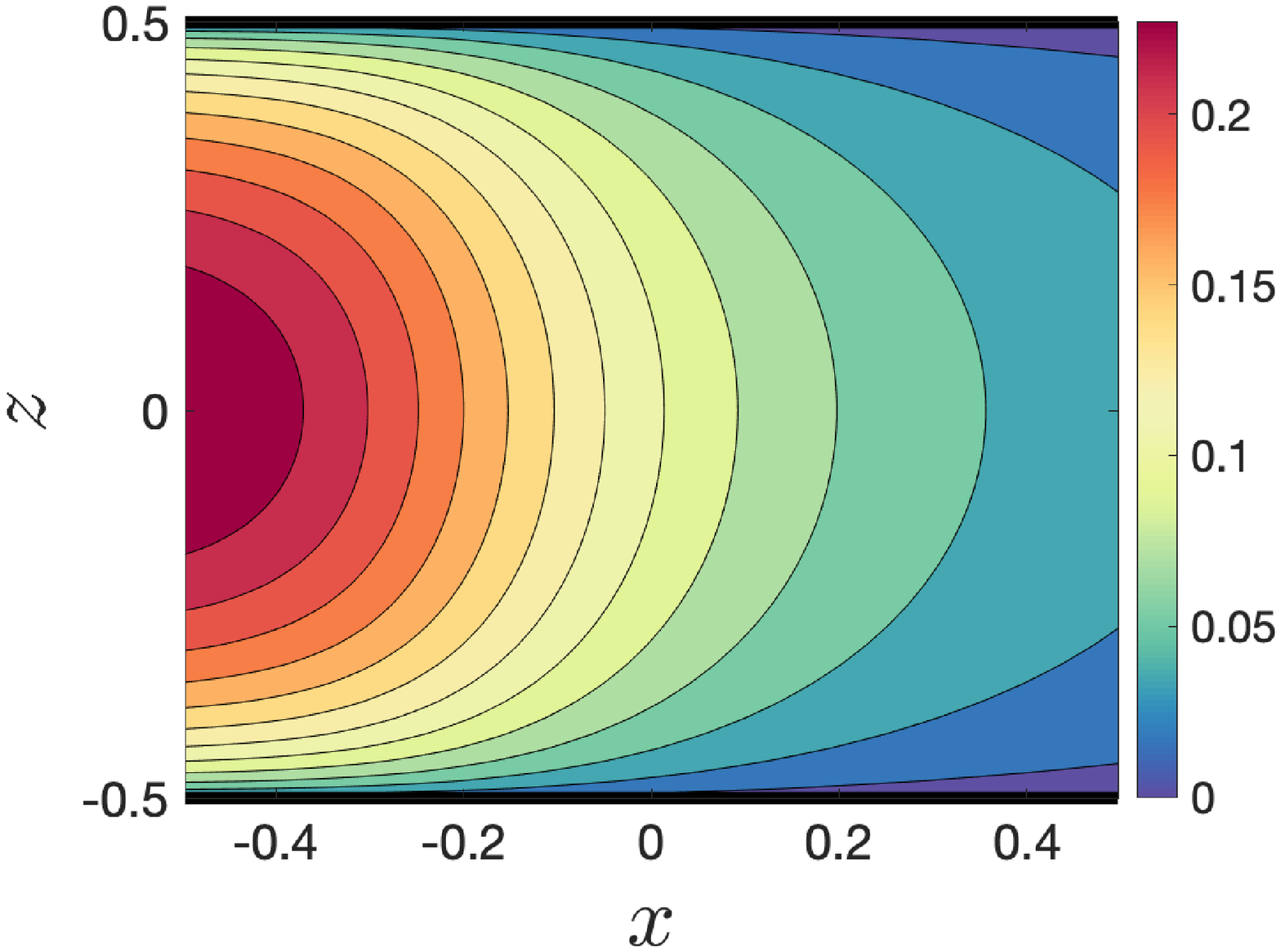} \hspace{.8cm} \includegraphics[width=.425\textwidth]{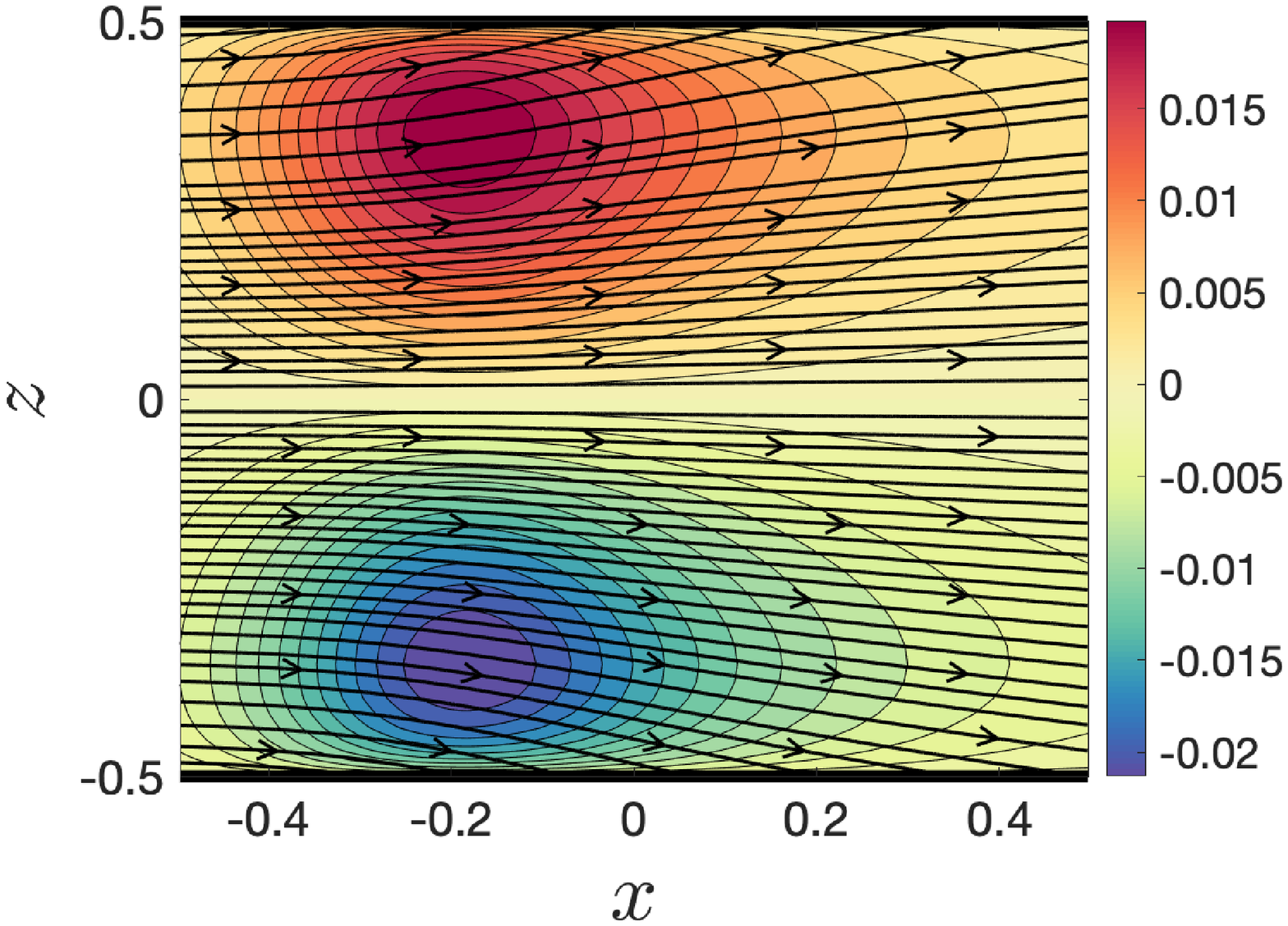}
\caption{
Contour maps of the flow field in the moderate-exchange problem, for $\alpha = 0.1$, $\beta = 1$, $\gamma = 0.2$, $\delta = 0.1$, $\nu = 10^{-4}$, $\phi_x = 0.5$, $\phi_z = 0.5$ and $P_z = 0.5$, corresponding to 50\% normalized drag reduction given by the star in figure \ref{fig:diff}(\textit{a}).
(\textit{a}) Leading-order streamwise velocity $u_0$ and (\textit{b}) leading-order wall-normal velocity $v_1$ with $(v_1, \, w_1)$ streamlines at the centre of the plastron, $x=0$. 
(\textit{c}) Leading-order streamwise velocity $u_0$ and (\textit{d}) leading-order transverse velocity $w_1$ with $(u_0, \, w_1)$ streamlines at the interfaces, $y=0$ or $2$.
The thick black lines in (\textit{a--d}) represent the solid regions of the SHS.
}
\label{fig:ff_2}
\end{figure}

Videos of $u_0$, $v_1$, $w_1$ and $\Gamma_{0x}$ are given in supplementary movie 2.
Again, we choose an example in figure \ref{fig:ff_2} for which ${DR}_0=0.5$, shown with a star in \ref{fig:diff}(\textit{a}).
Figure \ref{fig:ff_2}(\textit{a}) shows that $u_0$ has a similar structure to the strong-exchange problem at $x=0$ (see figure \ref{fig:ff_1}\textit{a}). 
However, when comparing figure \ref{fig:ff_2}(\textit{c}) to figure \ref{fig:ff_1}(\textit{c}), $u_0$ exhibits different behaviour along $\mathcal{I}$.
The streamwise velocity decreases slowly at the upstream end of $\mathcal{I}$, $\Gamma_{0x}$ and $\Gamma_{0xx}$ increase slowly with $x$, and there is a uniform distribution of surfactant at the interface which is almost shear-free.
The streamwise velocity decreases rapidly at the centre of $\mathcal{I}$, where there is a sudden change of surfactant gradient at the interface. 
The streamwise velocity then decreases at the downstream end of $\mathcal{I}$, where there is a more linear distribution of surfactant at the interface which is almost no-slip.
The cross-channel flow in figure \ref{fig:ff_2}(\textit{b}) closely resembles that observed in figure \ref{fig:ff_1}(\textit{b}).
In supplementary movie 2 and figure \ref{fig:ff_2}$(d)$, the magnitudes of both $v_1$ and $w_1$ increase until they attain a maximum at the start of the low-slip region, after which, they decrease towards $x=\phi_x$.
The streamwise location of the maximum of $v_1$ and $w_1$ is approximately the same as the streamwise location of the maximum of $\Gamma_{0xx}$ on $\mathcal{I}$, the ``corner'' in the surfactant field.

\subsection{Comparison with numerical simulations}

\begin{figure}
\centering
    (\textit{a}) \hfill (\textit{b}) \hfill \hfill \hfill \\
    \vspace{-.5cm} \includegraphics[width=.425\textwidth]{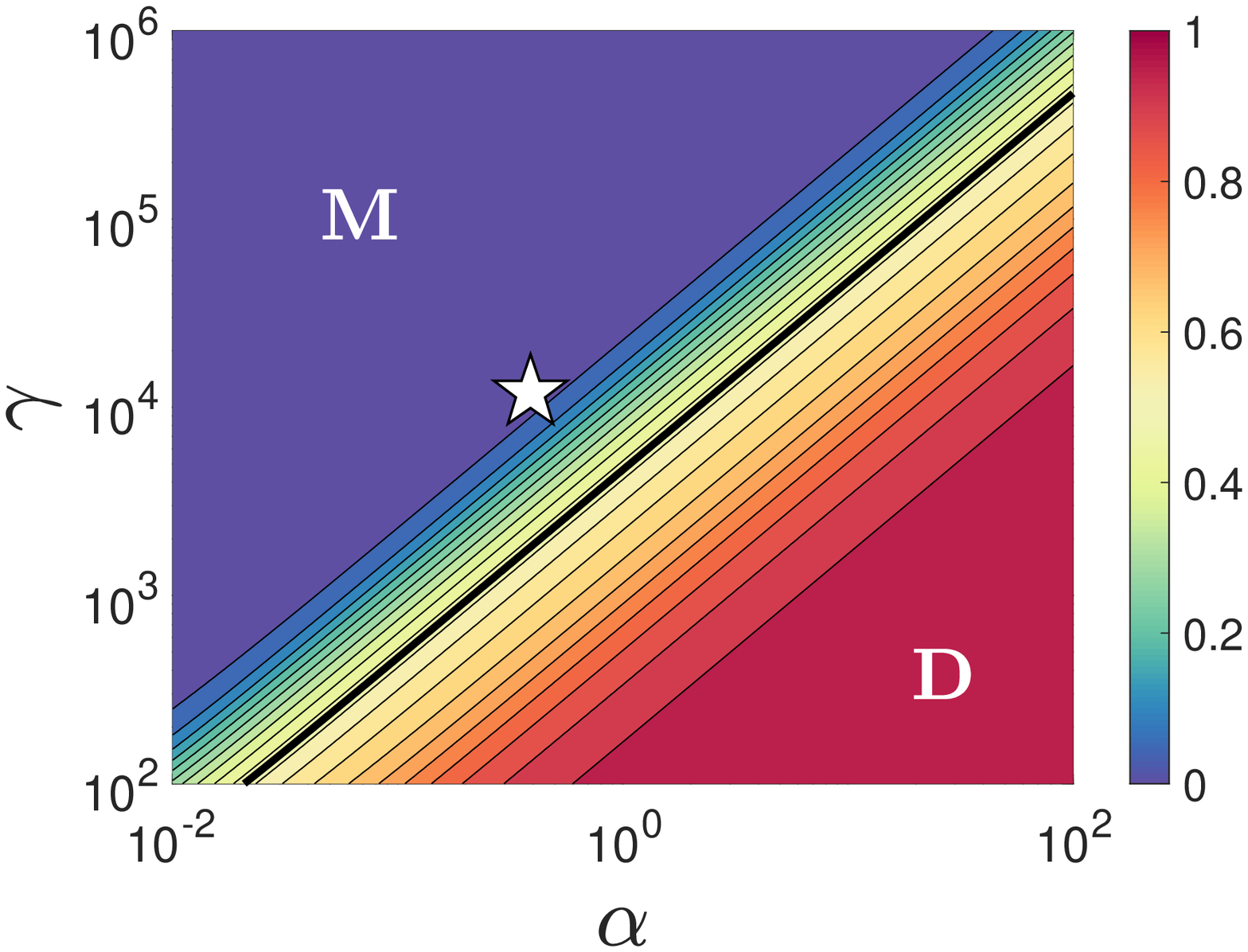} \hspace{.8cm} \includegraphics[width=.42\textwidth]{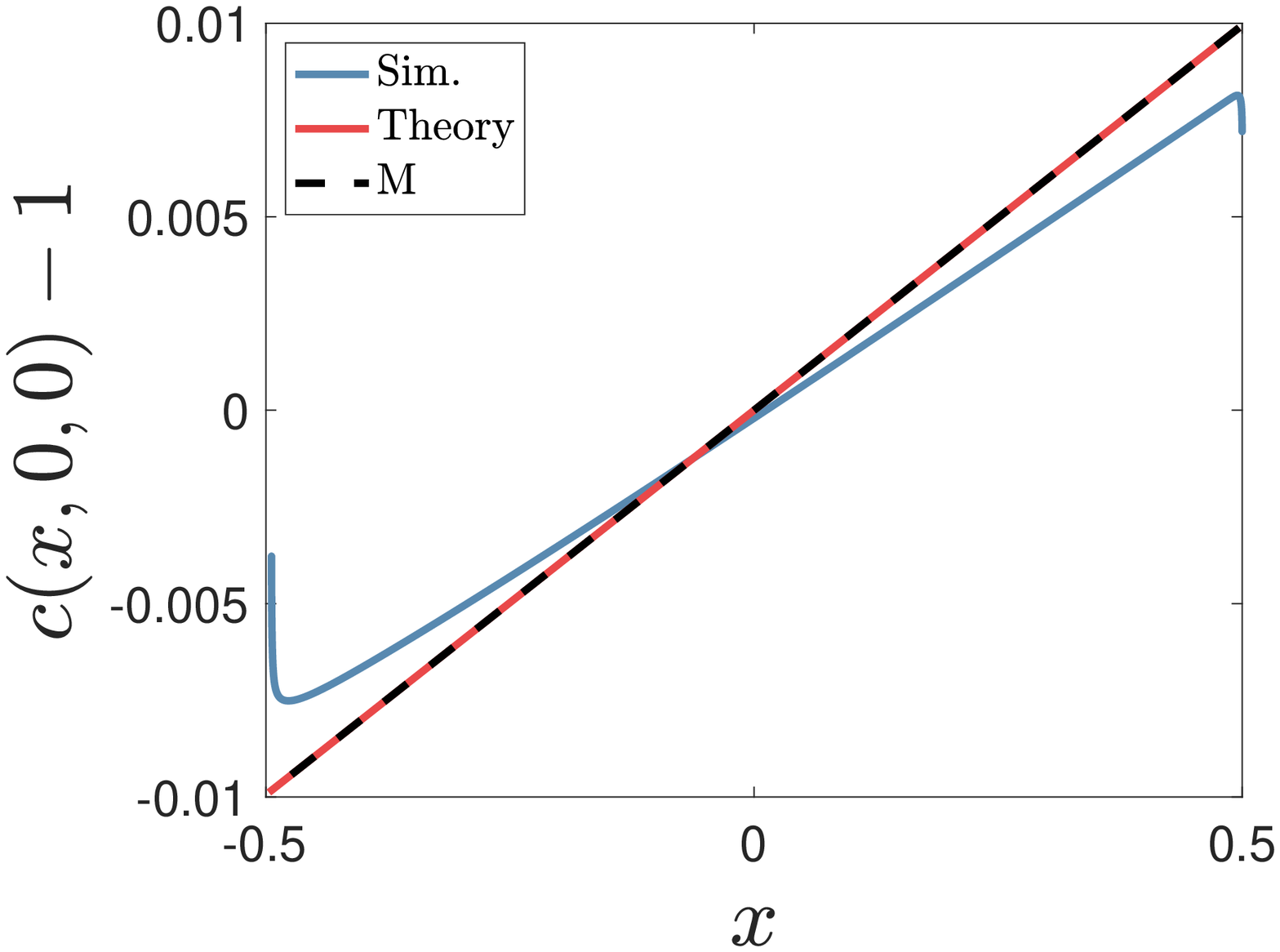} \\
    (\textit{c}) \hfill (\textit{d}) \hfill \hfill \hfill \\
    \vspace{-.5cm} \includegraphics[width=.425\textwidth]{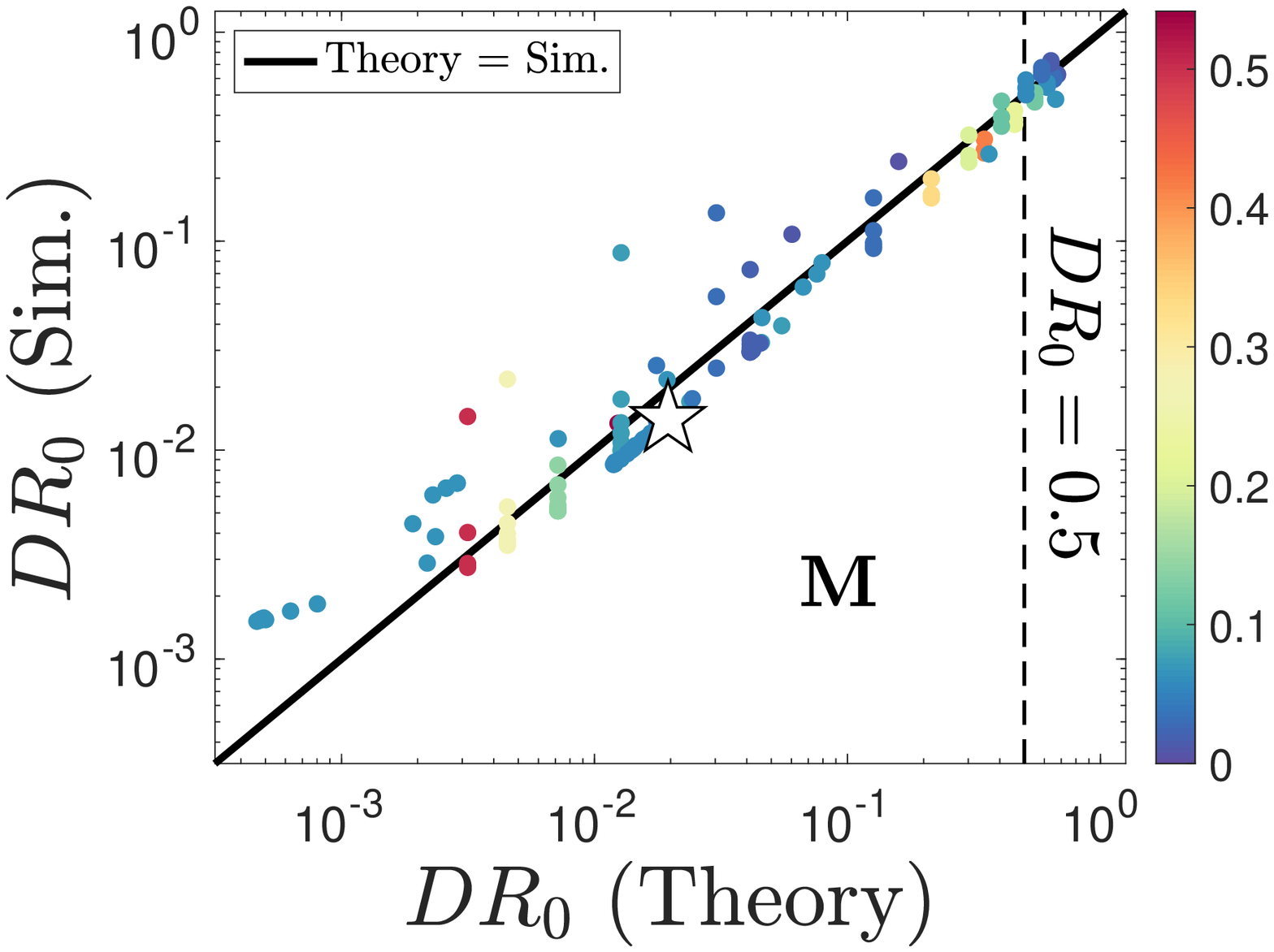} \hspace{.8cm} \includegraphics[width=.425\textwidth]{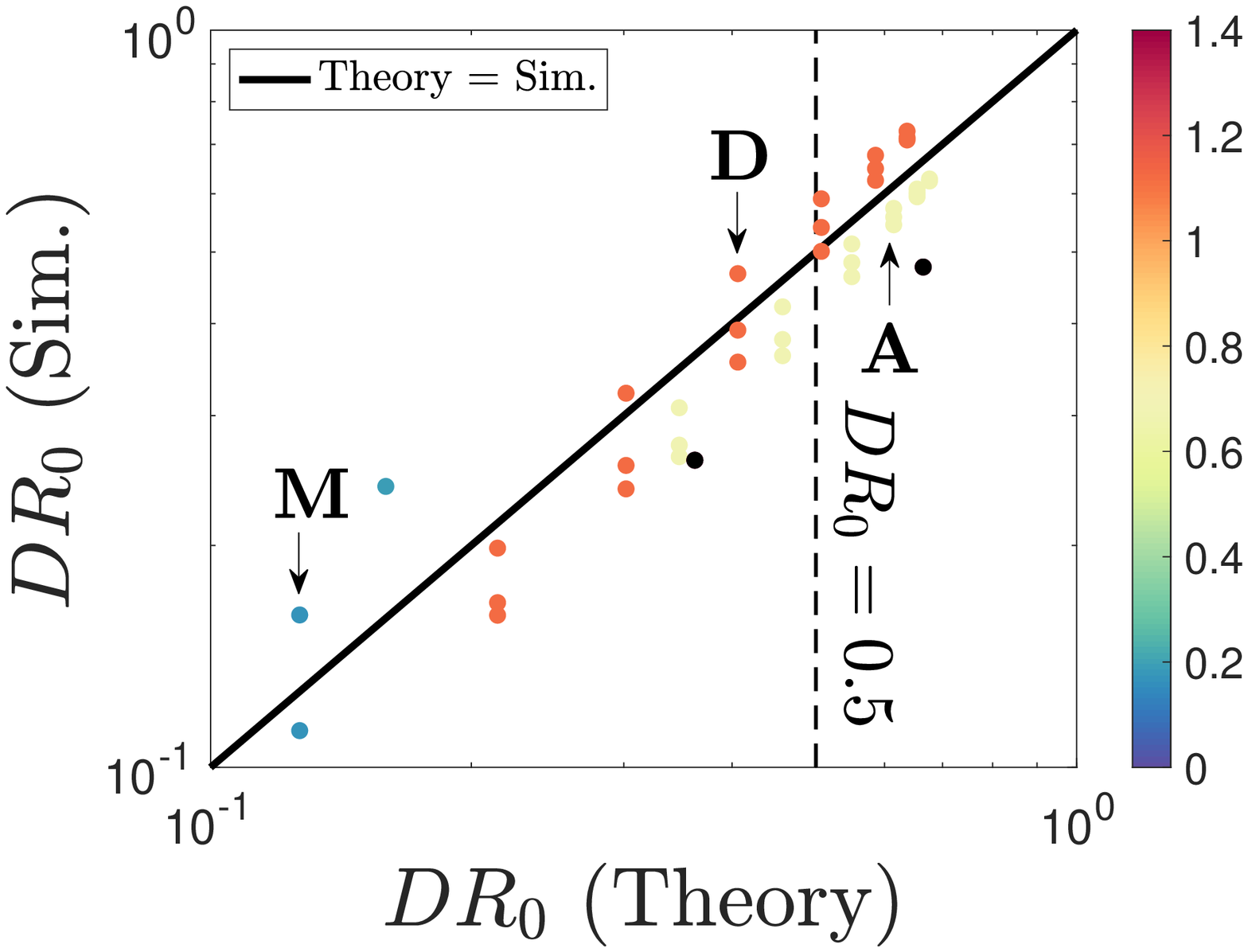}
\caption{
The drag reduction (${DR}_0$) and bulk surfactant distribution evaluated at the SHS ($c_0(x,0,0)$), using the theory presented here and numerical simulations provided by \citet{temprano2023single}.
(\textit{a}) Contours of ${DR}_0$, where ${DR}_0 = 0$ exhibits a no-slip SHS and ${DR}_0 = 1$ exhibits a shear-free SHS, for $\beta = 38.6$, $\delta = 1$, $\nu = 0.2$, $\phi_x = 0.99$, $\phi_z = 2/3$ and $P_z = 1$.
The Marangoni (M) and diffusion-dominated (D) regions are separated by the black line (along which ${DR}_0 = 0.5$).
The advection-dominated (A) region appears for $\alpha < 0.01$ and $\gamma < 100$.
(\textit{b}) Plot of $c_0$, where $\alpha = 0.4$, $\beta = 38.6$, $\gamma = 1.2\times 10^4$, $\delta = 1$, $\nu = 0.2$, $\phi_x = 0.99$, $\phi_z = 2/3$ and $P_z = 1$ (corresponding to the stars in panel \textit{a} and \textit{c}), computed using \eqref{eq:composite_equation} (red) and \eqref{eq:zeta_gg_1} (dashed and black).
(\textit{c}) Scatter plot of ${DR}_0$ using our theory (\eqref{eq:dr_def} and \eqref{eq:model1}--\eqref{eq:model3}) and the 159 numerical simulations detailed in \citet{temprano2023single}, the colorbar gives the magnitude of $\epsilon$ for given data point and the dashed line is where ${DR}_0=0.5$.
(\textit{d}) Same data as in (\textit{c}): scatter plot of ${DR}_0 \in [0.1,\, 1]$. 
The colourbar gives the magnitude of $\alpha$ for given data point and the black points have $\alpha \gg 1.5$.
The arrows indicate the region of parameter space for a given data point: orange points are in region D, yellow points are in region A and blue points are in region M.
}
\label{fig:dns}
\end{figure}

Finally, we compare our model predictions with the numerical simulations detailed in tables SI and SII and figures S1 and S2 in \citet{temprano2023single}, which were designed to be representative of microchannel applications. 
They used finite-element simulations to solve the steady 3D Stokes equations, which were coupled to advection--diffusion equations for bulk and interfacial surfactant.
The equation of state and adsorption-desorption kinetics, which link the velocity field to the bulk and interfacial concentrations, were derived from the nonlinear Frumkin isotherm \citep{chang1995adsorption}.
The authors considered a channel with only one SHS at the bottom wall ($y=0$), so our theory in \S\ref{sec:model} is adjusted accordingly for a solid surface at the top wall ($y=2$). 
This amounts to reevaluating $\tilde{U}$, $\bar{U}$ and $\breve{U}$, as well as modifying the surface surfactant flux term in \eqref{eq:leading_order_surfactant_flux} to account for the  contribution of a single interface at $y=0$.
Parameters are detailed in table SI in \citet{temprano2023single}.

In figure \ref{fig:dns}(\textit{a,\,b}) we compare our model to a representative example from numerical simulations performed in \citet{temprano2023single}, with parameters $\hat{H}=6\times10^{-5}\,$m, $\hat{P}_z=6\times10^{-5}\,$m and $\phi_z = 2/3$, from which we can evaluate $P_z$, $\tilde{Q}$, $\bar{Q}$, $\tilde{q}$ and $\bar{q}$ using (\ref{eq:bulkfluxes}, \ref{eq:surfacefluxes}).
Using \eqref{eq:coefficients} (appropriately adjusted for a single SHS), we can calculate the transport coefficients: $\alpha = 0.4$, $\beta = 38.6$, $\gamma = 1.2 \times 10^4$, $\delta = 1$, $\epsilon = 0.02$, $\nu = 0.2$ and $\phi_x=0.99$; with $\Pen = \Pen_I = 20.6$, $\Da=30.1$, $\Ma = 1.2\times 10^4$ and $\Bi = 0.2$, as in \citet{temprano2023single}.
Note that $\Ma$ is defined differently to \citet{temprano2023single}; the value provided here uses our definition of $\Ma$.
Figure \ref{fig:dns}(\textit{a}) shows that the example lies within the Marangoni-dominated region M, where ${DR}_0$ is close to zero and the liquid--gas interface is immobilised. 
Using this information, in figure \ref{fig:dns}(\textit{b}), we compare the numerically simulated bulk surfactant concentration $c_0(x,0,0)$ from \citet{temprano2023single} (blue solid curve) with our prediction for $c_0$ given by the solution to the 1D ODE model solving \eqref{eq:composite_equation} subject to (\ref{eq:model3}\textit{a,\,b}) (red solid line), and the (indistinguishable) asymptotic solution in M \eqref{eq:zeta_gg_1} (black dashed line).
The gradient of $c_0$ at the centre of the plastron computed using our theory ($\text{d} c_0 / \text{d} x(0) \approx 0.19$) and the numerical simulation ($\text{d} c / \text{d} x(0,\,0,\,0) \approx 0.16$) of \citet{temprano2023single} are similar, noting that our theoretical prediction does not require any fitting parameters. However, the numerical simulation reveals thin boundary layers close to the upstream and downstream contact lines. 
These boundary layers are not captured by our long-wave theory. Within these inner regions, the 3D Stokes and surfactant transport equations govern the flow and surfactant field near the no-flux stagnation points at the upstream and downstream ends of the interface. This comparison suggests that the inner layers at the upstream and downstream ends are not needed to estimate the leading-order drag reduction in region M, but capturing the inner layers may be necessary for more accurate predictions.

In figure~\ref{fig:dns}(\textit{c}), we  compare the drag reduction results from all the 3D numerical simulations of \citet{temprano2023single} with the leading-order drag reduction predictions from our 1D model. 
These simulations span the whole parameter space characteristic of realistic microchannel applications. 
In order to approximate the drag reduction from the streamwise slip length results provided by \citet{temprano2023single}, we integrate the streamwise velocity field when $\phi_z=0$ for the solid-walled and SHS flows.
For the flow over a single SHS, we replace the mixed boundary conditions at $y=0$ with $\lambda_e u_{0y} - u_0 = 0$, to find $u_0$ as a function of $y$ and $\lambda_e$.
We then use ${DR}_0 = (\Delta p_R - \Delta p_0) / (\Delta p_R - \Delta p_U)$ to relate the leading-order drag reduction and streamwise slip length, ${DR}_0 = 1 - 1 / (2 \lambda_e + 1)$, in a similar manner to \cite{landel2020theory}.
There is no obvious correlation between the scatter and the size of $\epsilon$ (shown in colour), which varies up to $\epsilon\approx 0.5$ as highlighted by the colorbar in figure \ref{fig:dns}(\textit{c}).
The scatter could be due to unresolved cross-channel concentration gradients or streamwise boundary layers at the ends of the plastron; both hypotheses require further testing against targeted numerical simulations. 
Nonetheless, our theory compares well with simulations for ${DR}_0 > 0.1$ in figure~\ref{fig:dns}(\textit{d}); the root-mean-squared error between the simulated data from \citet{temprano2023single} and our theory is $0.07$.
The majority of the data points have ${DR}_0 < 0.5$ and lie in region M. 
Those that have ${DR}_0 > 0.5$ are classified as belonging to region A when $\alpha < 1$ or D when $\alpha > 1$, as highlighted by the colorbar in figure \ref{fig:dns}(\textit{d}).
None of the simulations with high drag reduction (${DR}>0.5$) had a bulk diffusion strength small enough to lie on the boundary of region G ($\alpha = O(\epsilon^2)$), where cross-channel concentration gradients become important and shear dispersion decreases the drag. Hence, the effect of shear dispersion we have described cannot be investigated from the simulations of \citet{temprano2023single}.

\section{Discussion} \label{sec:discussion}

The drag-reducing potential of superhydrophobic surfaces (SHSs) may be compromised by trace amounts of surfactant \citep{peaudecerf2017traces}.
In this paper, we have derived an asymptotic theory for 3D laminar flow, in a plane-periodic channel with SHSs made of an array of long but finite-length longitudinal grooves along both walls, which has been contaminated with soluble surfactant.
The mass, momentum and surfactant equations are solved in the Stokes flow limit, where the adsorption--desorption kinetics and equation of state are linearised.
We have investigated regimes where cross-channel concentration gradients are small, developing a long-wave theory that accounts for a rapidly equilibrating surfactant-driven transverse flow. 
This results in a 1D model \eqref{eq:model1}--\eqref{eq:model3} for surfactant transport, which incorporates advection, diffusion, Marangoni effects and exchange between the bulk and the interface. 
No parameter fitting is required, in contrast to the 3D theory outlined in \citet{temprano2023single} and the 2D theory in \citet{landel2020theory}.
Using this theory, we gain access to parts of the parameter space that are unavailable using models that assume uniform shear stress at the interface \citep{landel2020theory}, and we make asymptotic predictions for the drag reduction and surfactant concentration distribution that complement expensive numerical simulations of the 3D flow and surfactant equations \citep{temprano2023single}. 

\setlength{\tabcolsep}{1.4em}
\begin{table}
\resizebox{\columnwidth}{!}{%
\centering
    \begin{tabular}{c c c c}
    \hline \\[-6pt]
    \multicolumn{3}{c}{\hspace{3cm} Strong exchange $\displaystyle \frac{\nu}{\epsilon^2} \gg \max(1,\,\alpha,\,\beta,\,\delta)$} \\[4pt]
    \hline
    Region & Parameter space & ${DR}_0$ & Regime \\[6pt]
    $\text{M}$ & $\displaystyle \gamma \gg \max\left(1,\,\alpha,\,\beta,\,\delta,\,\frac{\epsilon^2}{\alpha}\right)$ & $\displaystyle \frac{1}{\gamma}\left(\alpha + \delta + \frac{\epsilon^2 s}{\alpha} +  \frac{\phi_x(E+1)}{\left(E-1\right)} \right)$ & ${DR}_0 \ll 1$ \\[10pt]
    $\text{M}_\text{D}$ & $\displaystyle \gamma \gg (\alpha,\, \delta) \gg \max\left(1,\,\beta,\,\frac{\epsilon^2}{\alpha}\right)$ & $\displaystyle \frac{\alpha} {\gamma(1 - \phi_x)} + \frac{\delta}{\gamma}$ & ${DR}_0 \ll 1$ \\[10pt]
    $\text{M}_\text{G}$ & $\displaystyle \gamma \gg \frac{\epsilon^2}{ \alpha} \gg \max(1,\,\alpha,\,\beta,\, \delta)$ & $\displaystyle \frac{\epsilon^2 (s_4\phi_x + s(1 - \phi_x))}{\gamma\alpha(1 - \phi_x)}$ & ${DR}_0 \ll 1$ \\[10pt]
    $\text{A}$ & $\epsilon^2 \ll \alpha \ll 1, \quad \gamma \ll \min(1,\,\beta)$ & $\displaystyle 1 -\frac{\gamma}{2\phi_x\left(\beta +1\right)}$  & $1-{DR}_0 \ll 1$\\[10pt]
    D & $\min(\alpha,\,\delta) \gg \max(1,\,\gamma)$ & $ \displaystyle 1 - \frac{(1 - \phi_x)\gamma}{(1 + \phi_x \beta)\alpha + (1 - \phi_x)\delta}$ & $1-{DR}_0 \ll 1$ \\
    \hline \\[-6pt]
    \multicolumn{3}{c}{\hspace{3cm} Weak exchange $\displaystyle \frac{\nu}{\epsilon^2} \ll \min(1,\,\alpha,\,\beta,\,\delta)$}\\[4pt]
    \hline
    Region & Parameter space & ${DR}_0$ & Regime \\[6pt]
    $\text{M}_\text{E}$ & $\gamma \gg \max(1,\,\alpha,\,\beta,\,\delta)$ & $\displaystyle \frac{\delta}{\gamma}$ & ${DR}_0 \ll 1$ \\[10pt]
    $\text{A}_\text{E}$, $\text{D}_\text{E}$ & $\gamma \ll \min(1,\,\alpha,\,\beta,\,\delta)$ & $\displaystyle 1 - \frac{\gamma}{\delta}$ & $1-{DR}_0 \ll 1$ \\
    \hline
    \end{tabular}%
    }
    \caption{Summary of the asymptotic predictions of the leading-order drag reduction $DR_0$ in the  main asymptotic regions analysed in the strong-exchange problem with: the Marangoni-dominated region (M) with sub-regions $\text{M}_\text{D}$ and $\text{M}_\text{G}$, the advection-dominated region (A) and the diffusion-dominated region (D); and their analogues in the weak-exchange problem: the $\text{M}_\text{E}$, $\text{A}_\text{E}$ and $\text{D}_\text{E}$ regions. The drag reduction $DR_0$ is expressed in terms of the transport coefficients $\alpha$, $\beta$, $\gamma$, $\delta$ and $\nu$ given in (\ref{eq:coefficients}) and (\ref{eq:coefficients_weak_exchange}) and constants ($s_1 > 0$, $s_2$, $s_3$ and $s_4 > 0$) given in \eqref{eq:s} where $s \equiv s_1 + s_2 + s_3$ and $E \equiv \exp(2\alpha(1-\phi_x)/(\alpha^2 + \epsilon^2 s_4))$.}
    \label{tab:1}
\end{table}
\setlength{\tabcolsep}{.9em}
\begin{table}
    \centering
    \begin{tabular}{c c c c c c c}
    Quantity & $\displaystyle \frac{\gamma}{\alpha}$ & $\displaystyle \frac{\alpha\gamma}{\epsilon^2}$ & $\displaystyle\frac{\gamma}{\beta}$ & $\displaystyle\frac{\gamma}{\alpha\beta}$ & $\displaystyle\frac{\gamma}{\delta}$ \\[12pt]
    Proportional to & $\displaystyle \frac{\hat{A}\hat{C} \hat{L}_d^2\hat{H}}{\hat{\mu} \hat{D}\hat{P}_z}$ & $\displaystyle \frac{\hat{A}\hat{C}\hat{L}_d^2 \hat{D} \hat{H} \hat{P}_z}{\hat{\mu} \hat{Q}^2}$ & $\displaystyle \frac{\hat{A}\hat{C}\hat{L}_d \hat{H}^3}{\hat{\mu}\hat{Q}\hat{P}_x }$ & $\displaystyle \frac{\hat{A}\hat{C}\hat{L}_d\hat{H}^2}{\hat{\mu}\hat{D}\hat{P}_z}$ &  $\displaystyle \frac{ \hat{A}\hat{C}\hat{L}_d\hat{H}^2}{\hat{\mu} \hat{D}_I \phi_z \hat{P}_z}$
    \end{tabular}
    \caption{Summary of the dimensionless ratios appearing in table \ref{tab:1} that affect the leading-order drag reduction, and their dependence on the dimensional quantities characterising the flow and surfactant properties  and the geometry (outlined in \S\ref{sec:formulation}).}
    \label{tab:2}
\end{table}

We have investigated the leading-order drag reduction (${DR}_0$) across the parameter space, varying the strength of surface advection ($\beta$), bulk ($\alpha$) and surface ($\delta$) diffusion, Marangoni effects ($\gamma$) and exchange between the bulk and the interface  ($\nu$), compared to bulk advection. 
When exchange of surfactant is strong, the bulk and interfacial concentration are in equilibrium at leading-order.
We derived and solved a composite equation \eqref{eq:composite_equation} that includes shear dispersion to qualitatively highlight where 3D effects become important (region G in figure \ref{fig:map_1}\textit{a}).
When exchange of surfactant is moderate, the bulk and interfacial concentrations decouple.
We have identified three primary regions of the parameter space in both the strong- and moderate-exchange problems (figures \ref{fig:map_1} and \ref{fig:map_3}). 
In the Marangoni-dominated (M) regime, the interface is immobilised and the drag reduction vanishes to leading-order (low drag reduction regime, ${DR}_0\ll1$).
The interfacial surfactant distribution is linear with a shallow gradient. 
In the advection (A) and diffusion-dominated (D) regimes, the interface is shear-free and the drag reduction is unaffected by the surfactant at leading order (high drag reduction regime, $1-{DR}_0\ll1$).
The interfacial surfactant distribution can be non-uniform in A, near the AM boundary, where we find both exponential and almost piecewise-linear stagnant cap profiles.
To clarify the underlying physics associated with these results, the dependence of the 3D velocity field on surfactant transport at the bulk and interface has been determined in both strong- and moderate-exchange limits (figures \ref{fig:ff_1} and \ref{fig:ff_2}; supplementary movies 1 and 2).

Table \ref{tab:1} summarises asymptotic approximations of the leading-order drag reduction $DR_0$ in regions M, A and D, and the parts of the parameter space that describe them.
Starting in region M, where there is no drag reduction and the interface is immobilised, we present approximations of the drag reduction when bulk diffusion ($\text{M}_\text{D}$), shear dispersion ($\text{M}_\text{G}$) or surface diffusion ($\text{M}_\text{E}$) allow small surface mobilisation.
The drag can be reduced by strengthening diffusion across the DM boundary (when $\min(\gamma/(\alpha\beta),\,\gamma/\alpha) \sim O(1)$), strengthening shear dispersion across the GM boundary (when $\alpha \gamma/\epsilon^2 \sim O(1)$) or by reducing the surfactant strength relative to advection across the AM boundary (when $\min(\gamma/\beta,\,\gamma) \sim O(1)$).
The quantity  $\gamma/\alpha$, which affects $DR_0$ in regions M and D, is identified in \citet{temprano2023single} as the mobilisation length. 
The quantity $\gamma/\beta$, which affects $DR_0$ in regions A,  is identified in \citet{sundin2022slip} to determine whether the surfactant concentration is in the stagnant cap regime or not.

A number of dimensionless ratios appear in table \ref{tab:1} that increase drag: $\gamma/\alpha$, $\alpha\gamma/\epsilon^2$, $\gamma/\beta$, $\gamma/(\alpha \beta)$ and $\gamma/\delta$. We give these ratios in terms of dimensional parameters in table \ref{tab:2}.
All the ratios given in table \ref{tab:2} have the common factor $\hat{A}\hat{C}\hat{L}_d/\hat{\mu}$, where $\hat{A}$ is the surface activity, $\hat{C}$ is the bulk concentration scale, $\hat{L}_d$ is the depletion length and $\hat{\mu}$ is the dynamic viscosity.
Here, $\hat{C}\hat{L}_d$ measures the level of surfactant adsorbed on the plastron; $\hat{A}\hat{C}\hat{L}_d$ gives the corresponding surface tension reduction, making $\hat{A}\hat{C}\hat{L}_d/\hat{\mu}$ a velocity scale generated by interfacial Marangoni effects.
The factor $\hat{P}_z\hat{D}$, where $\hat{P}_z$ is the transverse pitch and $\hat{D}$ is the bulk diffusivity, decreases the drag across the DM boundary but increases the drag across the GM boundary; this reflects the smoothing effect of diffusion at the DM boundary and shear dispersion at the GM boundary.
When exchange is weak, the surface diffusivity $\hat{D}_I$ instead decreases the drag across the $\text{D}_\text{E}\text{M}_\text{E}$ boundary (figure \ref{fig:diff}).
The velocity flux $\hat{Q}$ decreases the drag across the AM and GM boundaries, quadratically in the latter case (table \ref{tab:2}).
The approximations of the leading-order drag reduction in table \ref{tab:2} can also be divided into those with a linear or quadratic dependence on $\hat{L}_d$. 
Recall from \S\ref{sec:formulation} that $\hat{L}_d/\hat{H}$ is the normalised surfactant depletion length in \S\ref{sec:formulation}, such that for $\hat{L}_d/\hat{H} \ll 1$ surfactant is essentially insoluble, and for $\hat{L}_d/\hat{H} \gg 1$ nearly all surfactant is adsorbed to the interface.
Accordingly, bulk diffusion and shear dispersion reduce drag through parameters that are quadratic in $\hat{L}_d$, requiring solubility for them to be effective.

A number of assumptions have allowed the present model to be derived.
When cross-channel diffusion is weak or the long wave parameter $\epsilon$ is not sufficiently small, the system lies in  region G (identified in \S\ref{sec:results}) where cross-channel concentration gradients become comparable to streamwise variation. In region G, our 1D model breaks down and the full 3D transport equations must be solved to resolve the flow and surfactant fields.
Furthermore, a host of higher-order physical effects associated with flows over SHSs may alter surfactant transport, e.g, interface curvature and the gas subphase, that have been neglected in our model but constitute important extensions \cite[for a detailed discussion of these and other effects, see][]{lee2016superhydrophobic}.
Another application of SHSs is in the thermal management of electronics, where streamwise and spanwise thermocapillary stresses arise due to temperature gradients at the liquid--gas interface \cite[as in considered in][]{kirk2020thermocapillary}.
With minor modifications it is possible that the theory outlined herein could also give insight into these diabatic flows.

To summarise, this paper highlights the range of physical balances that arise when one considers the effect of soluble surfactant in laminar flows bounded by SHSs.
In the appropriate regimes, our results provide a comprehensive analytical framework that can guide the design of surfactant-contaminated SHSs. 
Our simple closed-form theoretical predictions of the drag reduction can help minimize  drag  in realistic 3D SHS microchannels and other applications, where surfactant traces may be naturally present.

\section*{Acknowledgements}

We acknowledge support from CBET--EPSRC (EPSRC Ref. EP/T030739/1, NSF \#2054894), as well as partial support from ARO MURI W911NF-17-1-0306.
For the purpose of open access, the authors have applied a Creative Commons Attribution (CCBY) licence to any Author Accepted Manuscript version arising.
F. T-C. acknowledges support from a distinguished postdoctoral fellowship from the Andlinger Center for Energy and the Environment.

\section*{Declaration of interests}

The authors report no conflict of interest.

\appendix

\section{Numerical methods}\label{app:A}

\subsection{Transport equations} \label{subsec:transport}

We solve the transport equations, (\ref{eq:model1}, \ref{eq:model2}, \ref{eq:composite_equation}), subject to the boundary conditions, \eqref{eq:model3}, using Chebyshev collocation.
An in-depth discussion of the $(N+1)\times(N+1)$ differentiation matrix, $\boldsymbol{D}$, that forms the basis of Chebyshev collocation technique can be found in \citet{trefethen2000spectral}. 
Briefly, it allows us to approximate $f_\xi$ using $\boldsymbol{D} \boldsymbol{f}$, where $\boldsymbol{f} = (f(0),\,..., f(N))^T$ is the solution vector on a grid defined by $\xi(i) = \cos(i \pi/ N)$ for $i=0,\,1,\,...,\,N$.
We map domains $\mathcal{D}_1$, for $x \in [-\phi_x, \,\phi_x]$, and $\mathcal{D}_2$, for $x \in [\phi_x, \,2-\phi_x]$, to a discrete space $\xi(i) \in [-1,\,1]$ for $i=0,\,1,\,...,\,N$. 
We solve the transport equations using differentiation matrices and then map the numerical solution back to physical space. 

The linearised transport equations can be solved analytically to obtain an initial guess (see, e.g., Appendix \ref{subsec:a_weak_surfactant_strength} and \ref{subsec:a_we_sd_d}). We concatenate the solution in both domains as $\boldsymbol{c}_0 = (c_{0,\,1}(0),\,...,\,c_{0,\,1}(N),\,c_{0,\,2}(0),\,...,\,c_{0,\,2}(N))^T$, where $\boldsymbol{c}_{0,\,i} = (c_{0,\,i}(0),\,...,\,c_{0,\,i}(N))^T$ is the  solution in $\mathcal{D}_i$ for $i=1,\,2$.
The initial guess evolves to the nonlinear state by solving
\begin{subequations}
\begin{align}
    &\mathsfbi{L}_1(\boldsymbol{c}_{0}^{\text{old}}) \boldsymbol{c}_{0,\,1}(1:N-1) = \boldsymbol{1} \quad \text{where} \quad \boldsymbol{c}_{0,\,1}(0) = \boldsymbol{c}_{0,\,2}(N), \\ 
    &\mathsfbi{L}_2(\boldsymbol{c}_{0}^{\text{old}}) \boldsymbol{c}_{0,\,2}(1:N-1) = \boldsymbol{1} \quad \text{where} \quad \boldsymbol{c}_{0,\,2}(0) = \boldsymbol{c}_{0,\,1}(N),
\end{align}
\end{subequations}
where $\mathsfbi{L}_i$ are discrete approximations to (\ref{eq:composite_equation}, \ref{eq:model1}, \ref{eq:model2}) in $\mathcal{D}_i$ for $i=1,\,2$. 
We update the solution via $\boldsymbol{c}_{0}^{\text{new}} = \boldsymbol{c}_{0}^{\text{old}} + r(\boldsymbol{c}_{0} - \boldsymbol{c}_{0}^{\text{old}})$ where $r$ is a relaxation factor, until $\tau_i = \|\mathsfbi{L}_i(\boldsymbol{c}_{0,\,i}^{\text{new}} - \boldsymbol{c}_{0,\,i}^{\text{old}})\|_\infty/r$ falls below a specified tolerance for $i=1,\,2$. 

\subsection{Leading-order flow field} \label{subsec:flow_field}

Problems (\ref{eq:tildeu_bvp}, \ref{eq:baru_bvp}) were solved by modifying the framework introduced in \citet{game2017physical}. 
We consider the domain $\mathcal{A}_s = \{y\in[0,\,1]\}\times\{z\in[0,\,P_z]\}$, constructing the rest of the solution using symmetry arguments.
The domain is decomposed into two parts, separated by $z=\phi_z$, where continuity of the variable and its first derivative is enforced. 
The domains are transformed to facilitate Chebyshev collocation discretisations and the PDEs are transformed to discrete space using techniques outlined in \citet{trefethen2000spectral}.
The discontinuous boundary conditions at $z=\phi_z$, (\ref{eq:tildeu_bvp}\textit{b--e})--(\ref{eq:baru_bvp}\textit{b--e}), introduce integrable stress singularities into the problem.
The leading-order contributions of these singularities are subtracted to produce less singular problems. 
The unknown strengths of the singularities are determined by imposing a regularity condition, as follows.

Introducing a local polar coordinate system centred at the contact line $(r,\,\theta)=((z-\phi_z)^2 + y^2)^{1/2},\,\text{tan}^{-1}(y,\,z-\phi_z))$ and assuming $r \ll 1$, (\ref{eq:tildeu_bvp}\textit{a--c}, \ref{eq:baru_bvp}\textit{a--c}) can be used to evaluate the singular part of the solution ($U^s$) via
\refstepcounter{equation} \label{eq:sing_eq}
\begin{equation}
    U^s_{rr} + \frac{U^s_{r}}{r} + \frac{U^s_{\theta\theta}}{r^2} = b, \quad \text{such that} \quad U^s(r,\,0) = 0, \quad U^s_\theta(r,\,\pi) = d, 
    \tag{\theequation\textit{a--c}}
\end{equation}
where $b$ are $d$ are constants that can be chosen to construct (\ref{eq:tildeu_bvp}, \ref{eq:baru_bvp}). 
The solution must remain bounded as $r \rightarrow 0$ and \eqref{eq:sing_eq} can be solved to give
\begin{equation}  \label{eq:sing_sol}
    U^s = B_1^s U_1^s + O(r^{3/2}) = B_1^s r^{1/2} \sin(\theta/2) + O(r^{3/2}),
\end{equation}
where $B_1^s$ is an singularity strength that must be evaluated as part of the solution.
We substitute $U = B_1^s U_1^s + U^r$ into (\ref{eq:tildeu_bvp}\textit{a}, \ref{eq:baru_bvp}\textit{a}) where $U^r$ is the residual solution with the singularity in \eqref{eq:sing_sol} removed, to get
\begin{equation} \label{eq:singrem_eq}
    B_1^s \nabla_\perp^2 U_1^s + \nabla_\perp^2 U^r = b.
\end{equation}
The system is completed with a condition which requires regularity in first derivatives,
\begin{equation} \label{eq:singrem_reg}
    U^r_z(0, \, \phi_z) \quad \text{is constant},
\end{equation}
where this constant can be set to zero arbitrarily \citep{game2017physical}.
Equations (\ref{eq:singrem_eq}, \ref{eq:singrem_reg}) are combined into a matrix problem which determines the singularity strength as part of the solution.
The numerical convergence is improved significantly when compared to a single-domain solution without singularity removal. 

\section{Strong exchange and moderate cross-channel diffusion}

\subsection{Derivation of shear dispersion terms} \label{app:B}

We assume here that $\Pen^{-1} = \Pen_I^{-1} = \Bi = O(\epsilon^{2})$ for $\epsilon \ll 1$ and rescale $\Pen^{-1} =\epsilon^2 \mathscr{P}^{-1}$, $\Pen^{-1}_I =\epsilon^2 \mathscr{P}_I^{-1}$ and $\Bi =\epsilon^2 \mathscr{B}$, such that cross-channel diffusion is weak and exchange is strong, with $\Da = O(1)$. 
This scaling is chosen so that we can investigate the transition from weak to strong cross-diffusion problems, i.e. for moderate diffusion, where we first deviate from a well-mixed bulk surfactant concentration.
We substitute \eqref{eq:asymptotic_expansion} into \eqref{eq:nondimensional_equations}--\eqref{eq:nondimensional_drag} and take the $O(1)$ system. 
In domains $\mathcal{D}_1$ and $\mathcal{D}_2$,
\refstepcounter{equation} \label{eq:weak_diff_leading_order}
\begin{equation}
	\boldsymbol{\nabla}_{\perp} \cdot \boldsymbol{u}_{\perp 0} = 0, \quad \nabla^2_{\perp} \boldsymbol{u}_0 - \boldsymbol{\nabla} p_{0} = 0, \quad \boldsymbol{u}_{\perp 0} \cdot \boldsymbol{\nabla}_{\perp} c_0 =0,
    \tag{\theequation\textit{a--c}}
\end{equation}
on the interface $\mathcal{I}$,
\refstepcounter{equation} \label{eq:weak_diff_leading_order_interface_bcs}
\begin{multline}
	\boldsymbol{n}\cdot \boldsymbol{\nabla} u_0 - \Ma \Gamma_{0x} = 0, \quad \boldsymbol{n}\cdot \boldsymbol{\nabla} w_0 - \Ma \Gamma_{0z} = 0, \quad v_0 = 0, \\ \boldsymbol{n}\cdot \boldsymbol{\nabla} c_0 - \Da(c_0 - \Gamma_0) =0, \quad (w_0 \Gamma_0)_{z} = 0,
    \tag{\theequation\textit{a--e}}
\end{multline}
on the interface contour $\partial \mathcal{I}$,
\begin{equation}
\label{eq:weak_diff_leading_order_interface_boundary_bcs} 
	w_0\Gamma_0 = 0 \quad \text{at} \quad z = \pm \phi_z P_z,
\end{equation}
between domains $\mathcal{D}_1$ and $\mathcal{D}_2$ we have \eqref{eq:leading_order_periodicity}, on the ridge $\mathcal{R}$ and solid $\mathcal{S}$ we have \eqref{eq:leading_order_solid_bcs}, and at the ends of the transverse period we have \eqref{eq:leading_order_transverse_symmetry_bcs}.
The total fluid flux is given by \eqref{eq:leading_order_fluid_flux} and total surfactant flux is given by
\begin{equation} \label{eq:weak_diff_leading_order_surfactant_flux}
    \int_{z=- P_z}^{P_z} \int_{y=0}^{2} u_0 c_0 \, \text{d}y \, \text{d}z + \frac{2 \Da}{\mathscr{B} \mathscr{P}}\int_{z=-P_z}^{P_z} u_0 \Gamma_0 \, \text{d}z = 1,
\end{equation}
and the drag reduction becomes \eqref{eq:leading_order_drag}.

Similar to \S\ref{subsec:strong_diffusion}, from \eqref{eq:weak_diff_leading_order}--\eqref{eq:weak_diff_leading_order_surfactant_flux}, \eqref{eq:leading_order_periodicity}--\eqref{eq:leading_order_fluid_flux}, \eqref{eq:leading_order_solid_bcs}--\eqref{eq:leading_order_transverse_symmetry_bcs} and \eqref{eq:weak_diff_leading_order_surfactant_flux}, we have that $\Gamma_0 = \Gamma_0(x)$ where $\Gamma_0 = c_0(x,\,0,\,z) = c_0(x,\,2,\,z)$, $p_0 = p_0(x)$, and $v_0=w_0=0$. 
The streamwise velocity field is given by $u_0 = \tilde{U} p_{0x} + \Ma \bar{U} \Gamma_{0x}$ in $\mathcal{D}_1$ and $u_0 = \breve{U} p_{0x}$ in $\mathcal{D}_2$, where $\tilde{U}$, $\bar{U}$ and $\breve{U}$ are given by \eqref{eq:tildeu_bvp}--\eqref{eq:hatu_bvp}. 
Substituting $u_0$ into \eqref{eq:leading_order_fluid_flux}, we recover the velocity flux constraints satisfying \eqref{eq:bulk_flux_constraints}, where $\tilde{Q}$, $\bar{Q}$, $\tilde{q}$ and $\bar{q}$ are defined in (\ref{eq:bulkfluxes}, \ref{eq:surfacefluxes}).  

We use the $O(\epsilon^2)$ bulk surfactant equation to determine $c_0$.
In $\mathcal{D}_1$ and $\mathcal{D}_2$, 
\begin{equation} 
\label{eq:weak_diff_first_order}
	\mathscr{P}^{-1} \nabla^2_{\perp} c_0 - u_0 c_{0x} - \boldsymbol{u}_{\perp 1} \cdot \boldsymbol{\nabla}_{\perp} c_0 = 0.
\end{equation}
The third term in \eqref{eq:weak_diff_first_order} involves the first-order velocity components $v_1$ and $w_1$. 
However, we will shortly assume that cross-channel gradients are small, which means that $v_1$ and $w_1$ are not required to evaluate $c_0$.

Decomposing the bulk surfactant field into a cross-channel average and residual component, we write 
\begin{equation} \label{eq:bulk_decomp}
    c_0(x,\,y,\,z) = \langle c_0 \rangle(x) + c'_0(x,\,y,\,z),
\end{equation}
where $\langle c'_0 \rangle \equiv 0$. 
We can then evaluate the shear-dispersion contributions
\refstepcounter{equation} \label{eq:bulk_fluxes}
\begin{equation}
    \langle \tilde{U} c_0 \rangle = \tilde{Q} \langle c_0 \rangle + \langle \tilde{U} c'_0 \rangle, \quad \langle \bar{U} c_0 \rangle = \bar{Q} \langle c_0 \rangle + \langle \bar{U} c'_0 \rangle, \quad \langle \breve{U} c_0\rangle =\breve{Q} \langle c_0\rangle + \langle \breve{U} c'_0 \rangle.
\tag{\theequation\textit{a--c}}
\end{equation}
Substituting \eqref{eq:bulk_fluxes} into \eqref{eq:weak_diff_leading_order_surfactant_flux} and using \eqref{eq:u_def}, the surfactant flux constraints become
\begin{subequations} \label{eq:weak_diff_total_flux}
    \begin{align}
        \left(\frac{2 \Da}{\mathscr{B} \mathscr{P}}  \tilde{q} \Gamma_0 + \tilde{Q} \langle c_0 \rangle + 4 P_z \langle \tilde{U} c'_0 \rangle \right) p_{0x} \hspace{4cm} \nonumber \\ + \left(\frac{2 \Da}{\mathscr{B} \mathscr{P}} \bar{q} \Gamma_0 + \bar{Q} \langle c_0 \rangle + 4 P_z \langle \bar{U} c'_0 \rangle \right) \Ma \Gamma_{0x} &= 1 \quad \text{in} \quad \mathcal{D}_1, \\
        \left(\breve{Q} \langle c_0\rangle + 4 P_z \langle \breve{U} c'_0 \rangle \right)p_{0x} &= 1 \quad \text{in} \quad \mathcal{D}_2.
    \end{align}
\end{subequations}
In this approach, we are not accounting for variations in $\Gamma_0$ driven by $c_0'$. 
This will bring a level of refinement that we will address elsewhere.

Substituting \eqref{eq:bulk_decomp} into (\ref{eq:weak_diff_first_order}, \ref{eq:weak_diff_leading_order_interface_bcs}\textit{d}, \ref{eq:leading_order_solid_bcs}\textit{d}, \ref{eq:leading_order_transverse_symmetry_bcs}\textit{d}) and assuming $|c'_0 | \ll \langle c_0 \rangle$, i.e. the magnitude of the cross-channel concentration gradients is small, we have 
\refstepcounter{equation}\label{eq:cdash1}
\begin{multline}
    \mathscr{P}^{-1} \nabla^2_{\perp} c'_0 = u_0 \langle c_{0}\rangle_x, \quad \text{subject to} \quad \mathcal{D}^+ c'_{0}(0,\,z_s)=0, \quad c'_{0y}(0,\,z_{ns})=0, \\ \mathcal{D}^- c'_{0}(2,\,z_s)=0, \ \ c'_{0y}(2,\,z_{ns})=0, \ \ c'_{0z}(y,\,-1)=0, \ \ c'_{0z}(y,\,1)=0 \quad \text{in} \quad \mathcal{D}_1,
    \tag{\theequation\textit{a--g}}
\end{multline}
where $\mathcal{D}^\pm \equiv \pm \partial_y - \Da$ and
\refstepcounter{equation}\label{eq:cdash2}
\begin{multline}
    \mathscr{P}^{-1} \nabla^2_{\perp} c'_0 = u_0 \langle c_{0}\rangle_x, \quad \text{subject to} \quad c'_{0y}(0,\,z)=0, \quad c'_{0y}(2,\,z)=0, \\ c'_{0z}(y,\,-1)=0, \quad c'_{0z}(y,\,1)=0 \quad \text{in} \quad \mathcal{D}_2.
    \tag{\theequation\textit{a--e}}
\end{multline}
Using superposition, we may then write
\refstepcounter{equation} \label{eq:cdash_def}
\begin{equation}
	    c'_0 = \mathscr{P} \tilde{C} p_{0x} \langle c_{0}\rangle_x + \mathscr{P} \Ma \bar{C} \Gamma_{0x} \langle c_{0}\rangle_x \quad \text{in} \quad \mathcal{D}_1, \quad 
	    c'_0 = \mathscr{P} \breve{C} p_{0x} \langle c_{0}\rangle_x \quad \text{in} \quad \mathcal{D}_2.      \tag{\theequation\textit{a,\,b}}
\end{equation}
Substituting \eqref{eq:cdash_def} into (\ref{eq:cdash1}, \ref{eq:cdash2}), 
in order to obtain $c_0'$ we require the solution to the following three boundary-value problems: shear dispersion due to flow driven by a pressure gradient over the plastron,
\refstepcounter{equation} \label{eq:atildec1}
\begin{multline}
    \nabla^2_{\perp} \tilde{C} = \tilde{U}, \quad \text{subject to} \quad \mathcal{D}^+\tilde{C}(0,\,z_s) =0, \quad \tilde{C}_{y}(0,\,z_{ns})=0, \\ \mathcal{D}^-\tilde{C}(2,\,z_s) =0, \quad \tilde{C}_{y}(2,\,z_{ns})=0, \quad \tilde{C}_{z}(y,\,-1)=0, \quad \tilde{C}_{z}(y,\,1)=0;
    \tag{\theequation\textit{a--g}}
\end{multline}
shear dispersion due to flow driven by surface shear,
\refstepcounter{equation} \label{eq:atildec2}
\begin{multline} 
    \nabla^2_{\perp} \bar{C} = \bar{U}, \quad \text{subject to} \quad \mathcal{D}^+\bar{C}(0,\,z_s) =0, \quad \bar{C}_y(0,\,z_{ns})=0, \\ \mathcal{D}^-\bar{C}(2,\,z_s) =0, \quad \bar{C}_y(2,\,z_{ns})=0, \quad \bar{C}_z(y,\,-1)=0, \quad \bar{C}_z(y,\,1)=0;
    \tag{\theequation\textit{a--g}}
\end{multline}
and dispersion due to the flow in $\mathcal{D}_2$,
\refstepcounter{equation} \label{eq:ahatc}
\begin{equation}
    \breve{C}_{yy} = \breve{U}, \quad \text{subject to} \quad \breve{C}_{y}(0)=0, \quad \int_{y=0}^2 \breve{C} \text{d}y =0.
    \tag{\theequation\textit{a--c}}
\end{equation}
The solutions to (\ref{eq:atildec1},\,\ref{eq:atildec2}) are found numerically, following the procedure described in Appendix \ref{app:A}. 
Equation \eqref{eq:ahatc} can be integrated directly to give $\breve{C} = 1/5 - y^3/6 + y^4/24$.
The shear dispersion contributions become
\refstepcounter{equation}
\label{eq:new_flux_contributions}
\begin{multline} 
    \langle \tilde{U} c'_0 \rangle = \mathscr{P} \langle \tilde{U} \tilde{C} \rangle p_{0x} \langle c_{0}\rangle_x  + \mathscr{P} \langle \tilde{U} \bar{C} \rangle \Ma \langle c_{0}\rangle_x^2, \\ 
    \langle \bar{U} c'_0 \rangle = \mathscr{P} \langle \bar{U} \tilde{C} \rangle p_{0x} \langle c_{0}\rangle_x  + \mathscr{P} \langle \bar{U} \bar{C} \rangle \Ma \langle c_{0}\rangle_x^2, \quad \langle \breve{U} c'_0 \rangle = \mathscr{P} \langle \breve{U} \breve{C} \rangle p_{0x} \langle c_{0}\rangle_x. \tag{\theequation\textit{a--c}}
\end{multline}

\begin{figure}
    \centering
    (\textit{a}) \hfill (\textit{b}) \hfill (\textit{c}) \hfill \hfill \hfill \\
    \hfill \includegraphics[width=.32\textwidth]{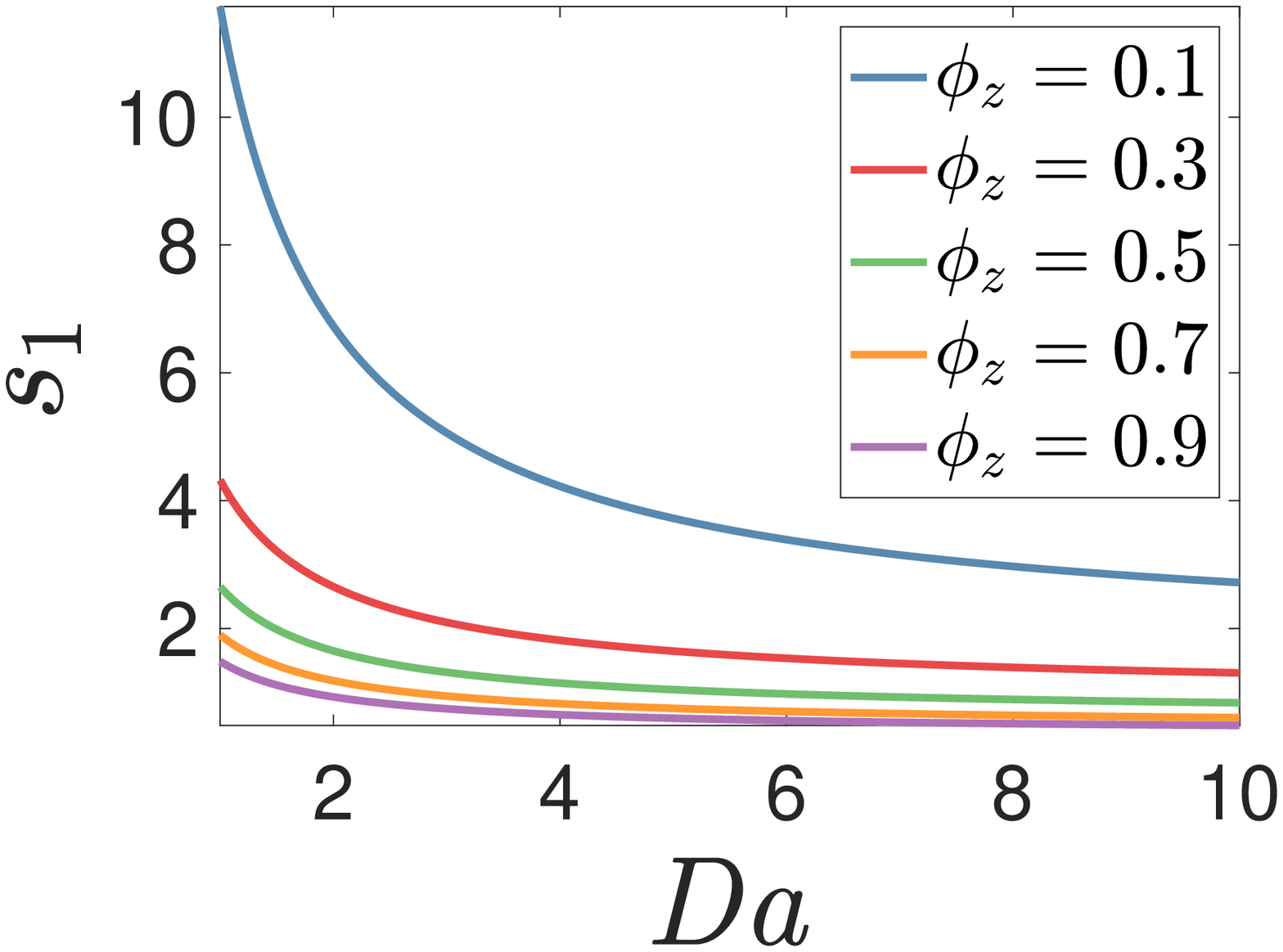} \hfill
    \includegraphics[width=.32\textwidth]{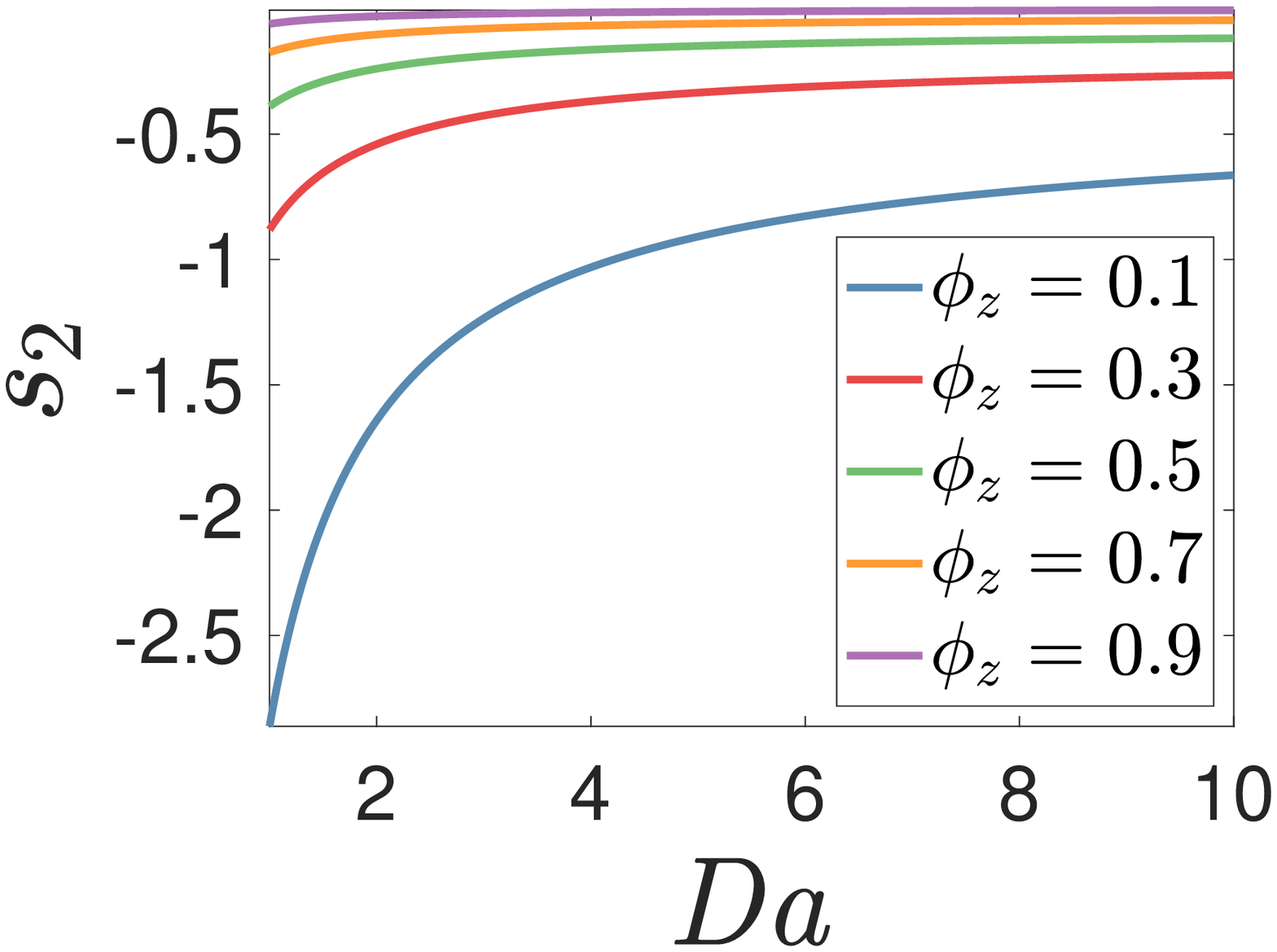} \hfill
    \includegraphics[width=.32\textwidth]{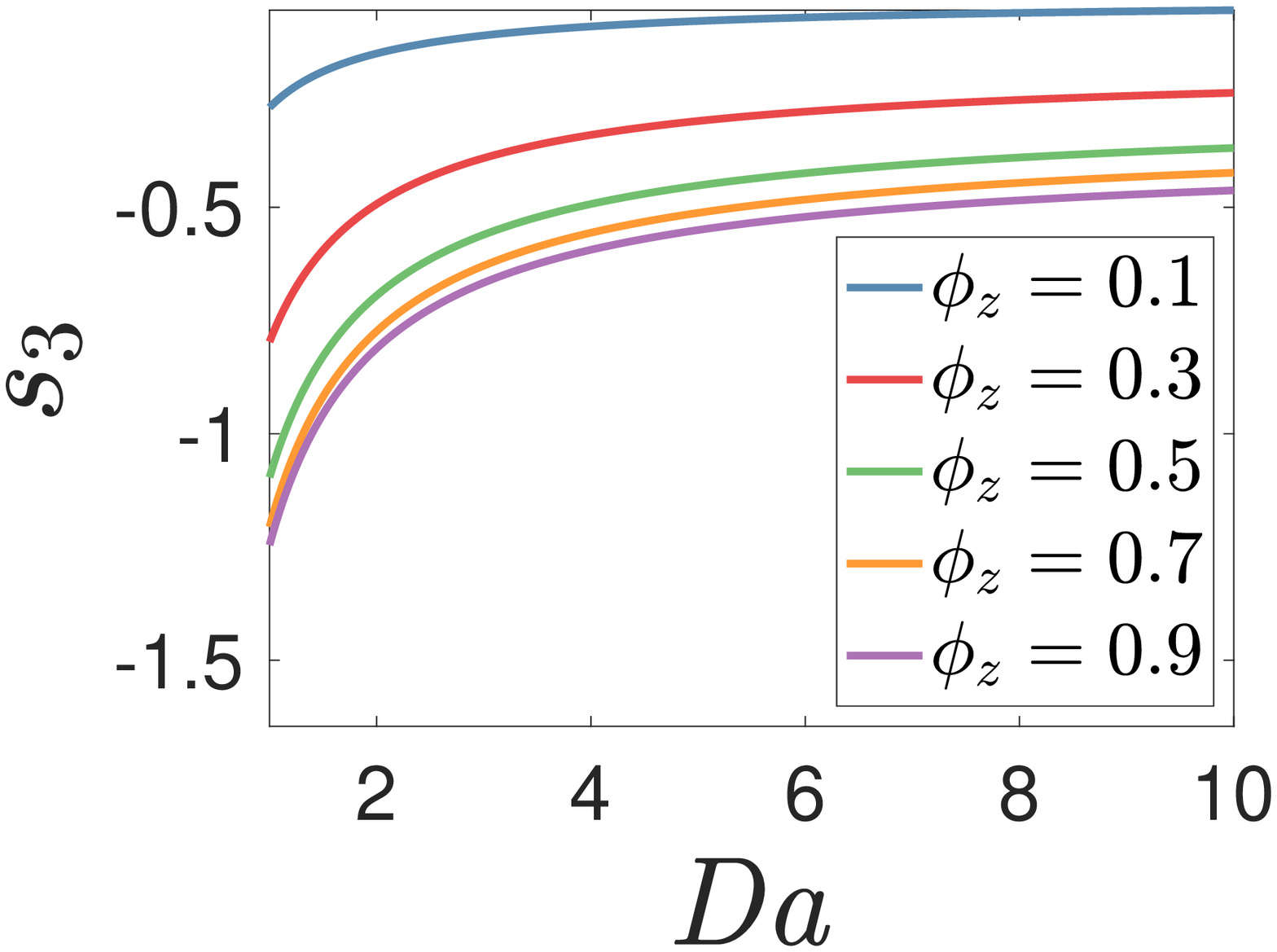} 
    \caption{Plots of (\textit{a}) $s_1$ defined in (\ref{eq:s}\textit{a}), (\textit{b}) $s_2$ defined in (\ref{eq:s}\textit{b}) and (\textit{c}) $s_3$ defined in (\ref{eq:s}\textit{c}), the coefficients multiplying the shear dispersion terms in the moderate cross-channel diffusion and strong-exchange problem, for varying $\phi_z$ and $\Da$ with $P_z = 1$.
    }
    \label{fig:s_i}
\end{figure}

Substituting \eqref{eq:new_flux_contributions} into \eqref{eq:weak_diff_total_flux} with $p_{0x}$ from (\ref{eq:bulk_flux_constraints}\textit{a}) in $\mathcal{D}_1$ and (\ref{eq:bulk_flux_constraints}\textit{c}) in $\mathcal{D}_2$, yields the steady integrated surfactant transport equations and boundary conditions (\ref{eq:model3}\textit{a}, \ref{eq:model3}\textit{b}, \ref{eq:weak_diffusion}), with $\langle c_0 \rangle$ replaced by $c_0$ and the coefficients $s_1$, $s_2$, $s_3$ and $s_4$ defined by \eqref{eq:s}. 
The coefficients $s_1$, $s_2$ and $s_3$ are plotted as functions of $\phi_z$ and $\Da$ in figure \ref{fig:s_i}; the coefficient $s_4 = 3/35$ for all $\phi_z$ and $\Da$, where $P_z = 1$.  

\subsection{Validity of the shear dispersion approximation}
\label{app:D}

When bulk diffusion is not sufficiently strong, concentration gradients normal to the SHS become comparable to the streamwise variation, the long-wave theory outlined in \S\ref{sec:model} breaks down, and 3D numerical simulation of the governing equations is required in order to resolve the coupled flow and surfactant fields.
Here, we compare the size of the cross-channel-averaged surfactant concentration $\langle c_0\rangle$ to the cross-channel variation $c_0'$, introduced in Appendix \ref{app:B}, assuming that gradients normal to the SHS become important when $c_0' = O(\langle c_0 \rangle)$.
This analysis gives an approximate boundary for the range of validity of our model. 

We assume throughout this section that $\beta = O(1)$ (the expressions for $\beta\ll1$ and $\beta\gg1$ can be derived using similar arguments as discussed below). 
To calculate the boundary between regions M and G, expand the cross-channel-averaged surfactant concentration as follows (see Appendix \ref{subsec:a_strong_surfactant_strength})
\begin{equation} \label{eq:exp1}
    \langle c_0\rangle = 1 + \frac{\beta}{\gamma}\left(x - \frac{\phi_x(E+1)}{(E-1)}\right) + ..., \quad \text{for} \quad \gamma \gg \max\left(1, \, \alpha, \, \delta, \, \frac{\epsilon^2}{\alpha}\right).
\end{equation} 
Substituting \eqref{eq:exp1} into \eqref{eq:cdash_def}, using $p_{0x} = 1/\tilde{Q} - \Ma \bar{Q} \Gamma_{0x}/\tilde{Q}$ from (\ref{eq:bulk_flux_constraints}\textit{a}) and noting that $\Gamma_{0x}=\langle c_0\rangle_x=O(1/\gamma)$ in the strong-exchange regime, we find that 
$c_0'= O(\mathscr{P}/ \gamma))$ because $\Ma = O(\gamma)$.
As $\mathscr{P} = O(\epsilon^2/\alpha)$, we have $c_0'= O(\epsilon^2/(\alpha \gamma))$, so that $c_0' = O(\langle c_0 \rangle)=O(1)$ when $\gamma = O \left(\epsilon^2/\alpha\right)$ in region M. 
Hence, $\gamma = O \left(\epsilon^2/\alpha\right)$ defines the boundary of validity of region M near the region G, where shear dispersion terms become important (see figure~\ref{fig:map_1}).

Similarly, to calculate the boundary between regions A and G, expand as follows (see Appendix \ref{subsec:a_weak_surfactant_strength}):
\begin{multline}
\label{eq:exp3}
    \langle c_0\rangle = \frac{1}{\beta +1}  + \frac{\beta}{(\beta+1)}\exp\left(\frac{(1+\beta)(x-\phi_x)}{\alpha +\delta + \epsilon^2 s_1 / \alpha }\right) + ..., \\ \text{for} \quad \epsilon^2 \ll \alpha \ll 1, \quad \gamma \ll \min(1,\,\beta).
\end{multline} 
Substituting \eqref{eq:exp3} into \eqref{eq:cdash_def} and using $p_{0x} = 1/\tilde{Q} - \Ma \bar{Q} \Gamma_{0x}/\tilde{Q}$, we find that $c_0'= O(\mathscr{P})$.
As $\mathscr{P}=O(\epsilon^2 / \alpha)$, we have that $c_0' = O(\langle c_0 \rangle)$ when $\alpha = O(\epsilon^2)$ in region A.

\section{Asymptotic solutions for strong exchange}\label{app:C}

\subsection{Strong Marangoni effect: region M} \label{subsec:a_strong_surfactant_strength}

Assuming that $\beta=O(1)$ and $\gamma \gg \max(1,\,\alpha,\,\delta,\,\epsilon^2/\alpha)$, but retaining the effects of diffusion and shear dispersion, we expand the concentration field from \S\ref{subsec:strong_exchange} using $c_0 = c_{00} + c_{01}/\gamma + c_{02}/\gamma^2 + ...$
in (\ref{eq:model3}\textit{a}, \ref{eq:model3}\textit{b}, \ref{eq:composite_equation}).
At $O(\gamma^2)$, we have
\refstepcounter{equation} \label{eq:marangoni_limited_leading_order}
\begin{multline} 
    c_{00x}^3 = 0 \quad \text{in} \quad \mathcal{D}_1, \quad c_{00} - \alpha c_{00x} - \frac{\epsilon^2 s_4}{\alpha} c_{00x} = 1 \quad \text{in} \quad \mathcal{D}_2, \\
    \text{subject to} \quad
    c_{00}(\phi_x^{-}) = c_{00}(\phi_x^{+}), \quad c_{00}(-\phi_x) = c_{00}(2 - \phi_x).
    \tag{\theequation\textit{a--d}}
\end{multline}
The system in \eqref{eq:marangoni_limited_leading_order} requires $c_{00}=1$ in $\mathcal{D}_1\cup\mathcal{D}_2$.
At $O(1)$ and $O(\gamma)$, we have
\refstepcounter{equation} \label{eq:marangoni_limited_first_order}
\begin{multline}
     c_{01x} = \beta \quad \text{in} \quad \mathcal{D}_1, \quad c_{01} - \alpha c_{01x} - \frac{\epsilon^2 s_4}{\alpha} c_{01x} = 0 \quad \text{in} \quad \mathcal{D}_2,
     \\ \text{subject to} \quad
    c_{01}(\phi_x^{-}) = c_{01}(\phi_x^{+}), \quad c_{01}(-\phi_x) = c_{01}(2 - \phi_x).
    \tag{\theequation\textit{a--d}}
\end{multline}
We can integrate \eqref{eq:marangoni_limited_first_order} to show that $c_{01} = \beta (x - \phi_x (E + 1)/(E - 1))$ in $\mathcal{D}_1$, where $E \equiv \exp(2\alpha(1-\phi_x)/(\alpha^2 + \epsilon^2 s_4))$, giving \eqref{eq:zeta_gg_1}.
At $O(\gamma^{-1})$, we have that 
\begin{equation} \label{eq:marangoni_limited_second_order}
    c_{02x} = (\beta+1) c_{01} - c_{01} c_{01x} - (\alpha + \delta) c_{01x} - \frac{\epsilon^2}{\alpha}\left(s_1 c_{01x} + \frac{s_2}{\beta} c_{01x}^2 + \frac{s_3}{\beta^2} c_{01x}^3\right) \ \text{in} \ \mathcal{D}_1.
\end{equation}
From (\ref{eq:marangoni_limited_first_order}) and (\ref{eq:marangoni_limited_second_order}) we can compute the concentration increase over $\mathcal{D}_1$, 
\begin{equation} \label{marangoni_limited_concentration_gradient}
    \Delta c_0 = \frac{2 \phi_x \beta}{\gamma} + \frac{2 \phi_{x} \beta }{\alpha\gamma^2\left(1 - E\right)}\big(\alpha(\phi_x - \alpha - \delta) - \epsilon^2 s +(\alpha(\phi_x +\alpha+\delta) + \epsilon^2 s)E\big),
\end{equation}
where $s \equiv s_1 + s_2 + s_3 > 0$, which is substituted into \eqref{eq:dr_def} to give \eqref{eq:drs_m}. 
The corrections to $c_0\approx c_{00}=1$ and ${DR}_0=0$ are small provided $\gamma \gg \max(1,\,\alpha,\,\delta,\,\epsilon^2/\alpha)$, defining the boundaries of the asymptotic region M.

\subsection{Strong Marangoni effect and strong shear dispersion: the GM boundary} 
\label{subsec:a_weak_diff_and_surf}

At the GM boundary, assume that $\beta = O(1)$ and $\gamma \gg 1$.
Rescale $\epsilon^2/\alpha = \gamma/ a$ where $a=O(1)$. 
Expand the concentration field from \S\ref{subsec:strong_exchange}, using $c_0 = c_{00} + c_{01}/\gamma + ...$ in (\ref{eq:model3}\textit{a}, \ref{eq:model3}\textit{b}, \ref{eq:composite_equation}).
At $O(\gamma^3)$, shear dispersion dominates with
\refstepcounter{equation}
\label{eq:combined_s_limited_leading_order}
\begin{equation} 
     c_{00x}^3 = 0 \quad \text{in} \quad \mathcal{D}_1, \quad c_{00x} = 0 \quad \text{in} \quad \mathcal{D}_2,
    \tag{\theequation\textit{a,\,b}}
\end{equation}
subject to \eqref{eq:marangoni_limited_leading_order}(\textit{c,\,d}).
In order to calculate $c_{00}$ we must proceed to the next order.
At $O(1)$--$O(\gamma^2)$, Marangoni effects and advection enter
\refstepcounter{equation} 
\label{eq:combined_s_limited_first_order}
\begin{multline} 
     (\beta + 1) c_{00} - c_{00} c_{01x} - \frac{1}{a}\left( s_1 c_{01x} + \frac{s_2}{\beta}c_{01x}^2 + \frac{s_3}{\beta^2} c_{01x}^3\right) = 1 \quad \text{in} \quad \mathcal{D}_1, \\ c_{00} - \frac{s_4}{a} c_{01x} = 0 \quad \text{in} \quad \mathcal{D}_2,
    \tag{\theequation\textit{a,\,b}}
\end{multline}
subject to \eqref{eq:marangoni_limited_first_order}(\textit{c,\,d}). 
As $c_{01x} = \beta$ in M (see Appendix \ref{subsec:a_strong_surfactant_strength}), we expect $s_2 c_{01x}^2/(a\beta) \approx s_2 c_{01x}/a$ and $s_3 c_{01x}^3/(a\beta^2) \approx s_3 c_{01x}/a$ as we approach the GM boundary, such that we instead solve the linearised problem
\refstepcounter{equation}
\label{eq:combined_s_limited_first_order_2}
\begin{equation} 
     (\beta + 1) c_{00} - c_{00} c_{01x} + \frac{s}{a} c_{01x} \approx 1 \quad \text{in} \quad \mathcal{D}_1, \quad c_{00} - \frac{s_4}{a} c_{01x} = 1 \quad \text{in} \quad \mathcal{D}_2.
    \tag{\theequation\textit{a,\,b}}
\end{equation}
Integrating \eqref{eq:combined_s_limited_first_order_2} over the period gives $c_{00}$ as the solution to
\begin{equation} \label{combined_s_concentration_gradient}
    (c_{00} - 1)(\phi_x - 1)(s  - a c_{00}) = s_4 \phi_x(1 - c_{00}(\beta + 1)).
\end{equation}
By solving \eqref{combined_s_concentration_gradient}, we can then integrate (\ref{eq:combined_s_limited_first_order_2}\textit{a}) in $\mathcal{D}_1$ to find $\Delta c_0$, which is substituted into \eqref{eq:dr_def} to give \eqref{eq:smboundary}.

\subsection{Strong diffusion and strong Marangoni effect: the DM boundary} \label{subsec:a_strong_diff_and_surf}

At the DM boundary, assume that $\beta=O(1)$ and $\gamma \gg 1$.
Rescale $\alpha = a \gamma$ and $\delta = d \gamma$ where $a \sim d \sim O(1)$.
Expand the concentration field using $c_0 = c_{00} + c_{01}/\gamma + ...$ in (\ref{eq:model3}\textit{a}, \ref{eq:model3}\textit{b}, \ref{eq:composite_equation}).
At $O(\gamma)$, Marangoni effects, diffusion and shear dispersion dominate with
\refstepcounter{equation} 
\label{eq:combined_d_limited_leading_order}
\begin{equation} 
     -c_{00} c_{00x} - (a + d)c_{00x} - \frac{\epsilon^2 s_3}{a \beta^2} c_{00x}^3 = 0 \quad \text{in} \quad \mathcal{D}_1, \quad  c_{00x} = 0 \quad \text{in} \quad \mathcal{D}_2,
    \tag{\theequation\textit{a,\,b}}
\end{equation}
subject to \eqref{eq:marangoni_limited_leading_order}(\textit{c},\,\textit{d}). 
Hence, $c_{00x}=0$ in both $\mathcal{D}_1$ and $\mathcal{D}_2$.
In order to calculate $c_{00}$ we must proceed to the next order.
At $O(1)$, the advection terms enter
\refstepcounter{equation} 
\label{eq:combined_d_limited_first_order}
\begin{equation} 
     (\beta + 1) c_{00} - c_{00} c_{01x} - (a + d)c_{01x} = 1 \quad \text{in} \quad \mathcal{D}_1, \quad c_{00} - a c_{01x} = 1 \quad \text{in} \quad \mathcal{D}_2,
    \tag{\theequation\textit{a,\,b}}
\end{equation}
subject to \eqref{eq:marangoni_limited_first_order}(\textit{c},\,\textit{d}). 
Integrating \eqref{eq:combined_d_limited_first_order} over $\mathcal{D}_1$ gives $c_{00}$ as the solution to
\begin{equation} \label{combined_d_concentration_gradient}
    (c_{00} - 1)(\phi_x - 1)(a + d + c_{00}) = a\phi_x(c_{00}(\beta + 1) - 1).
\end{equation}
By solving \eqref{combined_d_concentration_gradient}, we can then integrate (\ref{eq:combined_d_limited_first_order}\textit{a}) in $\mathcal{D}_1$ to find $\Delta c_0$, which is substituted into \eqref{eq:dr_def} to give \eqref{eq:mdboundary}.

\subsection{Strong diffusion: region D} \label{subsec:a_strong_diffusion}

Assume that $\beta = O(1)$ and $\min(\alpha,\,\delta)\gg\max(1,\,\gamma)$. 
Let $\delta = d \alpha$, where $d = O(1)$. 
Expand the concentration field from \S\ref{subsec:strong_exchange} using $c_0 = c_{00} + c_{01}/\alpha + ...$ in (\ref{eq:model3}\textit{a}, \ref{eq:model3}\textit{b}, \ref{eq:composite_equation}).
At $O(\alpha)$, diffusion dominates with 
\refstepcounter{equation} \label{eq:diffusion_limited_leading_order}
\begin{equation}
    c_{00x} = 0 \quad \text{in} \quad \mathcal{D}_1, \quad c_{00x} = 0 \quad \text{in} \quad \mathcal{D}_2, \tag{\theequation\textit{a,\,b}}
\end{equation}
subject to (\ref{eq:marangoni_limited_leading_order}\textit{c,\,d}).
In order to calculate $c_{00}$ we must proceed to the next order.
At $O(1)$, advection enters
\refstepcounter{equation} \label{eq:diffusion_limited_first_order}
\begin{equation}
    (\beta + 1) c_{00} - (1 + d) c_{01x} = 1 \quad \text{in} \quad \mathcal{D}_1, \quad c_{00} - c_{01x} = 1 \quad \text{in} \quad \mathcal{D}_2, \tag{\theequation\textit{a,\,b}}
\end{equation}
subject to (\ref{eq:marangoni_limited_first_order}\textit{c,\,d}).
Integrating \eqref{eq:diffusion_limited_first_order} over $\mathcal{D}_1 \cup \mathcal{D}_2$ and imposing periodicity, $c_{01x}(-\phi_x) = c_{01}(2-\phi_x)$, $c_{00}$ is given in \eqref{eq:alpha_gg_1}.
Using $c_{01x}$, we can obtain the jump in $c_0$ over $\mathcal{D}_1$,
\begin{equation} \label{diffusion_limited_concentration_gradient}
    \Delta c_0 \approx \frac{2\phi_x}{\alpha}c_{01x} =  \frac{2 \phi_x (1 - \phi_x)\beta}{(1 - \phi_x)(\alpha + \delta) + \phi_x\alpha(\beta + 1)},
\end{equation}
which is substituted into \eqref{eq:dr_def} to give \eqref{eq:drs_d}.
The corrections to $c_0=c_{00}$ and ${DR}_0=1$ are small provided $\min(\alpha,\,\delta) \gg \max(1,\,\gamma,\beta)$, defining the boundaries of the asymptotic region D.

\subsection{Weak Marangoni effect: region A} \label{subsec:a_weak_surfactant_strength}

Assume that $\beta=O(1)$ and $\gamma \ll 1$. Expand the concentration field from \S\ref{subsec:strong_exchange} using $c_0 = c_{00} + \gamma c_{01} + ...$
in (\ref{eq:model3}\textit{a}, \ref{eq:model3}\textit{b}, \ref{eq:composite_equation}).
At $O(1)$, advection, diffusion and shear dispersion dominate with
\refstepcounter{equation} \label{eq:marangoni_weak_first_order}
\begin{equation}
     (\beta+1) c_{00} - \left(\alpha + \delta + \frac{\epsilon^2 s_1}{\alpha}\right) c_{00x} = 1 \ \text{in} \ \mathcal{D}_1, \ c_{00} - \left(\alpha + \frac{\epsilon^2 s_4}{\alpha}\right) c_{00x} = 1 \ \text{in} \ \mathcal{D}_2, \tag{\theequation\textit{a,\,b}}
\end{equation}
subject to \eqref{eq:marangoni_limited_leading_order}(\textit{c},\,\textit{d}). 
Assuming $\epsilon^2 \ll \alpha \ll 1$, we integrate \eqref{eq:marangoni_weak_first_order} to get \eqref{eq:zeta_ll_1}.
In either case, we have that
\begin{equation}
\label{marangoni_weak_concentration_gradient}
    \Delta c_{0} \approx \frac{\beta}{\beta + 1}
\end{equation}
which is substituted into \eqref{eq:dr_def} to give \eqref{eq:drs_a}.
The corrections to $c_0=c_{00}$ and ${DR}_0=1$ are small provided $\epsilon^2 \ll \alpha \ll 1$ and $\gamma \ll 1$, defining the boundaries of the asymptotic region A.
Note that \eqref{eq:marangoni_weak_first_order} is linear and has a general solution that can be used to determine the AD boundary.
However, as this does not change the drag reduction to leading-order (the interface remains shear free for $\gamma \ll \min(1,\, \beta)$), we do not investigate this limit here.

\subsection{Strong advection and strong Marangoni effect: the AM boundary} \label{subsec:a_strong_diff_strong_surfactant_strength}

Assume that $\beta = O(\gamma)$, $\gamma \gg \max(\alpha,\, \delta,\, \epsilon^2/\alpha)$ and $\epsilon^2 \ll \alpha \ll 1$.
Rescale $\alpha = a/\gamma$, $\beta = b\gamma$, $\delta = d/\gamma$ and $\epsilon^2/\alpha = e / \gamma$ where $a$, $b$, $d$ and $e$ are positive $O(1)$ constants.
Expand the concentration field from \S\ref{subsec:strong_exchange}, using $c_0 = c_{00} + c_{01}/\gamma + ...$ in (\ref{eq:model3}\textit{a}, \ref{eq:model3}\textit{b}, \ref{eq:composite_equation}). 
At $O(\gamma)$, Marangoni effects and advection dominate with \refstepcounter{equation}
\label{eq:combined_am_leading_order}
\begin{equation} 
     c_{00}(b - c_{00x}) = 0 \quad \text{in} \quad \mathcal{D}_1, \quad c_{00} = 1 \quad \text{in} \quad \mathcal{D}_2,
    \tag{\theequation\textit{a,\,b}}
\end{equation}
subject to (\ref{eq:marangoni_limited_leading_order}\textit{c,\,d}).
The system in \eqref{eq:combined_am_leading_order} gives a linear profile  $c_{00} = \beta(x - \phi_x)/\gamma +1 $ for $\gamma/\beta > 2\phi_x$ in $\mathcal{D}_1$, such that $c_{00} \geq 0$ for all $x \in [-\phi_x,\, \phi_x]$. 
For $\gamma/\beta > 2\phi_x$, we find that $\Delta c_0 = 2\phi_x\beta/\gamma$ and ${DR}_0=0$ at leading-order, hence we proceed to $O(1)$ where we find that $\Delta c_0 = 2\phi_x\beta/\gamma + 2 \phi_x / \gamma + \ln(1 - 2 \beta \phi_x / \gamma ) / \beta$ provided $1 - 2\phi_x \beta / \gamma \gg \exp{(-\beta)}$.
For $\gamma/\beta \leq 2\phi_x$ a piecewise-linear solution exists with $c_{00} = 0$ for all $-\phi_x\leq x\leq\phi_x-\gamma/\beta$ and $c_{00} = \beta(x - \phi_x)/\gamma +1 $ for all $\phi_x-\gamma/\beta \leq x \leq \phi_x$.
This nonlinear profile typically represents the emergence of the stagnant cap profile \citep{he1991size}, where surfactant is swept to the downstream end of the plastron where it gives rise to a strong gradient.
For $\gamma/\beta \leq 2\phi_x$, we find that $\Delta c_0 = 1$, which is substituted into \eqref{eq:dr_def} to give \eqref{eq:largebetaflat}.

\section{Asymptotic solutions for weak exchange}\label{app:E}

\subsection{Strong Marangoni effect: region $\text{M}_\text{E}$} \label{subsec:a_we_strong_marangoni}

Assume that $\beta = O(1)$, $\gamma \gg \max(1,\,\alpha,\,\delta)$ and $ \nu/\epsilon^2 \ll \min(1,\,\alpha,\,\delta)$. 
Rescale $ \nu/\epsilon^2 = n/\gamma$ where $n = O(1)$.
Expand the concentration field from \S\ref{subsec:weak_exchange} using $\Gamma_0 = \Gamma_{00} + \Gamma_{01}/\gamma + \Gamma_{02}/\gamma^2 + ...$ and $c_0 = c_{00} + c_{01}/\gamma + ...$ into \eqref{eq:model1}--\eqref{eq:model3}.
At $O(\gamma)$, Marangoni effects, advection and diffusion dominate with
\refstepcounter{equation} 
\label{eq:we_marangoni_limited_leading_order}
\begin{multline} 
    (c_{00} - \alpha c_{00x})_x = 0, \quad \Gamma_{00} \Gamma_{00x} = 0 \quad \text{in} \quad \mathcal{D}_1, \quad c_{00} - \alpha c_{00x} = 1 \quad \text{in} \quad \mathcal{D}_2, \\ \text{subject to} \quad c_{00}(\phi_x^{-}) = c_{00}(\phi_x^{+}), \quad c_{00}(-\phi_x) = c_{00}(2 - \phi_x), \\
    c_{00}(\pm \phi_x) - \alpha c_{00x}(\pm \phi_x) = 1, \quad  \Gamma_{00}(\pm \phi_x) \Gamma_{00x}(\pm \phi_x) = 0.
    \tag{\theequation\textit{a--g}}
\end{multline}
The system in \eqref{eq:we_marangoni_limited_leading_order} requires $\Gamma_{00}$ constant on $\mathcal{I}$. 
At $O(1)$, surface advection enters 
\refstepcounter{equation} 
\label{eq:we_marangoni_limited_first_order}
\begin{multline} 
    (c_{01} - \alpha c_{01x})_x = 0, \quad \beta - \Gamma_{01x} = 0 \quad \text{in} \quad \mathcal{D}_1, \quad c_{01} - \alpha c_{01x} = 0 \quad \text{in} \quad \mathcal{D}_2, \\ \text{subject to} \quad c_{01}(\phi_x^{-}) = c_{01}(\phi_x^{+}), \quad c_{01}(-\phi_x) = c_{01}(2 - \phi_x), \\
    c_{01}(\pm \phi_x) - \alpha c_{01x}(\pm \phi_x) = 0, \quad \beta - \Gamma_{01x}(\pm \phi_x) = 0.
    \tag{\theequation\textit{a--g}}
\end{multline}
We can integrate \eqref{eq:we_marangoni_limited_first_order} to show that $\Gamma_{01} = \beta x + C_1$ on $\mathcal{I}$, using the net flux condition $\int_{x=-\phi_x}^{\phi_x} (\Gamma_{01} - c_{01}) \, \text{d} x = 0$ from \eqref{eq:dimensionless_net_flux} to find $C_1=0$, which gives \eqref{eq:we_m}.
At $O(\gamma^{-1})$, we have
\begin{equation}
\label{eq:we_marangoni_limited_second_order}
    \Gamma_{02x} = \beta \Gamma_{01} - \Gamma_{01}\Gamma_{01x} - \delta \Gamma_{01x} + c_{01} - \alpha c_{01x} \quad \text{on} \quad \mathcal{I}.
\end{equation}
Integrating (\ref{eq:we_marangoni_limited_first_order}, \ref{eq:we_marangoni_limited_second_order}) on $\mathcal{I}$, altogether we have that $\Delta \Gamma_0 = 2 \phi_x \beta/\gamma - 2 \phi_{x} \beta \delta/\gamma^2$, which is substituted into \eqref{eq:dr_def} to give \eqref{eq:we_m}.
The corrections to $\Gamma_0=\Gamma_{00}=1$ and ${DR}_0 = 0$ are small provided $\gamma \gg \max(1,\,\alpha,\,\delta)$ and $ \nu/\epsilon^2 \ll \min(1,\,\alpha,\,\delta)$, defining the boundaries of the asymptotic region $\text{M}_\text{E}$.

\subsection{Strong Marangoni effect and strong diffusion: the $\text{M}_\text{E}\text{D}_\text{E}$ boundary} \label{subsec:a_we_strong_marangoni_diffusion}

At the $\text{M}_\text{E}\text{D}_\text{E}$ boundary, assume that $\beta = O(1)$ and $\gamma \gg 1$. 
Rescale $\alpha = a \gamma$, $\delta = d \gamma$ and $ \nu/\epsilon^2 = n / \gamma$ where $a$,  $d$ and $n$ are positive $O(1)$ constants. 
Expand the concentration field from \S\ref{subsec:weak_exchange} using $\Gamma_0 = \Gamma_{00} + \Gamma_{01}/\gamma + ...$ and $c_0 = c_{00} + c_{01}/\gamma ...$ into \eqref{eq:model1}--\eqref{eq:model3}.
At $O(\gamma)$, Marangoni effects and diffusion dominate with
\refstepcounter{equation} 
\label{eq:we_marangoni_limited_leading_order_strong_diff}
\begin{multline} 
    c_{00xx} = 0, \quad a c_{00x} + \Gamma_{00} \Gamma_{00x} + d \Gamma_{00x} = 0 \quad \text{in} \quad \mathcal{D}_1, \quad c_{00x} = 0 \quad \text{in} \quad \mathcal{D}_2, \\ \text{subject to} \quad c_{00}(\phi_x^{-}) = c_{00}(\phi_x^{+}), \quad c_{00}(-\phi_x) = c_{00}(2 - \phi_x), \\
    c_{00x}(\pm \phi_x) = 0, \quad  \Gamma_{00}(\pm \phi_x) \Gamma_{00x}(\pm \phi_x) + d \Gamma_{00x}(\pm \phi_x) = 0.
    \tag{\theequation\textit{a--g}}
\end{multline}
The system in \eqref{eq:we_marangoni_limited_leading_order_strong_diff} requires $\Gamma_{00}$ and $c_{00}$ are constant. 
At $O(1)$, advection appears with 
\refstepcounter{equation} 
\label{eq:we_marangoni_limited_first_order_strong_diff}
\begin{multline} 
    c_{01xx} = 0, \quad c_{00} - a c_{01x} + \beta\Gamma_{00} - (\Gamma_{00}+d) \Gamma_{01x} = 1
    \quad \text{in} \quad \mathcal{D}_1, \quad c_{01x} = 0 \quad \text{in} \quad \mathcal{D}_2, \\ \text{subject to} \quad c_{01}(\phi_x^{-}) = c_{01}(\phi_x^{+}), \quad c_{01}(-\phi_x) = c_{01}(2 - \phi_x), \\
    c_{00} - a c_{01x}(\pm \phi_x) = 1, \quad \beta\Gamma_{00} - (\Gamma_{00}+d) \Gamma_{01x}(\pm \phi_x) = 0.
    \tag{\theequation\textit{a--g}}
\end{multline}
The system in \eqref{eq:we_marangoni_limited_first_order_strong_diff} requires that $c_{00}=1$ and $\Gamma_{01x} = (\beta \Gamma_{00})/(d+\Gamma_{00})$. 
Using information at $O(1/\gamma)$, we can show that $\Gamma_{00}=1$.
We can then evaluate $\Delta \Gamma_0 \approx 2 \phi_x \beta / (\gamma(d+1))$ and substitute it into \eqref{eq:dr_def} to find ${DR}_0 \approx 0.5$ when $\gamma = \delta$, as in \eqref{eq:we_50_1}.

\subsection{Strong Marangoni effect, advection and diffusion: the $\text{M}_\text{E}\text{AD}_\text{E}$ boundary} \label{subsec:a_we_strong_marangoni_o1diff}

At the $\text{M}_\text{E}\text{AD}_\text{E}$ boundary, assume that $\beta = O(1)$ and $\nu/\epsilon^2\ll 1$. 
Expand the concentration field from \S\ref{subsec:weak_exchange} using $\Gamma_0 = \Gamma_{00} + \epsilon^2 \Gamma_{01}/\nu + ...$ and $c_0 = c_{00} + \epsilon^2 c_{01} /\nu ...$ into \eqref{eq:model1}--\eqref{eq:model3}.
At $O(1)$, advection, diffusion and Marangoni effects dominate with
\refstepcounter{equation} 
\label{eq:we_marangoni_limited_leading_order_o1diff}
\begin{multline} 
    (c_{00} - \alpha c_{00x})_x = 0, \quad \beta \Gamma_{00} - \gamma \Gamma_{00} \Gamma_{00x} - \delta \Gamma_{00x} = 0 \quad \text{in} \quad \mathcal{D}_1, \\ c_{00} - \alpha c_{00x} = 1 \quad \text{in} \quad \mathcal{D}_2, \quad \text{subject to} \quad c_{00}(\phi_x^{-}) = c_{00}(\phi_x^{+}), \\ c_{00}(-\phi_x) = c_{00}(2 - \phi_x), \quad
    c_{00}(\pm \phi_x) - \alpha c_{00x}(\pm \phi_x) = 1, 
    \\ \beta\Gamma_{00}(\pm \phi_x) - \gamma \Gamma_{00}(\pm \phi_x) \Gamma_{00x}(\pm \phi_x) - \delta \Gamma_{00x}(\pm \phi_x) = 0.
    \tag{\theequation\textit{a--g}}
\end{multline}
We can integrate \eqref{eq:we_marangoni_limited_leading_order_o1diff} directly over $\mathcal{I}$ using the no-net-flux condition $\int_{x=-\phi_x}^{
\phi_x}\Gamma_{00} \, \text{d} x = 2\phi_x$ as $c_{00} = 1$ to derive
\begin{equation}\label{eq:int_o1diff}
    2\phi_x\beta - \gamma\Delta\Gamma_{00}(\Gamma_{00}(-\phi_x) + \Gamma_{00}(\phi_x))/2 = \delta \Delta \Gamma_{00}. 
\end{equation}
From \eqref{eq:int_o1diff} we can evaluate $\Delta \Gamma_0$ and substitute it into \eqref{eq:dr_def} to get 
\begin{equation}
    {DR}_0 = \frac{\delta + \gamma((\Gamma_{00}(-\phi_x) + \Gamma_{00}(\phi_x))/2 - 1)}{\delta + \gamma(\Gamma_{00}(-\phi_x) + \Gamma_{00}(\phi_x))/2}.
\end{equation}
As long as $(\Gamma_{00}(-\phi_x) + \Gamma_{00}(\phi_x))/2 \approx 1$ (as in figure \ref{fig:map_3}\textit{d}), then ${DR}_0 \approx 1 - \gamma/(\delta + \gamma)$ and ${DR}_0 = 0.5$ when $\gamma = \delta$. 
Note that we recover \eqref{eq:drs_m_we} for $\gamma \gg \delta$ in $\text{M}_\text{E}$ and \eqref{eq:we_dr_d} for $\gamma \ll \delta$ in $\text{A}_\text{E}$, $\text{D}_\text{E}$ and $\text{AD}_\text{E}$.

\subsection{Strong Marangoni effect and strong advection: the $\text{M}_\text{E}\text{A}_\text{E}$ boundary} \label{subsec:a_we_strong_marangoni_weak_diffusion}

At the $\text{M}_\text{E}\text{A}_\text{E}$ boundary, assume that $\beta = O(1)$ and $\alpha \ll 1$. 
Rescale, $\delta = d \alpha$ and $ \nu/\epsilon^2 =n \alpha$ where $d$ and $n$ are $O(1)$ constants. 
Expand the concentration field from \S\ref{subsec:weak_exchange} using $\Gamma_0 = \Gamma_{00} + \alpha \Gamma_{01} + ...$ and $c_0 = c_{00} + \alpha c_{01} ...$ into \eqref{eq:model1}--\eqref{eq:model3}.
At $O(1)$, advection and Marangoni effects dominate with
\refstepcounter{equation} 
\label{eq:we_marangoni_limited_leading_order_weak_diff}
\begin{multline} 
    c_{00x} = 0, \quad \beta \Gamma_{00} - \gamma \Gamma_{00} \Gamma_{00x} = 0 \quad \text{in} \quad \mathcal{D}_1, \quad c_{00} = 1 \quad \text{in} \quad \mathcal{D}_2, \\ \text{subject to} \quad c_{00}(\phi_x^{-}) = c_{00}(\phi_x^{+}), \quad c_{00}(-\phi_x) = c_{00}(2 - \phi_x), \\
    c_{00}(\pm \phi_x) = 1, \quad  \beta\Gamma_{00}(\pm \phi_x) - \gamma \Gamma_{00}(\pm \phi_x) \Gamma_{00x}(\pm \phi_x) = 0.
    \tag{\theequation\textit{a--g}}
\end{multline}
The system in \eqref{eq:we_marangoni_limited_leading_order_weak_diff} requires $\Gamma_{00} = 0$ for $-\phi_x \leq x \leq x_0$ and $\Gamma_{00} = \beta(x - x_0)/\gamma$ for $x_0 \leq x \leq 
\phi_x$, for $\gamma\leq\beta\phi_x$. 
This nonlinear profile typically denotes the presence of the stagnant cap profile \citep{he1991size}.
The constant $x_0 = \phi_x - 2(\phi_x \gamma/\beta)^{1/2}$ is evaluated by the no-net-flux condition $\int_{x = -\phi_x}^{\phi_x} \Gamma_{00} \, \text{d} x = 2\phi_x$ as $c_{00}=1$. 
We then evaluate $\Delta \Gamma_0 = 2 (\phi_x \beta / \gamma)^{1/2}$ and substitute it into \eqref{eq:dr_def} to find ${DR}_0 \approx 0.5$ when $\gamma = \beta \phi_x/4$, as in \eqref{eq:we_50_2}.

To extend the region of validity of the above solution, we retain the surface diffusion term at leading-order, such that $c_{00} = 1$ in $\mathcal{D}_1\cup \mathcal{D}_2$ but now (\ref{eq:we_marangoni_limited_leading_order_weak_diff}\textit{b,\,g}) become
\refstepcounter{equation} 
\label{eq:we_marangoni_limited_leading_order_weak_diff_corr}
\begin{multline} 
    \beta \Gamma_{00} - \gamma \Gamma_{00} \Gamma_{00x} - \delta \Gamma_{00x} = 0 \quad \text{in} \quad \mathcal{D}_1, \\ \text{subject to} \quad \beta\Gamma_{00}(\pm \phi_x) - \gamma \Gamma_{00}(\pm \phi_x) \Gamma_{00x}(\pm \phi_x) - \delta \Gamma_{00x}(\pm \phi_x) = 0.
    \tag{\theequation\textit{a--b}}
\end{multline}
We can integrate \eqref{eq:we_marangoni_limited_leading_order_weak_diff_corr} term by term, using $\int_{x=-\phi_x}^{
\phi_x}\Gamma_{00} \, \text{d} x = 2\phi_x$ and the fact that weak surface diffusion makes $\Gamma_{00}(-\phi_x)$ exponentially small (recall that $\Gamma_{00}(-\phi_x)=0$ when there was no surface diffusion), to calculate $\Delta \Gamma_0$, substitute this into \eqref{eq:dr_def} to get 
\begin{equation}
    {DR}_0 = 1 + \frac{\delta}{2\phi_x \beta} - \left(\frac{\gamma}{\phi_x \beta} + \left(\frac{\delta}{2\phi_x \beta}\right)^2\right)^{1/2}.
\end{equation}
We therefore have that ${DR}_0=0.5$ when $\gamma = \beta\phi_x/4 + \delta / 2$, as in \eqref{eq:we_50_2}; we have now evaluated the leading and first-order correction in the limit where Marangoni effects are strong and diffusion is weak.

\subsection{Weak Marangoni effect: region $\text{A}_\text{E}$} \label{subsec:a_we_sd_d}

Assume that $\beta = O(1)$, $\gamma \ll \min(1,\,\alpha,\,\delta)$ and $ \nu/\epsilon^2 \ll \min(1,\,\alpha,\,\delta)$.
Rescale $ \nu/\epsilon^2 = n \gamma$ where $n = O(1)$.
Expand the concentration field from \S\ref{subsec:weak_exchange} using $\Gamma_0 = \Gamma_{00} + \gamma \Gamma_{01} + ...$ and $c_0 = c_{00} + \gamma c_{01} + ...$ into \eqref{eq:model1}--\eqref{eq:model3}.
At $O(1)$, 
\refstepcounter{equation} 
\label{eq:we_marangoni_weak_leading_order}
\begin{multline} 
    (c_{00} - \alpha c_{00x})_x = 0, \quad \beta \Gamma_{00} - \delta \Gamma_{00x} + c_{00} - \alpha c_{00x} = 1 \quad \text{in} \quad \mathcal{D}_1, \\ c_{00} - \alpha c_{00x} = 1 \quad \text{in} \quad \mathcal{D}_2, \quad \text{subject to} \quad c_{00}(\phi_x^{-}) = c_{00}(\phi_x^{+}), \quad c_{00}(-\phi_x) = c_{00}(2 - \phi_x), \\
    c_{00}(\pm \phi_x) - \alpha c_{00x}(\pm \phi_x) = 1, \quad \beta \Gamma_{00}(\pm \phi_x)  - \delta \Gamma_{00x}(\pm \phi_x) = 0.
    \tag{\theequation\textit{a--g}}
\end{multline}
Integrating \eqref{eq:we_marangoni_weak_leading_order} on $\mathcal{I}$ and making use of the no net-flux condition, $\int_{x=-\phi_x}^{
\phi_x}\Gamma_{00} \, \text{d} x = 2\phi_x$, we obtain \eqref{eq:we_d}. 
Therefore, $\Delta \Gamma_0 \approx 2 \beta \phi/\delta$, which is substituted into \eqref{eq:dr_def} to give \eqref{eq:we_dr_d}.
The corrections to $\Gamma_0=\Gamma_{00}$ and ${DR}_0 = 1$ are small provided $\gamma \ll \min(1,\,\alpha,\,\delta)$ and $\nu/\epsilon^2 \ll \min(1,\,\alpha,\,\delta)$, defining the boundaries of the asymptotic region $\text{A}_\text{E}$ and $\text{D}_\text{E}$.

\bibliographystyle{jfm}
\bibliography{jfm-instructions}

\end{document}